# Wide-Field InfraRed Survey Telescope
# WFIRST
# Final Report

## Science Definition Team

J. Green[1], P. Schechter[2]

C. Baltay[3], R. Bean[4], D. Bennett[5], R. Brown[6], C. Conselice[7], M. Donahue[8], X. Fan[9], B. S. Gaudi[10], C. Hirata[11], J. Kalirai[6], T. Lauer[12], B. Nichol[13], N. Padmanabhan[3], S. Perlmutter[14], B. Rauscher[15], J. Rhodes[16], T. Roellig[17], D. Stern[16], T. Sumi[18], A. Tanner[19], Y. Wang[20], D. Weinberg[10], E. Wright[21], N. Gehrels[15], R. Sambruna[22], W. Traub[16]

Consultants
J. Anderson[6], K. Cook[23], P. Garnavich[5], L. Hillenbrand[11], Z. Ivezic[24], E. Kerins[25], J. Lunine[4], P. McDonald[14], M. Penny[10], M. Phillips[26], G. Rieke[9], A. Riess[27], R. van der Marel[6]

R.K. Barry[15], E. Cheng[28], D. Content[15], R. Cutri[29], R. Goullioud[16], K. Grady[15], G. Helou[29], C. Jackson[30], J. Kruk[15], M. Melton[15], C. Peddie[15], N. Rioux[15], M. Seiffert[16]

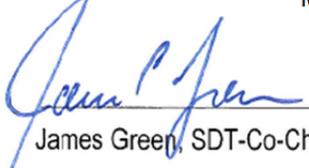

James Green, SDT Co-Chair    Date    Aug 15, 2012

Paul Schechter

Paul Schechter, SDT Co-Chair    Date

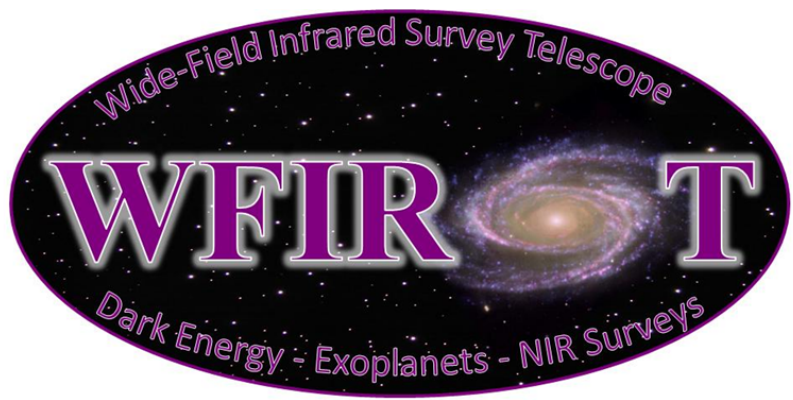

1  University of Colorado/Center for Astrophysics and Space Astronomy
2  Massachusetts Institute of Technology
3  Yale University
4  Cornell University
5  University of Notre Dame
6  Space Telescope Science Institute
7  University of Nottingham
8  Michigan State University
9  University of Arizona
10 Ohio State University
11 California Institute of Technology
12 National Optical Astronomy Observatory
13 University of Portsmouth
14 University of California Berkeley/Lawrence Berkeley National Laboratory
15 NASA/Goddard Space Flight Center
16 Jet Propulsion Laboratory/California Institute of Technology
17 NASA/Ames Research Center
18 Osaka University
19 Mississippi State University
20 University of Oklahoma
21 University of California Los Angeles
22 NASA Headquarters
23 Lawrence Livermore National Laboratory
24 University of Washington
25 University of Manchester
26 Las Campanas Observatory
27 Johns Hopkins University
28 Conceptual Analytics
29 Infrared Processing and Analysis Center/ California Institute of Technology
30 Stinger Ghaffarian Technologies

# EXECUTIVE SUMMARY

The Wide Field Infrared Survey Telescope (WFIRST) is the highest ranked recommendation for a large space mission in the New Worlds, New Horizons (NWNH) in Astronomy and Astrophysics 2010 Decadal Survey. The most pressing scientific questions in astrophysics today require a very wide field survey in order to be answered, and existing telescopes such as the Hubble Space Telescope (HST) or the Keck Telescope cannot make these kinds of observations due to their optical designs and narrow fields of view. The first generation of wide field surveys from the ground (*e.g.*, SDSS) has resulted in significant advances in astronomy and astrophysics, and the value of such surveys is recognized by the parallel decadal recommendation of LSST, a ground-based telescope designed for survey work. Translating wide field design principles into space will revolutionize astronomical surveys in much the same way that the HST revolutionized the imaging of individual astronomical objects and galaxies. The absence of atmospheric distortion and absorption, the darkness of space, and the opportunity for continuous observations enable space-based surveys to go deeper, with more precision and accuracy, than can ever be possible from the ground. WFIRST will be the most sensitive infrared space telescope designed for precision wide-field survey work, and will produce the most powerful and informative astronomical data set for probing the nature of dark energy, cataloguing the variety of exoplanet systems, and mapping the distribution of matter across cosmic time.

The WFIRST Science Definition Team (SDT) was formed to refine the science case for the mission, optimize the design and implementation scheme, and develop two Design Reference Missions (DRMs) to serve as the basis for further programmatic and technical review. The first DRM (DRM1) represents the completion of the Interim Design Reference Mission (IDRM), and was scoped to meet all of the scientific and observational requirements outlined in the decadal survey. It utilizes existing, proven technology, and could move into design and development immediately if a new start were to be approved.

DRM1 incorporates a 205K, 1.3m, unobstructed, three mirror anastigmat telescope. It utilizes current technology infrared sensors operating from 0.7 – 2.4 μm, and fits within a medium lift class launch vehicle fairing. DRM1 assumes that WFIRST will be deployed in an Earth-Sun L2 libration point orbit, and have a minimum operational lifetime of 5 years, but will have no design elements (*e.g.*, propellant supply) that intrinsically limit the lifetime to less than 10 years.

DRM2 was requested by NASA in response to the European selection of the Euclid dark energy mission, which is scheduled to launch in 2020. DRM2 is designed to be non-duplicative with the capabilities of the Euclid mission and/or planned ground based facilities such as LSST, and also to represent a lower cost alternative to DRM1, though possibly with a reduced science return compared to DRM1. A recent committee of the National Research Council examined the potential overlap in the WFIRST and Euclid missions and found that there was no duplication in their science capabilities (NRC 2012). However, to reduce mission costs, DRM2 employs a 1.1 m telescope (still a three mirror anastigmat), utilizes a smaller number of higher performance infrared sensors operating from 0.7 – 2.4 μm (H4RG-10's, which require further development before being ready for launch, but have four times as many pixels in a similar sized package), uses an observatory design with only selective redundancy, and fits within a lower cost, lower throw weight launch vehicle. DRM2 assumes that WFIRST will be deployed in an Earth-Sun L2 libration point orbit, and have a minimum operational lifetime of 3 years. With the improved sensors, DRM2 has similar sensitivity to DRM1 for a fixed observation time, and can achieve any, but not all, of the NWNH science objectives within its 3 year design lifetime. It is assumed that the actual observing plan for a WFIRST utilizing the DRM2 mission concept will be optimized based on the data available at that time.

**Both DRM1 and DRM2 represent exciting and viable mission concepts. DRM1 would accomplish all of the goals of the decadal survey, at a slightly lower cost than originally anticipated. DRM2 would accomplish a substantial fraction of these, and would be considerably less expensive. The WFIRST SDT recommends expedited development of both concepts, with a decision to go ahead with one or the other *prior* to the NRC's mid-decadal review. The SDT's immediate short term recommendation is to further improve the IR detector capability as the most cost effective way to improve the science return of either design, and to continue this effort as the top priority for WFIRST in the years 2013-2015.**

The SDT has refined the scientific objectives of WFIRST to four cornerstone goals, of equal importance:

1) Complete a statistical census of planetary systems in the Galaxy, from the outer habitable zone to free

floating planets, including analogs to all of the planets in our Solar System with the mass of Mars or greater.

2) Determine the expansion history of the Universe and the growth history of its largest structures in order to test possible explanations of its apparent accelerating expansion including Dark Energy and modifications to Einstein's gravity. This objective must be achieved through multiple techniques, and must include a galaxy redshift survey, a SN Ia survey, and a gravitational weak lensing survey.

3) Produce a deep map of the sky at near-infrared wavelengths, enabling new and fundamental discoveries ranging from mapping the Galactic plane to probing the reionization epoch by finding bright quasars at z>10.

4) Provide a general observer program utilizing a minimum of 10% of the mission minimum lifetime.

The SDT re-affirms, even after the selection of the Euclid mission, that WFIRST must be capable of executing any and all of its cornerstone science objectives, including all of the observational techniques listed above, which are necessary to achieve a highly reliable result. However, the SDT recognizes that in the intervening years between now and launch, the actual allocation of observing times to the various approaches will be determined in response to the work that will be carried out in space and on the ground before WFIRST begins its science operations.

Many of the most pressing questions in astrophysics today, such as the nature of dark energy, the mechanisms of galaxy evolution, and the demographics of planetary systems, require large, uniform statistical samples -- observations of thousands of square degrees of sky or rapid-cadence observations of a few square degrees. WFIRST is designed to provide these samples, taking advantage of the stability and dark infrared sky available only in space. It is a very different mission from the James Webb Space Telescope (JWST), which is designed to provide much more detailed information on small numbers of objects. No other existing or planned instrumentation can produce a wide field infrared sky survey that approaches the depth and resolution that WFIRST will achieve.

**Box 1**

"WFIRST is a wide-field-of-view near-infrared imaging and low-resolution spectroscopy observatory that will tackle two of the most fundamental questions in astrophysics: Why is the expansion rate of the universe accelerating? And are there other solar systems like ours, with worlds like Earth? In addition, WFIRST's surveys will address issues central to understanding how galaxies, stars, and black holes evolve. ..... WFIRST will settle fundamental questions about the nature of dark energy, the discovery of which was one of the greatest achievements of U.S. telescopes in recent years. It will employ three distinct techniques—measurements of weak gravitational lensing, supernova distances, and baryon acoustic oscillations—to determine the effect of dark energy on the evolution of the universe. An equally important outcome will be to open up a new frontier of exoplanet studies by monitoring a large sample of stars in the central bulge of the Milky Way for changes in brightness due to microlensing by intervening solar systems. This census, combined with that made by the Kepler mission, will determine how common Earth-like planets are over a wide range of orbital parameters. It will also, in guest investigator mode, survey our galaxy and other nearby galaxies to answer key questions about their formation and structure, and the data it obtains will provide fundamental constraints on how galaxies grow."

From the New Worlds, New Horizons Decadal Survey in Astronomy and Astrophysics

The SDT carefully considered the relationship between WFIRST, the Euclid mission, and the ground-based LSST. Euclid has two instruments: a high-resolution visible imager, and a NIR imaging/spectroscopy instrument. Euclid's design was optimized to maximize survey area, and other objectives (such as reaching diffraction-limited resolution in the NIR) were traded away to make this possible. Although its survey footprint will be smaller, WFIRST will observe 2 magnitudes deeper, achieve diffraction-limited resolution for a 1.1-1.3 m telescope, and include the K band.

The K band is a key instrument capability with science implications ranging from charting the expansion history of the Universe at z>2 to mapping the highly reddened Galactic Plane.

LSST is dedicated to imaging surveys in the optical bands (out to ~1 micron). WFIRST was specifically designed not to duplicate the optical survey capabilities of LSST. Instead, with the well-matched depths of the surveys (Figure 19), LSST optical imaging and WFIRST infrared imaging will provide an unprecedented panchromatic view of several thousands of square degrees of sky.

WFIRST will be one of the pre-eminent astronomical science machines of the 2020's if approved for development. While the cost of WFIRST, even if its DRM2 version is selected, is anticipated to be that of a large facility class mission, it will be a fraction of the cost of its brethren telescopes, HST and JWST. In telescope size and instrument complexity, WFIRST is more directly comparable to the recent, highly successful Kepler mission. It is the unanimous recommendation of the WFIRST SDT that the preliminary studies we have begun be allowed to proceed to development and flight as rapidly as programmatic realities allow, and the SDT reaffirms the Astro 2010 decision that development of WFIRST is the highest priority for space astronomy in the coming decade.

# 1 INTRODUCTION

## 1.1 Background

The acronym WFIRST stands for Wide Field Infra-Red Survey Telescope. The NRC's 2010 decadal survey of astronomy and astrophysics, "New Worlds, New Horizons" (henceforth NWNH) gave such a telescope the highest priority for a large space mission in the decade 2010-2020. In December 2010 the Director of the Astrophysics Division of NASA's Science Mission Directorate appointed a Science Definition Team (henceforth SDT) charged with producing a design reference mission (henceforth DRM) for WFIRST. This document fulfills that charge.

Part of the original charge was to produce an interim design reference mission by mid-2011. That document was delivered to NASA and widely circulated within the astronomical community. In late 2011 the Astrophysics Division augmented its original charge, asking for *two* design reference missions. The first of these, henceforth DRM1, was to be a finalized version of the interim DRM, reducing overall mission costs where possible. The second of these (henceforth DRM2), was to identify and eliminate capabilities that overlapped with those of NASA's James Webb Space Telescope (henceforth JWST), ESA's Euclid mission, and the NSF's ground-based Large Synoptic Survey Telescope (henceforth LSST), and again to reduce overall mission cost, while staying faithful to NWNH. This report presents both DRM1 and DRM2.

The NWNH science goals for WFIRST are quite broad. They include: tiered infrared sky surveys of unprecedented scope and depth; a census of cold (as opposed to hot) exoplanets using microlensing; measurements of the history of cosmic acceleration using three distinct but interlocking methods; a galactic plane survey, and a general observer/archival research program. What brought these very different science goals together was the realization, across the astronomical community, that recent advances in infrared detector technology have, for the first time, made it possible to launch a wide field infrared telescope with a very large number of diffraction limited "effective pixels" in the focal plane.

In October 2011 ESA selected Euclid (see Laureijs et al. 2011), as one of two Cosmic Vision medium class missions. Euclid is also a wide field telescope, but the majority of its detectors (36 of 52) work at optical rather than infrared wavelengths. While its optical pixels properly sample diffraction limited images, its infrared pixels are substantially bigger and give back the angular resolution that is gained by going into space. In January 2012 the NRC's *Committee on the Assessment of a Plan for US Participation in Euclid* recommended a modest contribution to Euclid, saying "This investment should be made in the context of a strong U.S. commitment to move forward with the full implementation of WFIRST in order to fully realize the decadal science priorities of the NWNH report."

Also in October 2011, the Nobel Prize in Physics was awarded to Saul Perlmutter (a member of the SDT), Adam Riess (a consultant to the SDT) and Brian Schmidt "for the discovery of the accelerating expansion of the Universe through observations of distant supernovae." The mechanism of that acceleration is unknown. Three of the mandated WFIRST observing programs are aimed at measuring the acceleration history of the Universe over a wide range of redshifts, so as to narrow the range of possible mechanisms.

In June 2012, NASA and ESA signed an MOU for US participation in Euclid. The US is to provide near-infrared detectors, and ESA is to provide a slot for a US scientist on the Euclid Science Team and up to 40 slots for US scientists in the Euclid Consortium.

| Box 2 | |
|---|---|
| **Key Events for WFIRST** | |
| Aug. 2010 | NWNH report from NRC |
| Dec. 2010 | SDT appointed by NASA |
| June 2011 | IDRM report delivered by SDT |
| Oct. 2011 | Euclid selected by ESA |
| Dec. 2011 | 2 DRMs requested from SDT by NASA |
| Jan. 2012 | NRC affirms that WFIRST is non-duplicative of Euclid and is needed |
| June 2012 | MOU for US participation in Euclid signed by NASA and ESA |
| July 2012 | DRM1 and DRM2 report delivered by SDT; SDT disbanded |

## 1.2 Design Reference Mission #1

Consistent with its charge, the SDT herein presents two design reference missions. DRM1 has evolved from the interim DRM in two important ways. First, the IDRM carried out slitless spectroscopy using two spectroscopic channels, each of which had its own focal plane array and its own tertiary in addition to those for the imaging channel. These have been eliminated in DRM1, leaving only a single focal plane array and tertiary. In their stead, dispersing elements can be inserted and withdrawn from the beam to carry out the needed slitless spectroscopy. This produces gains both in simplicity and in flexibility.

Second, the limiting wavelength of the detectors in DRM1 is 2.4 microns rather than 2 microns. This carries advantages for many of the observing programs proposed. The combination of these two changes allows a deeper baryon acoustic oscillation survey (albeit over a smaller solid angle), measuring the acceleration history of the universe at higher redshifts than was possible with the IDRM or than is possible with Euclid.

JDEM-Omega was an axisymmetric 1.5 meter wide field infrared telescope described at length in one of the responses to the Astro2010 "requests for information." NWNH singled it out among the various similar proposed telescopes as the template for WFIRST. The IDRM itself differs significantly from JDEM-Omega in that it has an unobstructed 1.3 meter telescope, which offers a cleaner and more compact point spread function than an obstructed 1.5 meter, with no loss in collecting area and no added cost. This is especially important for the measurement of the cosmic acceleration history using weak lensing. Both DRM1 and DRM2 are likewise unobstructed.

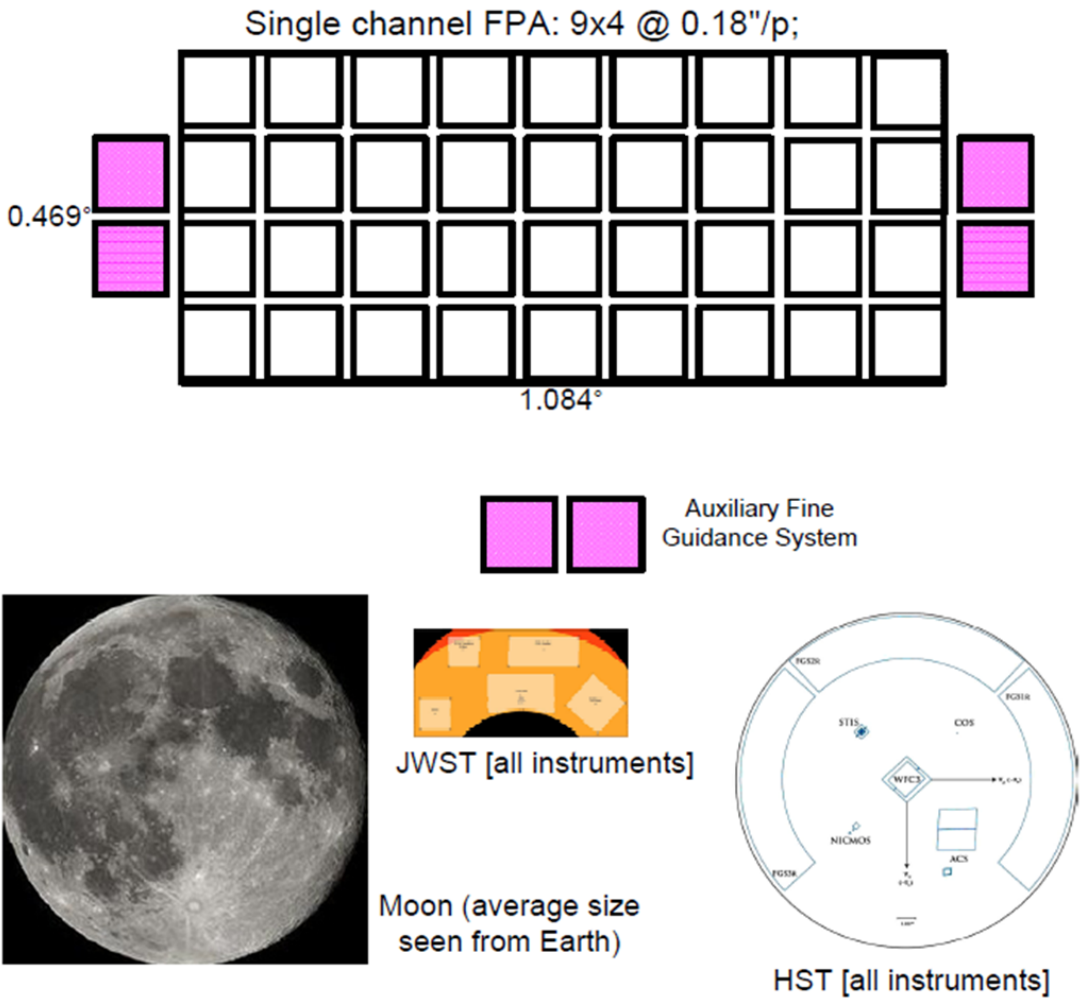

**Figure 1: Comparison of WFIRST DRM1 field of view with the Moon, JWST and HST**

## 1.3    Design Reference Mission #2

In producing the second DRM, the SDT debated at considerable length which, if any, of the capabilities of WFIRST duplicated those of other upcoming astronomical efforts. LSST is also a wide-field telescope, but it works at wavelengths shorter than 1 micron, generally (if inaccurately) characterized as "optical" wavelengths. It has a much larger primary mirror but its images suffer from atmospheric "seeing," resulting in much lower spatial resolution and image stability.

JWST has 18 primary mirror segments, each of which is bigger than WFIRST's primary mirror. But whereas the imager on JWST has 33 million pixels, WFIRST's DRM1 has 150 million pixels. JWST has a much smaller diffraction limit, and will go much deeper than WFIRST, which is designed for breadth rather than depth. JWST is a telephoto lens to WFIRST's wide angle lens, with less than 1% of the total instantaneous sky coverage of WFIRST (see Figure 1).

The Euclid Definition Study Report, better known as the Red Book, Laureijs et al. (2011), describes a mission in which the first 6.25 years of operation are devoted entirely to the study of cosmic acceleration, using two of the same techniques that WFIRST would use: weak gravitational lensing and baryon acoustic oscillations (henceforth BAO). But in the case of weak lensing, WFIRST works in three infrared bands rather than a single wide optical band, and in the case of BAO, WFIRST works at higher redshift. The NRC Euclid committee wrote:

"Euclid's and WFIRST's measurements are *not duplicative* (emphasis added) and the combinations will be more powerful than any single measurement. Combining WFIRST with Euclid and ground-based data sets, such as that expected from LSST, should further enable astronomers to address the systematic challenges that previous ground-based weak lensing measurements have experienced. These combined data sets will likely overcome systematic limitations and realize the full potential of this powerful technique."

Euclid can, in principle, discover exoplanets using microlensing and study cosmic acceleration history using supernovae. Both of these are described in the Red Book as options for a possible extension of the mission beyond the initial 6.25 years. But they are subordinated to weak lensing, which must be carried out as quickly as possible following launch to mitigate the effects of charged particle damage to Euclid's optical CCDs. Moreover, both exoplanet microlensing and supernova cosmography are more efficiently and effectively carried out in the infrared, as is the case with WFIRST. Euclid has no general observer program. Euclid will produce an infrared sky survey, but its "fat" infrared pixels do not deliver nearly diffraction limited NIR images. Additionally, the fast, read noise limited survey mode required to achieve Euclid's sky coverage results in a survey 2 magnitudes shallower than that planned for WFIRST.

The SDT debated whether any of six NWNH-mandated components of the WFIRST observing program might be expendable in light of the above mentioned efforts:

1) Exoplanet microlensing;

2) Galactic plane infrared survey;

3) General observer/archival research (GO/AR) program;

4) Cosmic acceleration history via supernovae;

5) Cosmic acceleration and structure growth via BAO and redshift-space distortions; and

6) Cosmic acceleration history and structure growth via weak lensing. The imaging survey conducted for weak lensing also produces the high-latitude infrared sky survey applicable to a wide range of extragalactic and Galactic science investigations.

Strenuous and cogent objections were raised by at least several SDT members to declaring any one of these to be duplicative. Furthermore, the cost savings from dropping any one of these capabilities from the mission appear tiny relative to the overall mission cost.

The cost of a wide field infrared survey mission is driven largely by the number of photons collected, rather than by the observing program. The number of photons collected is proportional to the collecting area and the mission duration. DRM2 cuts costs by reducing both of these, with a 1.1 m primary and a three year, rather than five year, mission. While DRM2 can carry out each of the above programs, it cannot accomplish as much as DRM1 given its reduced aperture and shortened mission life time. We describe a version of DRM2 that essentially multiplies the DRM1 observing allocations by 0.6 across-the-board, but the decision as to whether to cut all programs equally or to give priority to one or another is one that can and should be made by a program committee on a date closer to launch. DRM2 cannot carry out all the science called for in NWNH in its prime mission, but it would be able to do so in an extended mission. It is consistent in spirit with the conclusion of NWNH that a wide field infrared survey telescope would be broadly useful to many different constituencies within the astronomical community.

The enabling technology, without which DRM2 could not accomplish even its more limited goals, is the incorporation of 4096 x 4096 pixel H4RG-10 detector arrays, rather than the 2048 x 2048 H2RG-18 detector arrays baselined in the IDRM and in DRM1. By packing 4x as many pixels into about the same detector array area (see Figure 2), H4RG-10 detectors permit coverage of a larger solid angle at reduced cost per unit solid angle. But as they are not yet at an appropriate technical readiness level, DRM2 can only go forward if a timely effort is made to bring the H4RG-10 devices to maturity.

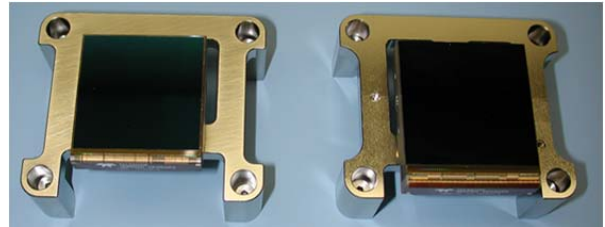

**Figure 2: Comparison of size of an H2RG (2K x 2K pixels, 18 μm pixel size) on left to an H4RG (4K x 4K pixels, 10 μm pixel size).**

| Box 3 | | | | | | |
|---|---|---|---|---|---|---|
| **WFIRST Design Incarnations** | | | | | | |
| Version | CATE date | Primary Mirror Diameter (m) | Obstructed Telescope | Red Limit (μm) | Science Focal Planes | Science Detectors |
| JDEM-Ω | 2010 | 1.5 | Yes | 2.0 | 3 | 36 H2RG-18 |
| IDRM | 2011 | 1.3 | No | 2.0 | 3 | 36 H2RG-18 |
| DRM1 | N/A | 1.3 | No | 2.4 | 1 | 36 H2RG-18 |
| DRM2 | 2012 | 1.1 | No | 2.4 | 1 | 14 H4RG-10 |

| WFIRST Science Program |
| --- |
| Complete the statistical census of planetary systems in the Galaxy |
| Determine the expansion history of the Universe and the growth history its largest structures |
| Perform a deep NIR survey of the Galactic and extra-Galactic sky |
| Execute a General Observer Program |

| Baseline Survey Characteristics[1] | | | | | |
| --- | --- | --- | --- | --- | --- |
| Survey | Bandpass | Area (deg$^2$) | Depth[2] | Duration | Cadence |
| Exoplanet Microlensing | Y,W | 3.38 | n/a | 1.2 years (72 days x 6) | W:15 min Y:12 hrs |
| Galactic Plane | Y,J,H,K | 1240 | 25.1 | 0.45 years | n/a |
| High Latitude Survey (HLS)[3] | Y,J,H,K | 3400 | 26.0 | 2.4 years | n/a |
| | GRS Prism | 3400 | $1.0 \times 10^{-16}$ | | n/a |
| Supernova (SN) Survey | J,H,K | 6.5 / 1.8 (wide/deep) | 28.1 / 29.6 27.6 / 28.5 | 0.45 years (in 1.8 year interval) | 5 days |
| | SNe Prism | | | | |

| Payload | | | | |
| --- | --- | --- | --- | --- |
| Telescope | Aperture 1.3m | Form Unobstructed TMA | Focal Ratio 15.9 | Plate Scale 0.18"/pixel |
| Focal Plane | Detectors HgCdTe H2RG | Layout 9x4 [150 Mpix] | Detector Cutoff 2.5 μm | Active area 0.375 deg$^2$ |
| Filters (μm) | Z 0.73-0.962 | Y 0.92-1.21 | J 1.156-1.52 | H 1.453-1.91 | K 1.826-2.4 | W 0.92-2.40 |
| Prisms[4] | SN Ia | | Galaxy Redshift Survey (GRS) | |
| | R=75 | 0.6-2.0 μm | R=600 | 1.5-2.4 μm |

| Spacecraft | | | |
| --- | --- | --- | --- |
| Orbit | Sun-Earth L2 halo | | |
| Pointing control | Stability 40 mas, rms/axis | Revisit accuracy 25 mas, rms/axis | Slew+settle time[5] <30 sec |
| Station keeping, Momentum management | hydrazine thrusters | | |
| Communications | CMD, RT TLM S-band | Data downlink Ka band 150Mbps | Network DSN |
| Data Volume | SSR Storage 1.1 Tb | Daily Data Volume 1.3 Tb | Contacts per Day 2 |
| Power | Solar Array Output 2250 W EOL | Average Draw 1150W | Battery 80 A-h |
| Mass | 2509 kg | | |

**Table 1: Characteristics of the WFIRST DRM1 hardware and observing programs**
[1] An additional 0.5 years is dedicated to a General Observer program
[2] Imaging: $5\sigma$ point source; H-band AB magnitudes are listed, other bands are similar; Galaxy redshift survey: $7\sigma$ minimum detectable line flux in erg/cm$^2$/sec; SN spectroscopy: continuum AB at which S/N=1 per pixel. Depths are cumulative over mission. Single-visit depths in SN Ia program are 2.5-3 mags brighter.
[3] High Latitude Survey
[4] R = mean point source resolving power for 2-pixel resolution element, single SN prism, 2 oppositely dispersing GRS prisms
[5] Slew and settle time for typical survey step

## 2    THE WFIRST SCIENCE PROGRAM

### 2.1    Science Overview

This science overview subsection begins with a narrative description of how WFIRST will use visible and near-infrared surveys of large parts of the sky to make dramatic progress in fundamental areas of astrophysics, providing new insights into the workings of our Universe, and giving the public a taste of the excitement of astronomy as well as a renewed appreciation of our astrophysical origins and destiny. It concludes with a selection of specific numerical examples of WFIRST's astrophysical capabilities.

### a) Exoplanet Microlensing

Planets orbiting stars produce minute Doppler shifts in the spectra of their host stars and, when they transit in front of them, tiny dips in the host's brightness. Measurements of these shifts and dips have been used with great success to discover large numbers of hot exoplanets -- those that are hotter than the Earth by virtue of being relatively close to their hosts. We have learned that exoplanetary systems are, in general, quite different from our own Solar System.

Doppler shifts and transits are far less effective at finding cold exoplanets -- those that are further from their hosts than the Earth. The Doppler shifts are small, the transit probabilities are small, and the periods are long. This is especially unfortunate because many (indeed probably most) planets are born cold, with some of these migrating into or being scattered into hotter orbits.

Fortunately a third technique, gravitational microlensing, is most sensitive to cold exoplanets. It involves high cadence observations of a very large number of distant stars. The gravitational field of the occasional planetary system that passes in front of one of these distant stars produces changes in that star's brightness that can be used to determine the mass of an exoplanet and its distance from its host.

Both a wide field and a high stellar density are needed to observe a large number of distant stars simultaneously. While the densest stellar fields are obscured by dust, this is a less serious problem in the infrared than in the optical. WFIRST will complete the census of exoplanets begun by Kepler, extending that census to planets at and beyond a star's Habitable Zone and extending to planets that were ejected beyond the gravitational pull of their original star.

### b) High Latitude Imaging and Spectroscopy

WFIRST will carry out three distinct infrared surveys at high latitude, looking above and below the plane of our Galaxy. The first is a wide area imaging survey in four different filters. The second is a slitless spectroscopic survey of the same area. The third is a smaller area survey for supernovae, which, by virtue of the large number of repeat visits, will also produce a very deep, co-added imaging survey. These three surveys enable the cosmic acceleration investigations discussed below, but they also enable an enormous range of other astronomical studies. The broad value of wide area surveys is evidenced by the huge impact of the Sloan Digital Sky Survey (SDSS) at visible wavelengths. WFIRST will play this role in the infrared, reaching much deeper into the universe than Sloan and with much higher angular resolution.

Among the many phenomena to be studied will be the assembly of galaxies and clusters of galaxies from a time when the universe was less than 10 percent of its present age up to today. WFIRST will study the growth of supermassive black holes at the centers of galaxies, back to when the universe was 5 percent of its present age, but they are sufficiently rare that one must search a wide area, and their radiation is shifted so far to the red that one must observe in the infrared.

### c) Low Latitude Survey

It is ironic that we know the structure of the Milky Way, the "low-latitude" part of our sky, far less well than that of distant galaxies. This is the consequence of dust that prevents visible light from reaching us, and of the atmospheric blurring of stellar images that causes them to overlap in high density fields. By virtue of providing diffraction limited images at infrared wavelengths, WFIRST will penetrate further into the depths of the Milky Way than has heretofore been possible, helping to map out the bar and spiral arms that have until now evaded decisive measurement.

### d) Cosmic Acceleration via Supernovae, Baryon Acoustic Oscillations, Weak Gravitational Lensing, and Redshift-Space Distortions

Many astronomers would argue that the discovery of the acceleration of cosmic expansion was the single most important astronomical discovery of the last third of the twentieth century (with the discovery of exoplanets being the strongest competition for this title). The 2011 Nobel Prize in Physics was awarded for its discovery, which was carried out using supernovae as standard candles. There are two principal competing explanations for this acceleration: a heretofore undetected constituent of the universe with a negative pressure, commonly referred to as "dark energy"; or a modification to Einstein's theory of General Relativity on large scales. By studying the history of cosmic acceleration and structure growth we can discriminate between the two and determine the properties of dark energy or the nature of the needed correction to General Relativity.

WFIRST will make precision measurements of cosmic acceleration over 75 percent of the age of the Universe. Four different methods will be used: supernovae, baryon acoustic oscillations (BAO), weak gravitational lensing, and redshift-space distortions. Each method is strongest at a different time in cosmic history, but they overlap, permitting careful checking for possible sources of error. Supernovae give the greatest precision at low redshifts and BAO are best at high redshifts. Weak lensing measurements overlap with both of these, and produce the strongest tests of General Relativity. Redshift-space distortions sharpen the precision of BAO distance measurements and provide an alternative test of General Relativity.

**e) General Observer Program**

Roughly six months of WFIRST observations will be reserved for executing observations proposed as part of a general observer program. We anticipate extensive observations of nearby galaxies looking for traces of their catastrophic ingestion of smaller galaxies. We expect programs to determine the ages of globular clusters with 1 percent precision, taking advantage of features of the Hertzsprung-Russell diagram that are peculiar to the infrared. We also anticipate programs to find Kuiper Belt Objects, the not-quite-planets of which Pluto is the most famous, far beyond the orbit of Pluto.

**f) Numerical Examples**

One illustration of WFIRST's remarkable reach is to compare it to the most ambitious survey programs from Hubble Space Telescope. As an example, Figure 3 shows the fields of the various components of the Multi-Cycle Treasury program CANDELS superposed on the WFIRST field of view. The entirety of CANDELS fits easily within one WFIRST pointing. The "wide" CANDELS survey[1] covers 688 arcmin$^2$ to IR depths of $H_{AB}$ = 27 in (approximately) Y, J, and H bands. The standard depth of the imaging portion of the High Latitude Survey (HLS), with exposure times of ≈ 750 sec/band, is 1 mag shallower, but it includes K-band, and the HLS is larger in area by a factor of 18,000! Achieving CANDELS "wide" depth over one WFIRST pointing (1350 arcmin$^2$) takes a total of 1 hour in each filter.

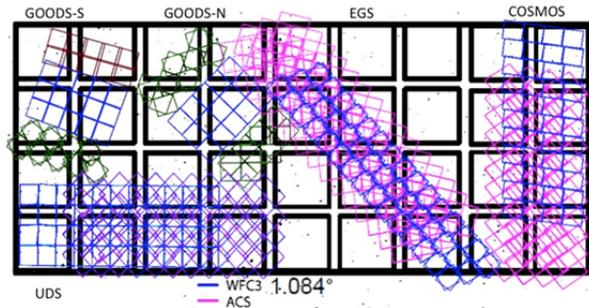

**Figure 3: CANDELS fields superimposed on the WFIRST field of view**

The WFIRST supernova survey is constructed in two tiers, with exposure times designed to measure the flux of a Type Ia SN near the peak of its light curve at S/N > 15 out to z = 0.8 (shallow, 300 sec exposures in J, H, K) or z = 1.7 (deep, 1500 sec exposures). The imaging observations, which are carried out every five days over a period of 1.8 years, can be co-added to achieve AB-magnitude depths of ≈ 28.1 (shallow) and 29.6 (deep). These limits are, respectively, 0.7 mag and 2.2 mag deeper than the IR depth of the CANDELS *deep* survey (area 120 arcmin$^2$), with survey areas that are larger by factors of ≈ 200 and ≈ 50, respectively.

The most direct comparison for the spectroscopic portion of the HLS is the 3D-HST survey[2], which is using the grism capability of WFC3 to obtain 1.1-1.7 μm spectra over 600 arcmin$^2$ of the CANDELS area (plus ACS grism spectra at shorter wavelengths). The 7σ line detection flux limit of $10^{-16}$ erg/cm$^2$/sec of the HLS is about a factor of 2 shallower than the corresponding limit for 3D-HST (~ 5-7 × $10^{-17}$ erg/cm$^2$/sec), but the area is larger by a factor of 20,000, and the longer wavelength range (1.5 – 2.4 μm) of the WFIRST spectroscopy allows it to reach higher redshift.

The HLS will be an incredible resource for extragalactic studies. It will detect more than a thousand quasars at redshift z>7, emitting less than a Gyr after the Big Bang, and provide probes of and to a time before the universe became reionized. WFIRST is even expected to detect quasars at z>10, corresponding to less than 500 Myr after the Big Bang. The HLS will also

---

probe the mass, environment, and star-formation rates for hundreds of millions of galaxies out to high redshifts, allowing an unprecedented probe of galaxy evolution across most of cosmic time. At the depth of the HLS imaging survey, one can image a 0.1L* star-forming galaxy at z=2 with S/N ≈ 5 and characterize its rest-frame optical morphology with 20-30 pixels within the optical radius; for a red sequence galaxy, one obtains comparable S/N and morphological resolution at 0.5L*. The co-added SN survey images probe 2-3.5 magnitudes deeper. The depth of the HLS spectroscopy survey is sufficient to measure the Hα emission-line redshift of a 1.1L* star-forming galaxy at z=1.5 and a 1.7L* star-forming galaxy at z=2.6. General observer programs can, of course, achieve greater depths over still substantial areas by adopting longer exposure times than the HLS (see Table 7).

WFIRST will provide major advances in our understanding of the content and structure of the Milky Way, including the detection of red clump giant stars, which are standard candles used to map Galactic structure, across the full disk of the Galaxy. The microlensing data set will also allow additional exoplanet discoveries through both near-infrared transits and astrometric searches. In the standard HLS mode, WFIRST achieves the sensitivity needed to detect (a) 10 Gyr old helium atmosphere white dwarfs at a distance of 2.5 kpc; (b) the bottom of the hydrogen burning main sequence at a distance of 15 kpc; (c) the main sequence turnoff of a 10 Gyr old stellar population at a distance of 300 kpc; (d) a red clump giant at a distance of 3.3 Mpc; and (e) the tip of the red giant branch at a distance of 26 Mpc.

WFIRST will enable the most comprehensive mapping of resolved stellar populations in the Milky Way and nearby galaxies. The combination of WFIRST's sensitivity, resolution, and field of view is unmatched by HST or JWST. WFIRST will be the first mission that can study nearby galaxies to significant near-infrared depths over their full spatial extent and at high spatial resolution. Such data will greatly advance our understanding of their star-formation histories, galaxy structure and substructure, and galaxy interactions and accretion. While the HLS will be very useful for finding high-redshift galaxy clusters, targeted galaxy cluster programs will also provide significant science, ranging from HLS-depth studies of interesting clusters identified by other programs (e.g., SZ surveys, WISE, eROSITA) outside the HLS field, to deep studies of the most extreme clusters. The latter data could provide wide-field weak lensing masses and wide-area near-IR prism red-shifts; the latter are particularly important for high-redshift, early-type galaxies, which are quite faint at optical wavelengths but bright in the near-IR.

WFIRST will carry out a high-cadence microlensing survey of ~300 million stars towards the Galactic bulge. Extrapolating from empirical constraints on the population of cold exoplanets from current surveys, WFIRST is expected to detect more than 2200 cold exoplanets in the mass range of 0.1 – 10,000 Earth masses, including over 260 planets with roughly the mass of Earth or smaller. If there is one free-floating Earth for every star in the Galaxy, then WFIRST will detect 30 of them (and much larger numbers for more massive free-floating planets). This census will provide unprecedented insights into the formation and evolution of planetary systems and the frequency of habitable worlds.

For the DRM1 dark energy program, the SN survey is expected to discover nearly 2000 supernovae and achieve statistical errors of 0.5-1.0% in luminosity distance in each of 16 $\Delta z = 0.1$ redshift bins from $z = 0.15$ to $z = 1.65$. Including systematic errors, the aggregate distance precision forecast for the supernova survey is 0.25-0.3%. The WL survey will measure 480 million galaxy shapes, most of them in three independent bands, achieving an aggregate precision of 0.3% in the amplitude of the WL power spectrum. The galaxy redshift survey will measure redshifts of 17 million galaxies in the range $1.3 < z < 2.7$, with BAO analysis yielding 1.5-2.5% measurements of the angular diameter distance $D_A(z)$ and Hubble expansion rate $H(z)$ in each $\Delta z = 0.1$ bin, with aggregate precision of 0.5-0.75%. Analysis of the full galaxy power spectrum can improve the precision of these constraints and also give direct measurements of the growth rate of perturbations at these redshifts, with aggregate precision (for a conservative assumption about theoretical modeling uncertainty) of about 1.5%.

## 2.2 Exoplanet Microlensing

*Primary Exoplanet Science Objective: Complete the statistical census of planetary systems in the Galaxy, from the outer habitable zone to free floating planets, including analogs of all of the planets in our Solar System with the mass of Mars or greater.*

### 2.2.1 Introduction/Retrospective

The first discovery of planetary companions to Sun-like stars was, along with the discovery of cosmic acceleration, one of the greatest breakthroughs in modern astronomy (Latham et al. 1989, Mayor & Queloz 1995, Marcy & Butler 1996). In the intervening years, hundreds of exoplanets have been discovered, mostly by the radial velocity technique, and over 2300 candidates have been detected by the Kepler space telescope (Batalha et al. 2012). From the very first discovery, it has been clear that nature hosts an enormous and unexpected diversity of exoplanetary systems, containing planets with physical properties and orbital architectures that are radically different from our own Solar System. Theories of planet formation and evolution, originally developed to explain our Solar System (*e.g.*, Lissauer 1987), have struggled to keep up with the flood of new planets. Our understanding of the origins of planetary systems both similar and very different from our own remains poor. Not surprisingly, the explosive and often surprising progress of the exoplanet field draws talented young researchers to careers in astronomy and planetary science.

Microlensing observations from space, when combined with results from the Kepler mission, offers the best possibility of fully understanding the frequency and origins of planetary systems. It will do this by fulfilling the primary science goal of "*Completing the statistical census of planetary systems in the Galaxy, from the outer habitable zone to free floating planets, including analogs of all of the planets in our Solar System with the mass of Mars or greater.*" In particular, WFIRST will be uniquely sensitive to three broad classes of planets: cold planets, free-floating planets, and very low-mass planets down to the mass of Mars.

**Cold planets**: WFIRST's exoplanet microlensing survey is a perfect complement to Kepler's survey, and indeed is the only way to discover the majority of the planets to which Kepler is insensitive. Kepler's intrinsic ability to detect planets is ten times smaller at separations of 1 AU than 0.1 AU. Kepler is also much less sensitive to Earth-size planets than Jupiter-sized planets. As a result, even though it was designed explicitly to detect Earth-size planets in the habitable zones of solar-type stars, this is close to the limit of its capabilities, and Kepler will be essentially insensitive to small planets with periods longer than one year. Simply put: Kepler is sensitive to "hot" and "warm" planets, but not to the "cold" planets in the outer regions of planetary systems. Thus Kepler is insensitive to analogs of all of the planets in our Solar System from Mars outward. In contrast, the exoplanet survey on WFIRST is sensitive to planets from roughly the outer habitable zone outwards, including rocky planets with the mass of Earth up to the largest gas giant planets, and analogs to the giant planets in our Solar System. Thus the sum of

Kepler plus WFIRST's exoplanet survey will yield a composite census of planets with planets the mass of Earth and greater on both sides of the habitable zone, overlapping at almost precisely that zone.

**Free-Floating Planets**: WFIRST's exoplanet program can detect "free-floating" planets in numbers sufficient to test planet-formation theories (Jurić & Tremaine 2008, Chatterjee et al. 2008), a task not possible from the ground. It will also extend the search for free-floating planets down to the mass of Earth and below, addressing the question of whether ejection of planets from young systems is a phenomenon associated only with giant planet formation or also involves terrestrial planets.

**Mars-mass embryos**: WFIRST is the only way to detect planets with masses as small as that of Mars in significant numbers. This is because microlensing's ability to detect planets is much less sensitive to planet mass than Kepler's transit technique is to planet size. Since Mars-mass bodies are thought to be the upper limit to the rapid growth of planetary "embryos", determining the planetary mass function down to a tenth the mass of the Earth uniquely addresses a pressing problem in understanding the formation of terrestrial-type planets.

The microlensing method is well developed and mature. Its capabilities have been amply demonstrated from the ground through several discoveries that have provided important new insights into the nature and diversity of planetary systems. These include the discovery of a substantial population of cold "Super-Earths" with masses between 5 and 20 times the mass of the Earth (Beaulieu et al. 2006, Bennett et al. 2008, Muraki et al. 2011, see Figure 4 for an example), the first discovery of a planetary system with an analog to our own Jupiter and Saturn (Gaudi et al. 2008, Bennett et al. 2010a) and the detection of a new population of Jupiter-mass planets loosely bound or unbound to any host star (Sumi et al. 2011). Importantly, ground-based microlensing surveys have determined that cold and free-floating planets are ubiquitous: on average, every star in the Galaxy hosts a cold planet (Cassan et al. 2012), and "free-floating" planets may outnumber the stars in our galaxy by two to one. Although these results are tantalizing, ground-based systems have intrinsic limitations that severely limit their sensitivity and efficiency, particularly to the low-mass planets with mass less than the Earth that are of such interest and importance to our understanding of planetary systems.

**Box 4**

**Estimated WFIRST DRM1 Exoplanet Discoveries**

For the assumptions described in Section 2.2.5, the WFIRST DRM1 design will detect **2200** total bound exoplanets in the range of 0.1-10,000 Earth masses, including **790** "Super-Earths" (roughly 10 times the mass of Earth), **240** Earth-mass planets, and **30** Mars-mass planets. This would enable the measurement of the mass function of cold exoplanets to better than 10% per decade in mass for masses >0.3 $M_{Earth}$, and an estimate of the frequency of Mars-mass embryos accurate to ~20%.

In addition, the DRM1 design would detect **30** free-floating Earth-mass planets, if there is one per star in the Galaxy.

These estimates are likely conservative. With better optimization of the target fields, and using recently obtained data with newly developed analyses that suggest higher microlensing event rates in the target fields, WFIRST may detect up to 2 times more planets.

**Comparison to IDRM Exoplanet Capabilities**

The estimated yields for DRM1 are about 30% larger than the yields of the original IDRM design for the same total observing time. On average ~20% of this gain comes from the larger field-of-view, and another ~10% comes from the wider bandpass.

**DRM2 Exoplanet Capabilities**

The yields for DRM2 are about 20% larger than the yields of DRM1 for the same total observing time, due to the larger field-of-view. For the baseline 3-yr DRM2 observing plan, in which two of the four microlensing seasons will be interrupted by SN observations every 5 days, the yields are roughly 75% of the DRM1 yields, with the larger FOV partially compensating for the decreased aperture, poorer resolution, and reduced observing time. The interrupted observations do not affect the total yields for a fixed total exposure time, but they do affect the planet parameter uncertainties.

**Comparison to Euclid**

WFIRST will detect **2-3** times more planets than a possible 300 day survey with an extended Euclid mission.

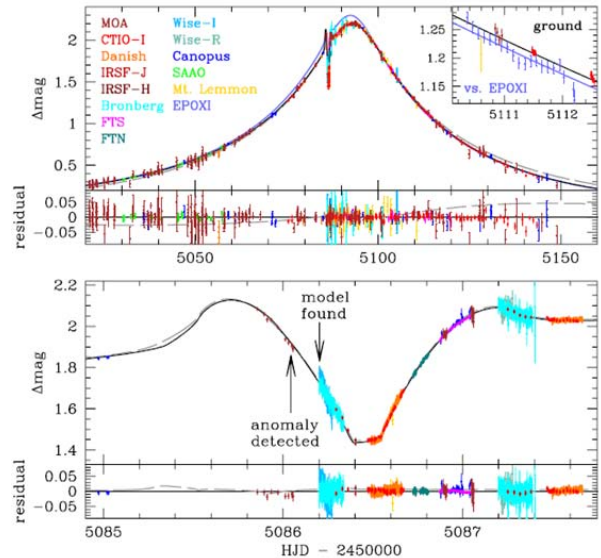

**Figure 4:** Data and models for microlensing event MOA-2009-BLG-266, which resulted in the discovery of a cold, Super-Earth. The large upper panel shows the primary event caused by an M-dwarf lens. The large lower panel shows focuses on the deviation caused by the ~10 $M_\oplus$ planetary companion to the lens with a projected separation of ~3.2 AU, roughly twice the snow-line distance of the host star. From Muraki et al. (2011).

While exoplanet microlensing from the ground has had well-documented successes, realizing the true potential of the microlensing method, and achieving the primary science goals outlined above, is only possible from space with an instrument like WFIRST. Microlensing requires monitoring a very large number of stars in very crowded, heavily reddened fields toward the Galactic bulge continuously for months at a time. Reaching the needed sensitivity and number of planet detections requires steady viewing in the near-infrared without interruptions for weeks at a time, as well as high angular resolution, stable images over very wide-area fields. In other words, it requires a wide-field infrared survey telescope in space. Without such a mission, we will not be sensitive to Mars-mass embryos, we will not determine the frequency of low-mass free-floating planets, and we will not complete the census of planets begun by Kepler.

In summary, a microlensing survey from space with WFIRST is a critical component of our efforts to understand the origins, evolution, and demographics of planetary systems. The Exoplanet Task Force (Lunine et al. 2008) recognized this, embedding into its strategy a medium-cost space-based microlensing mission to complement the census of close-in planets currently underway with spaceborne transit techniques. The De-cadal Survey for Astronomy and Astrophysics placed the science of exoplanets in the top two or three science areas of importance to astrophysics in the next decade. Space-based microlensing plays a pivotal role in the Decadal Strategy for exoplanets. As evidenced by the wide range of parameter space it will explore (see Figure 5), and the large number of expected discoveries (see Box 4), WFIRST is the means to understanding the range of planetary system architectures (masses, orbital parameters) and to determining how many Earth-mass worlds inhabit the crucially important region from the outer portion of the classically defined habitable zone outward.

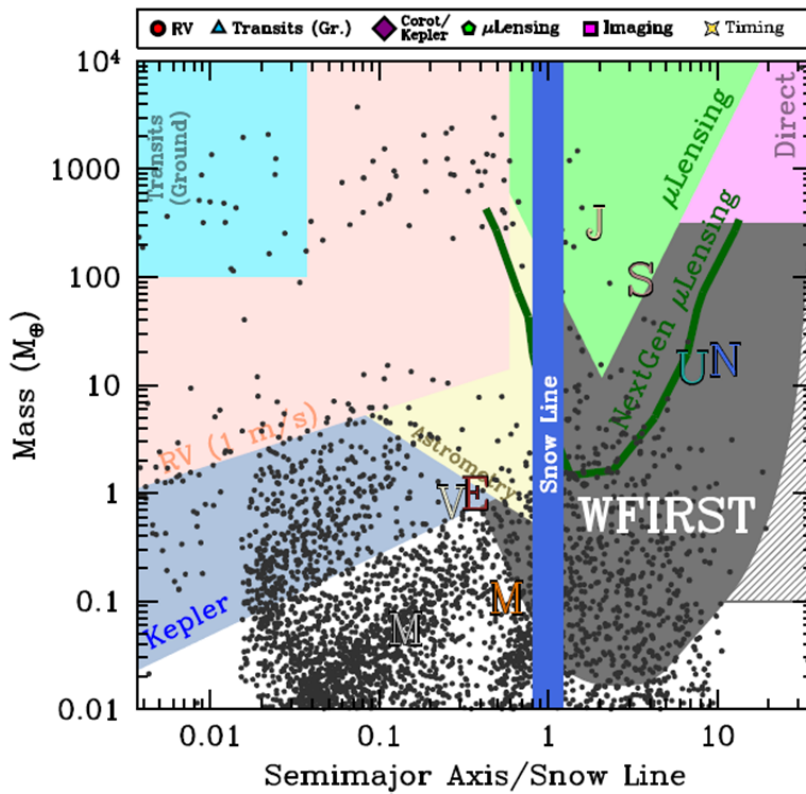

**Figure 5: Grey dots show the mass versus semi-major axis scaled to the location of the snow line for planets predicted by the semianalytic planet formation models of Ida & Lin (2004). The colored shaded regions show approximate regions of sensitivity for various missions. Current ground-based surveys with radial velocity, transits, and microlensing are unable to probe the large reservoirs of planets that are predicted by these and other theories of planet formation. While Kepler is probing the abundant, hot terrestrial planets, WFIRST is required to complete the statistical census of cold Super-Earths, cold Earth-mass planets, and cold Mars-mass embryos.**

### 2.2.2 The Importance of Exoplanet Demographics with WFIRST

After nearly two decades of exoplanet discovery, our empirical statistical census of exoplanets, as well as our theoretical understanding of the physical processes at work in the formation and evolution of the exoplanets, remains largely incomplete. The nearly ~700 confirmed exoplanets occupy a region of parameter space that is largely disjoint from both the planets in our Solar System, and from the largest reservoirs of planets that are predicted by many planet formation theories (Ida & Lin 2004, Mordasini et al. 2009). Kepler is partially ameliorating this problem by assaying the population of small, hot planets (Howard et al. 2012, Batalha et al. 2012), but without WFIRST's survey of low-mass, cold exoplanets, we will be left with an essential ignorance of the demographics of exoplanets in vast regions of parameter space, which include analogs of Mars, Saturn, Neptune, and Uranus. Without WFIRST, essential questions about the formation and evolution, as well as the habitability of exoplanets, will remain unanswerable.

*Exoplanet Survey Question #1: How do planetary systems form and evolve?*

In the most general terms, planet formation theories should describe all of the relevant physical processes by which micron-sized grains grow through 13-14 orders of magnitude in size and 38-41 orders of magnitude in mass to become the terrestrial and gas-giant planets we see today. These physical processes are ultimately imprinted on the final distributions of planet frequencies, compositions, and orbits (e.g. Ida & Lin 2004, Mordasini et al. 2009). Thus by measuring these distributions, i.e., by determining the demographics of exoplanets, it is possible to gain insight into the physical processes that drive planet formation.

The discovery of gas giant planets orbiting at periods of only a few days, as well as evidence for the migration of the giant planets in our own Solar System, have highlighted the fact that planet formation theories must also account for the possibility of large-scale rearrangement of planet positions through gravitational and gas dynamical effects during and after the epoch of planet formation (Lin et al. 1996, Rasio & Ford 1996), and thus must also track the planets through billions of years of evolution. Many of these theories also predict a substantial population of "free-floating" planets that have been ejected from their planetary systems through interactions with other planets (Jurić & Tremaine 2008, Chatterjee et al. 2008). Indeed, evidence for such a population was recently found using microlensing (Sumi et al. 2011). The interpretation of the final exoplanetary system architectures that we observe today must account for these dynamical processes.

The exoplanet microlensing survey of the WFIRST mission will provide an integral and essential component of a coordinated plan by the exoplanet community to answer Exoplanet Survey Question #1. In particular, WFIRST will provide the only way to complete the statistical census of planets begun by Kepler, by measuring the demographics of planets with masses larger than that of Mars and separations of greater than about 1 AU. This includes analogs to all the Solar System's planets except for Mercury, as well as most types of planets predicted by planet formation theories thus far. The number of such discoveries will be large with roughly 2200 bound and hundreds of free-floating planet discoveries expected. Whereas Kepler is sensitive to close-in planets but is unable to detect the more distant ones, WFIRST is less sensitive to close-in planets, but is more sensitive than Kepler beyond about 1 AU (see Figure 5). WFIRST is sensitive to unbound planets with masses as low as the Earth, offering the only possibility to constrain the frequency of these planets, which may have been ejected during the planet formation process. *Thus, WFIRST and Kepler complement each other, and together they cover the entire planet discovery space in mass and orbital separation and provide the comprehensive understanding of exoplanet demographics necessary to fully understand the formation and evolution of planetary systems.*

*Exoplanet Survey Question #2: What determines the Habitability of Earthlike Planets?*

The age-old question of whether or not there is life on other worlds has gained even more relevance and immediacy with the discovery of a substantial population of exoplanets. The first step in determining how common life is in the universe is to determine the frequency of potentially habitable worlds, commonly denoted $\eta_\oplus$. While of course interesting in its own right, an accurate measurement of $\eta_\oplus$ also provides a crucial piece of information that informs the design of direct imaging missions intended to characterize potentially habitable planets around nearby stars and search for biomarkers. Indeed, the primary goal of Kepler is to provide a robust measurement of $\eta_\oplus$. Early results from Kepler indicate that terrestrial planets are likely to be very common, at least for short periods of <50 d (Borucki et al. 2011, Howard et al. 2012), thus bolstering the case that $\eta_\oplus$ might be high. However, a robust measurement of $\eta_\oplus$ will only be possible at the end of the full extended mission, and will likely rely on extrapolation from shorter periods or larger planets, if $\eta_\oplus$ ends up being low. Furthermore, it will be difficult or impossible to measure the masses of the potentially habitable planets detected by Kepler because of the faintness of the host stars, and Kepler has already found that small planets in close orbits can have a wide range of densities (Lissauer et al. 2011).

WFIRST will improve the robustness of our estimate of $\eta_\oplus$. The sensitivity of WFIRST to Earth-mass planets in the habitable zone has significant dependence on a number of Galactic model parameters that are poorly known, but it is clear that all of the WFIRST designs have sensitivity to terrestrial planets just outside the habitable zone, as well as sensitivity to super-Earths in the habitable zone. Thus, WFIRST will be able to estimate $\eta_\oplus$ with only modest extrapolation, and the combination of WFIRST and Kepler data will make it possible to robustly interpolate into the habitable zone from regions just outside of it, even if the frequency of habitable planets turns out to be small. Furthermore, WFIRST will be sensitive to planet mass, as opposed to planet radius, thus providing complementary information to Kepler, and enabling the statistical determination of the densities and surface gravities of habitable terrestrial planets.

Importantly, Kepler will only measure the frequency of *potentially* habitable planets. The factors that determine what makes a *potentially* habitable planet *actually* habitable are not entirely understood. Certainly, the suitability of a planet for life depends on the average surface temperature, which defines if the planet resides in the habitable zone. In addition, it is likely that the amount of water is a contributing factor. However, the water content of planets in the habitable zone must have been delivered from beyond the snow line, with the number, masses, and orbits of planets beyond the snow line having a potentially dramatic effect on the water content of planets in the habitable zone (e.g. Raymond et al. 2004; Lissauer 2007). Thus the habitability of planets in the habitable zone cannot be considered in isolation: an understanding of the frequency of habitable planets (as opposed to potentially habitable planets) necessitates an understanding of the demographics beyond the snow line, which requires the WFIRST survey.

In summary, WFIRST will provide crucial empirical constraints on planetary systems that will allow us to address the two fundamental and interrelated questions: "*How do planetary systems form and evolve?*" and *"What determines the habitability of Earthlike planets?"*.

### Exoplanet Survey Question #3: What kinds of unexpected systems inhabit the cold, outer regions of planetary systems?

As is frequently the case in astronomy, it may be the unexpected and unpredictable returns of an exoplanet survey with WFIRST that will prove to be the most enlightening for our understanding of the formation, evolution, and habitability of exoplanets. One of the most important lessons from the past two decades of exoplanet research is that there exists an enormous diversity of exoplanetary systems, generally far exceeding theoretical expectations. Indeed, one hallmark of the field is the fact that, whenever new regions of parameter space are explored, the subsequent discoveries necessitate revisions of our planet formation theories. Kepler is currently revolutionizing our understanding of the demographics of small, short-period planets. Because it will open up a similarly broad expanse of parameter space, and because the expected yields are similarly large, WFIRST is essentially guaranteed to do the same. Furthermore, it will do so in a region of parameter space that is almost certainly critical for our understanding of planet formation and habitability.

These three survey questions directly address two of the science questions called out specifically by the Planet Systems and Star Formation (PSF) Science Frontiers Panel of the NWNH Decadal Survey. In particular, the broad picture of exoplanet demographics enabled by WFIRST is needed to answer PSF Question #3 "How diverse are planetary systems?" and the insight WFIRST will bring to bear on the frequency and habitability of potentially habitable worlds will provide an important input into the first part of PSF Question #4 "Do habitable worlds exist around other stars, and can we identify the telltale signs of life on an exoplanet?"

### 2.2.3  WFIRST Exoplanet Data Requirements

Microlensing events require extremely precise alignments between a foreground lens star and a back-ground source star, that are both rare and unpredictable. Furthermore, the probability that a planet orbiting the lens star in any given microlensing event will give rise to a detectable perturbation is generally much smaller than unity, ranging from a few tens of percent for a Jupiter-mass planet and a typical low-magnification event, to less than a percent for planets with mass less than that of the Earth. These planetary perturbations have amplitudes ranging from a few percent for the lowest-mass planets to many tens of percent for the largest perturbations, but are brief, ranging from a few days for Jupiter-mass planets to a few hours for Earth-mass planets (see Figure 4, Figure 6, and Figure 7 for examples). Also, the time of the perturbations with respect to the peak of the primary event is unpredictable. Thus detecting a large number of exoplanets with microlensing requires monitoring of a very large number of stars continuously with relatively short cadences and photometric precision of a few percent. Practically, a sufficiently high density of source and lens stars, and thus a sufficiently high microlensing event rate, is only achieved in lines of sight towards the Galactic bulge. However, these fields are also crowded, and this high star density means that high resolution is needed to resolve out the individual stars in order to achieve the required photometric precisions and to identify the light from the lens stars.

Quantitative predictions for the yields of a given realization of an exoplanet survey dataset require sophisticated simulations that incorporate models for the Galactic distribution of lenses and sources to simulate and evaluate the detectability of events with realistic photometric precision. In particular, the source star density and event rate are strong functions of Galactic coordinates, and the detection probability of a planet with a given set of properties depends sensitively on the detailed properties of the event (host star mass and distance, event duration, angular source size, photometric precision, cadence). We have used such simulations (described in more detail below) to determine a set of exoplanet survey data requirements necessary to meet a minimum survey yield.

### Exoplanet Data Requirements:

- Monitor >2 square degrees in the Galactic bulge for at least 250 total days.
- S/N of ≥100 per exposure for a J = 20.5 star.
- Photometric sampling cadence of ≤15 minutes.

- Better than 0.4" angular resolution to resolve the brightest main sequence stars.
- Monitor microlensing events continuously with a duty cycle of ≥80% for at least 60 days to measure basic light curve parameters.
- Sample light curves with W filter.
- Monitor fields with Y filter, 1 exposure every 12 hours to measure the color of the source stars.
- Separation of >2 years between first and last observing seasons to measure lens-source relative proper motion.

**Time-series photometry is combined to uncover light curves of background source stars being lensed by foreground stars in the disk and bulge.**

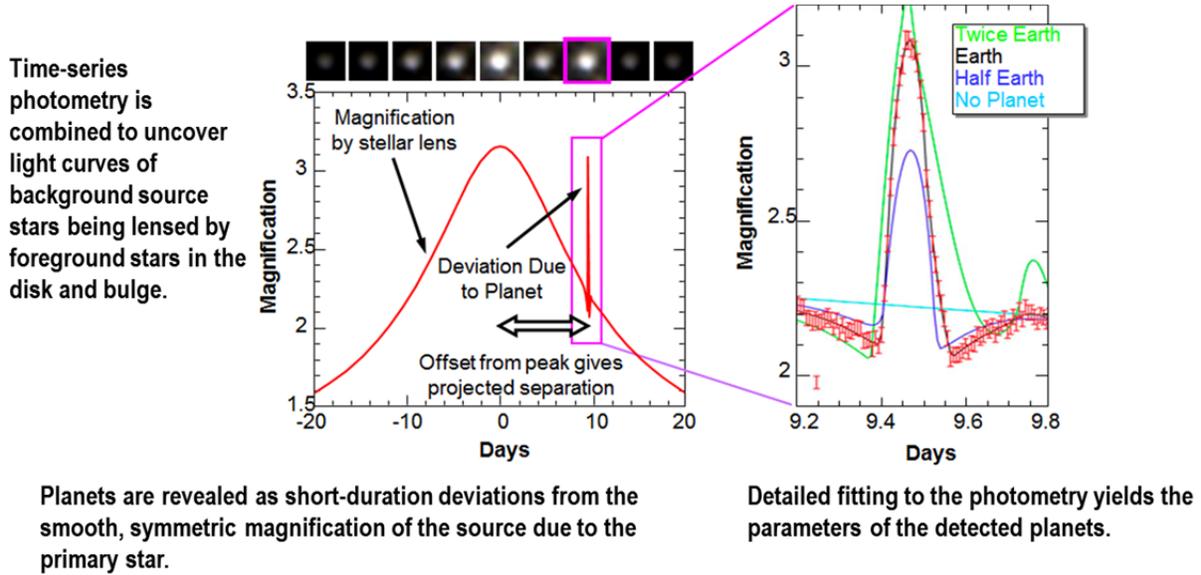

Planets are revealed as short-duration deviations from the smooth, symmetric magnification of the source due to the primary star.

Detailed fitting to the photometry yields the parameters of the detected planets.

**Figure 6: Schematic illustration of how WFIRST discovers signals caused by planetary companions in primary microlensing events, and how planet parameters can be extracted from these signals. The left panel shows a simulation primary microlensing event, containing a planetary deviation from an Earth-mass companion to the primary lens. The offset of the deviation from the peak of the primary event, when combined with the primary event parameters, is related to the projected separation of the planet. The right panel shows a detail of the planetary perturbation. The width and precise shape of the planetary deviation yield the mass of the companion relative to that of the primary host lens.**

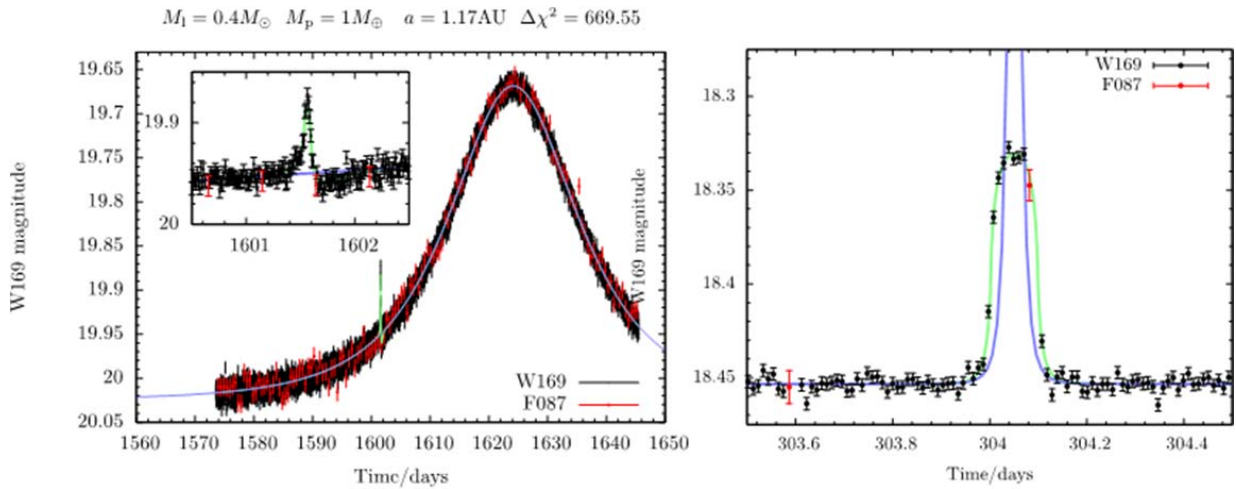

**Figure 7: Example simulated event light curves with detected planetary signals from simulations of a WFIRST exoplanet survey. The left panel shows a typical signal-to-noise ratio detection of an Earth-mass planet orbiting a 0.4 solar mass star with an intrinsic semi-major axis of 1.17 AU. The right panel shows a simulated free-floating Earth-mass planet detection.**

We can roughly understand the order of magnitude of these requirements. Consider, as a specific example, Earth-mass planets with separations of 2 AU, and assume that, if every star hosts such a planet, we would like to detect at least 200. The typical detection probability for a planet with $M_p = M_\oplus$ and $a = 2AU$ is ~1.5%, and thus ~200/0.015 ~ $10^4$ microlensing events must be monitored to detect ~200 such planets. The average microlensing event rate in the DRM1 fields is ~$5 \times 10^{-5}$ events/year/star, and thus $10^4/5 \times 10^{-5}$~200 million star-years must be monitored. The typical stellar density down to J=23 is ~$10^8$ stars per square degree, and thus at least ~2 square degrees must be monitored. As illustrated in Figure 6, in order to detect, fully sample, and accurately characterize the perturbations due to such planets, which typically last a few hours and have amplitudes of several percent, photometric precisions of a few percent and cadences of better than 15 minutes are needed. Finally, given the areal density of ~$10^8$ stars per square degree, an angular resolution of $10^{-4}$ degrees or 0.4 arcseconds is needed to resolve the faintest stars. The remaining three requirements above ensure the ability to accurately measure the parameters of the primary events, which typically last ~40 days, and allow one to infer the angular size of the source star from its color and magnitude, separate the light from the lens and source, and measure the relative lens-source proper motion, all three of which are required to measure the mass and distance to the primary lens (and thus measure the mass and separation of the detected planets) for the majority of events (Bennett et al. 2007). These data requirements then provide the constraints within which hardware and operations concept designs must operate.

### 2.2.4   WFIRST Exoplanet Performance Forecasts

Below we summarize the expected yields of WFIRST under several different mission designs. The original yields reported in the IDRM report were estimated using an updated version of the simulation described in Bennett & Rhie (2002). For this report, yields for all three WFIRST designs were also computed using two largely independent codes. These two codes use substantially different methodologies, and thus provide an important test of the robustness of the predictions. After calibrating these codes to common empirical constraints, we find that their predictions agree to better than a factor of 2 for the majority of the parameter space where WFIRST is sensitive. There remains sub-stantial disagreement for planets that are close to the limit of the sensitivity of the various WFIRST designs. We explore these issues further in Appendix A, where we describe the basic simulation ingredients and methodology, and identify sources of uncertainty and likely contributing causes to the remaining discrepancies. Given these significant discrepancies for planets near the edge of WFIRST's sensitivity, we therefore adopt a conservative approach and only quote results for those regions of peak sensitivity to exoplanets, where the agreement is generally better than a factor of ~2.

The baseline yields presented in Box 4, Section 2.2.6, and Table 2, Table 3, and Table 4 adopt the results from the simulation methodology described in Penny et al. (2012), after scaling these results to match empirical constraints on the number density of stars and published microlensing optical depths (which are then used to estimate event rates). This simulation was used to estimate the yields of the WFIRST IDRM, DRM1, and DRM2 designs, as well as Euclid. We expect the relative yields between different survey designs (given our assumptions) to be more robust than the absolute rates.

These baseline yields were not optimized relative to the number, location, and cadence of the target fields for each design. As a result, the relative yields between the different designs only reflect the effects of the hardware and scheduling, and further optimization would likely increase the relative yields. Furthermore, the choice of fields was constrained by the underlying model used to simulate the microlensing events (the Besançon model). As a result, we expect these results to be conservative. Therefore, we briefly consider the effects of field optimization in Section 2.2.9.

Finally, an analysis of MOA-II survey data (Sumi et al., in preparation) is, for the first time, able to measure the microlensing rate with a large number of events in and near the WFIRST fields. The preliminary findings from this analysis indicate a significantly higher event rate for these fields close to the Galactic center than we adopted for our baseline assumptions. In Section 2.2.9, we use the simulation methodology of Bennett & Rhie (2002), updated to apply to an IR survey (Bennett et al. 2010b), to estimate the yields using this higher event rate, as well as including the effects of field optimization. We find roughly double the number of detections.

### 2.2.5   Conservative Baseline Results

The basic results are summarized in Box 4, Table 2, Table 3, and Table 4, and illustrated in Figure 5, and are described in more detail below. All of the proposed

WFIRST designs will detect a large number of planets. Our best estimate is that of order 2200 bound planets will be detected with the DRM1 design, with masses in the range of 0.1-10,000 $M_\oplus$ and semi-major axes in the range of 0.03-30 AU. As illustrated in Figure 5, while massive planets can be detected over nearly this entire range of separations, lower-mass planets can only be detected over a narrower range of separations beyond the snow line. At least 10% of the detected bound planets will have mass less than three times the mass of the Earth and WFIRST will have significant sensitivity well below the mass of the Earth, with >30 Mars-mass planet detections expected. WFIRST will measure the mass function of cold planets to ~5-20% in 1-dex bins of planet mass down to the mass of Mars. If free-floating planets are common, WFIRST will detect a large number of them with masses down to that of the Earth and below. The DRM1 and DRM2 designs are intrinsically more capable than the IDRM design, resulting in roughly 30% and 60% more detections respectively for the same observing time. For a reduced total observation time of 266 days and including scheduling constraints with SN observations, the yield of DRM2 is ~75% of the DRM1 yield.

### 2.2.6 *Quantitative Metrics: IDRM Revisited*

In the original IDRM report, we focused on four metrics that we combined into a single Figure of Merit (FOM). The metrics were: (1) the number of detected planets with M = $M_\oplus$ and period P = 2 years, (2) the number of planets detected with M = $M_\oplus$ and period P = 2 yr for which the primary mass can be determined to 20%, (3) the number of free-floating $1M_\oplus$ planets detected, assuming one free floating planet per star, and (4) the number of habitable planets detected assuming every F, G and K star has one, where habitable means 0.1-10$M_\oplus$, and [0.7-2 AU]$(L/L_{Sun})^{1/2}$. After performing additional, mostly independent simulations and comparing the results of these simulations to the simulation used to predict the yields for the IDRM, we have discovered significant (factors of 3-10) discrepancies in the predicted yields for planets that are at the edges of the region of peak sensitivity of the various designs, including those with M = $M_\oplus$ and P = 2 years and habitable planets. These discrepancies are primarily due to differences in the Galactic models used for these calculations. In addition, we have not yet performed the analysis needed to estimate the uncertainty in the primary mass measurements, and thus do not have a quantitative estimate for the second metric. Furthermore, for this report we are no longer emphasizing the FOMs, and thus the utility of these metrics is less obvious.

Here we adopt a different approach. Specifically, we will report the predicted yields for planets of a range of masses, where for each mass we will average over a range of semi-major axes between 0.03-30 AU. The majority of these planets will be detected in regions near the peak sensitivity of WFIRST, for which the predictions are much more robust and the simulations agree to better than a factor of 2.

Of course, to estimate planet yields, we must adopt assumptions for the intrinsic frequency distribution of planets in mass and semi-major axis. While we have some constraints on the frequency of more massive planets in this range of semi-major axis probed from current microlensing surveys, as well as some information about the distribution of planet masses in this region, these constraints remain relatively poor, particularly in the low-mass regime. This ignorance is, of course, the primary justification for an exoplanet survey with WFIRST.

Therefore, we will quote the number of detections under two sets of assumptions, our raw yields assuming a simple uniform prior, as well as yields derived by adopting the distribution and normalization of cold planets inferred from current microlensing surveys. We will refer to the former as our 'raw' yields, which can be easily scaled to any assumed mass distribution function. We will refer to the latter as our `best estimate' of the yields; these are the results that are presented in Table 2, Box 4 and Section 2.2.5.

Specifically, the assumptions for our two sets of yields are as follows:

**"Raw Yields"** Yields: We assume that every star and white dwarf has one planet per dex[2] in mass and semi-major axis, with a uniform distribution in log mass and log semi-major axis.

**"Best Estimate"** Yields: We assume our best estimate of the planet distribution function from Cassan et al. 2012, which has the form

$$dN/d\log M_p d\log a = f (M_p/95M_\oplus)^\alpha$$

where f = 0.24/dex[2] and $\alpha$ = -0.73. We assume that this power-law distribution saturates at a constant value of 2/dex[2] for $M_p$<4.4 $M_\oplus$. This is likely conservative, as Kepler has found several multi-planet systems that have more than 2 planets per decade in semi-major axis (e.g., Lissauer et al. 2011, Gautier et al. 2012). Note that the exponent of this power-law distribution with planet mass is slightly steeper but of opposite sign to the exponent of the (roughly) power-law dependence of the detection efficiency on planet mass, and thus the number of expected planets is a relatively slowly rising function of decreasing mass in this model.

We also quote the expected yields for free-floating Earth-mass planets, assuming that there is one free-floating Earth-mass planet per star in the Galaxy. For free floating planets of mass $M_p$, the yields are expected to roughly scale as $(M_p/M_\oplus)^\beta$, with $\beta$~0.5 to 0.7 (Penny et al. 2012).

The predicted yields are summarized in Table 2 for our "Raw Yields" and Table 3 for "Best Estimate" yields. We show the results for the IDRM, DRM1, and two versions of the DRM2. The first assumes the same amount of time and same number of fields for the exoplanet survey in DRM2 as for the DRM1 and IDRM surveys (7 fields, 6 seasons of 72 days for a total of 432 days), in order to highlight the difference in the expected yields solely due to the hardware designs. The second assumes the straw man observing schedule proposed in Section 4.10: a total of 4 72-day microlensing seasons, two of which are interrupted by SN observations such that 1 day out of 5 is lost (for a total 266 days). In addition, the second set of DRM2 yields assume a total of 6 fields, the maximum number of fields allowed for a continuous microlensing survey by the maximum data downlink rate. We also show the yields for Euclid assuming a 300 day survey. It is important to emphasize that microlensing is not one of the primary science components of Euclid, and therefore such a survey is not being currently planned. Furthermore, it is not clear if the final Euclid design (*e.g.*, pointing constraints) will even allow a substantial microlensing component. However, we nevertheless include these yields in order to highlight the differences in intrinsic capability arising from the different hardware choices for the two missions.

| $M/M_\oplus$ | IDRM (432 days) | DRM1 (432 days) | DRM2 (432 days) | DRM2 (266 days) | Euclid-Extended (300 days) |
|---|---|---|---|---|---|
| 0.1 | 11 | 15 | 14 | 11 | 5 |
| 1 | 101 | 119 | 140 | 88 | 33 |
| 10 | 463 | 639 | 739 | 482 | 159 |
| 100 | 2032 | 2725 | 3170 | 2093 | 623 |
| 1000 | 6934 | 8529 | 10,252 | 6364 | 2036 |
| 10,000 | 16,089 | 19,984 | 24,813 | 15,078 | 5102 |
| Total | 25,630 | 32,011 | 39,128 | 24,116 | 7959 |

Table 2: Predicted "raw" yields for bound planets for various mission designs. We assume that every star and white dwarf has one planet per dex² in mass and semi-major axis, with a uniform distribution in log mass and log semi-major axis. We have adopted event rates extrapolated from observed rates for bright clump giant sources.

| $M/M_\oplus$ | IDRM (432 days) | DRM1 (432 days) | DRM2 (432 days) | DRM2 (266 days) | Euclid-Extended (300 days) |
|---|---|---|---|---|---|
| 0.1 | 21 | 30 | 29 | 21 | 10 |
| 1 | 202 | 239 | 279 | 176 | 66 |
| 10 | 576 | 794 | 918 | 599 | 197 |
| 100 | 470 | 630 | 733 | 484 | 144 |
| 1000 | 299 | 367 | 442 | 274 | 88 |
| 10,000 | 129 | 160 | 199 | 121 | 41 |
| Total | 1697 | 2221 | 2600 | 1676 | 546 |

Table 3: Predicted "best estimate" yields for bound planets for various mission designs. We have adopted the planet distribution function for cold exoplanets as measured from ground-based microlensing surveys by Cassan et al. (2012), and have used event rates extrapolated from observed rates for bright clump giant sources.

| IDRM (432 days) | DRM1 (432 days) | DRM2 (432 days) | DRM2 (266 days) | Euclid-Extended (300 days) |
|---|---|---|---|---|
| 23 | 33 | 40 | 27 | 5 |

**Table 4: Predicted yields for free-floating Earth-mass planets, assuming one free-floating Earth-mass planet per star in the Galaxy.**

### 2.2.7 DRM1 vs. IDRM

We find that the yields for DRM1 are 15-40% larger than the IDRM over the full range of masses considered, with an average improvement of ~27%.

This relative sensitivity of IDRM and DRM1 can roughly be understood as follows. The IDRM and DRM1 designs differ in two ways: the FOV and the passband. The FOV of DRM1 is 1.29 times larger than the imaging field of IDRM. The planet yield does not go linearly with the FOV because the additional FOV covers areas with lower event rate. For our choice of fields, the "effective" increase in FOV is 1.21. The wider passband of DRM1 leads to a 27% increase in the photon flux for a flat spectrum source. The amount that this increases the yields depends on the shape of the distribution of the cumulative number of detections with $\Delta\chi^2$, which in turn depends on the masses and semi-major axes of the planets being considered. For planets in the range of 1-1000 $M_\oplus$, we find that $N(>\Delta\chi^2) \propto (\Delta\chi^2)^a$ with a = -0.5 to -0.2. Thus the increased photon flux leads to yields that are larger by ~5-15%, with the largest gains coming from the smallest planets. Thus the total increase in planet yield is expected to be ~25-40% for fixed total observing time, roughly in line with the results of the detailed simulations. Again, it is likely possible to improve the relative yield between the two designs further by optimizing the target fields and cadences.

### 2.2.8 DRM2 vs. IDRM

Similarly, we have also performed simulations of the yields of DRM2 assuming the same total observing time as the DRM1. The results are listed in Table 2 and Table 3. We find yields for DRM2 that are 25-35% larger than the IDRM for most of the range of masses considered.

We can also understand the relative sensitivity of DRM2 and DRM1. DRM2 has the same bandpass as the DRM1, but has a still larger FOV due to the change to H4RG-10 detectors. However, it also has a smaller aperture (1.1m vs. 1.3m), and thus the overall photon rate is smaller by ~$(1.1/1.3)^2$ ~0.72. Using the scaling of the number of detections with $\Delta\chi^2$ described in the previous section, this leads to a 5-15% decrease in the yields. However, this is more than compensated for by the 1.55 times larger FOV of the DRM2 relative to the DRM1, which results in an effective increase in yield of ~1.44. Therefore, for fixed total observing time, we would predict that the DRM2 yields will be ~23-35% larger than the DRM1 yields. On one hand, differences between these analytic estimates and the detailed results from the simulations are likely due to the somewhat larger amount of background noise due to the larger PSF in the DRM2 design, which will more strongly affect the yields for lower-mass planets. On the other hand, the reasonably good agreement suggests that background noise is not the dominant effect for determining the detectability for these designs. However, this is only based on a few simulations and then only for the planets near the peak of the detection sensitivity. This should be studied in more detail in the future, particularly for planets near the limits of the detection sensitivity.

We also simulated the yield of an exoplanet DRM2 survey with a reduced total observing time. We considered three different scenarios (1) 5 seasons of 72 days, with no interruptions from SN observations, for a total 360 days; (2) 5 seasons of 72 days, with 2 seasons being interrupted by SN for one day out of every 5 days, for a total of 332 days, and (3) 5 seasons of 72 days, with all 5 seasons interrupted by SN for one day out of every 5 days, for a total of 290 days. In general, we find that the total number of detections scales very closely with the total observing time (to within ~5% for $M_p < 100 M_\oplus$) suggesting that `edge effects' do not substantially reduce the planet yields, at least for the smallest planets. However, we do expect that the parameter uncertainties will be affected.

The example observing program presented in Section 4.10 calls for four 72-day microlensing seasons, two of which are interrupted by SNe observations, corresponding to a total of 266 days of observing time. Since we have not simulated this exact program, the numbers we report in Table 2 and Table 3 have been scaled from the DRM2 scenario 2 above by the ratio of the total observing time, 266/332, ~80%. In addition, due to data downlink constraints, a maximum of 6 fields can be monitored continuously with the larger DRM2 field of view. We have estimated the yield for 6 fields by discarding the detections from the least sensitive field, and scaling the remaining detections in the other field by the amount expected based on the increased exposure time per field. The yields for this observing program are also given in Table 2 and Table 3. They are on average 75% of the yield for DRM1, with the largest effect on the yields relative to a 5 year, 7-field DRM2 survey arising from smaller total observing time.

One disadvantage of the shorter DRM2 mission lifetime that is likely to be significant, but that we have not yet quantified, is its effect on the ability to infer host masses by measuring the relative lens-source proper motion (Bennett et al. 2007). The accuracy of the primary mass measurement using this method depends either linearly or quadratically on the baseline of the observations (depending on the amount of information available), and thus the reduced mission duration may dramatically reduce the number of host stars for which mass measurements are possible. This is an important topic for future study.

#### 2.2.9 Field Optimization

The yields presented in the previous section were not optimized relative to the location of the target fields. Because there can be large gradients in the microlensing event rate, extinction, and source star surface density over the fields likely to be targeted by WFIRST, such optimization can have a significant effect on the yields, particularly for designs with larger FOVs. Unfortunately, the empirical data on these quantities is substantially incomplete, particularly for the faint source stars that will be targeted by WFIRST, and for the fields closer to the Galactic center, where the optical extinction is high. In Appendix A, we describe the models or measurements we use to estimate these quantities and so optimize our fields, but we note that substantial uncertainties remain, and further work is needed.

We selected 14 candidate fields for DRM1 and 12 candidate fields for DRM2 based on estimated source star densities, and then evaluated observing strategies involving a variable number of survey fields. We assumed a 15-minute observing cadence for each field and a conservative 40 seconds to slew and settle when moving between fields. For both DRM1 and DRM2, we found that the optimal number of planet detections was found for 10 fields, which would cover 3.75 deg$^2$ and 5.85 deg$^2$ for DRM1 and DRM2, respectively. However, the larger number of fields implies a higher data downlink rate, and the rate required for 10 fields exceeds the data downlink rate assumed in Section 3.6. The current data downlink constraints limit the exoplanet micro-lensing survey to 9 fields for DRM1 (3.38 deg$^2$) and 6 fields for DRM2 (3.51 deg$^2$).

The individual fields selected to maximize the planet detection rate are shown in Figure 8. We find that these fields will lead to yields that are of order 10-20% higher than our baseline estimates. It should be noted that this is just a crude attempt at field selection. This preliminary investigation indicates that we do quite well in the inner fields with somewhat higher extinction, so it would be worthwhile to repeat the field selection process with some candidate fields even closer to the Galactic plane.

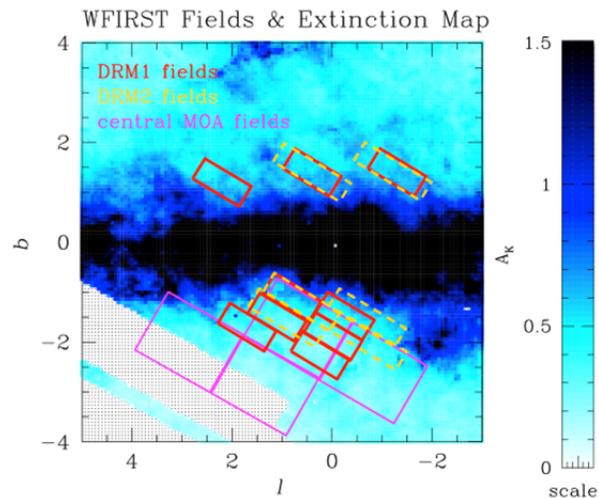

**Figure 8: The DRM1 (red) and DRM2 (gold) fields are laid out on top of the Dutra et al. (2003) bulge extinction map. The magenta rectangles are the MOA fields used for the new event rate estimate described in Section 2.2.10.**

#### 2.2.10 New Event Rate Estimates

The event rate estimates in the previous section were based on published microlensing optical depth estimates (Popowski et al. 2005; Sumi et al. 2006; Hamadache et al. 2006, Alcock et al. 2000, Sumi et al. 2003). These results cover a region of the Galactic bulge somewhat more distant from the Galactic plane than the preferred WFIRST fields, and thus a simple extrapolation of a linear fit to the outer fields was used to predict the optical depths in the WFIRST fields. We considered two sets of optical depth data when performing the linear fit: optical depths derived only from events that appeared to have red clump giant star sources (RCG optical depths), and optical depths derived from all events (all-star optical depths), which tend to include faint, unresolved sources. As discussed in Appendix A, all-star optical depth measurements tend to be higher than the RCG optical depth measurements by about 25% for the same fields. This is also true of these data sets, which cover fields several degrees off the Galactic plane. However, the gradients of both sets of data with Galactic latitude $|b|$ are such that, the predicted optical depths based on these two linear extrapolations differ by <10% when averaged over the WFIRST fields of interest.

The MOA team has recently performed a preliminary analysis of the 2006-2007 MOA-II data (Sumi et al. 2011), and used this to determine the optical depth and event rate toward each of the 22 MOA fields, each covering 2.2 deg². This MOA-II data set has about twice as many events as all the previously published data sets combined. Four of these fields, containing 215 microlensing events (45% of the total), overlap with or are very close to some of the proposed WFIRST fields, as shown in Figure 8. A linear fit in $|b|$ to all 22 MOA fields yields an average optical depth (and thus event rate) in the WFIRST fields that is consistent with previously published data for all-star samples (Alcock et al. 2000; Sumi et al. 2003), but ~25% higher than published results from RCG samples (Popowski et al. 2005; Sumi et al. 2006; Hamadache et al. 2006). Furthermore, the four inner fields (with low $|l|$ and $|b|$) have substantially higher optical depth. A modest extrapolation from these fields yields an average optical depth that is ~80% higher than the extrapolation based on published estimates. Because these optical depths are based on preliminary, unpublished MOA-II data and analysis, we will not use them for our baseline predictions. However, we note that they have much higher statistical weight than the published results, so if they are correct, our "best estimate" planet yields are, in fact, significantly underestimated.

Table 5 shows the estimated yields of DRM1 and DRM2, including both the effects of the optimized field selection, and the higher optical depth and thus event rate implied by a fit to the inner four MOA-II fields. Under these assumptions, we estimate WFIRST yields will be a factor of roughly 2x higher than our baseline predictions. As described in Appendix A, these yields were calculated using a different code and under somewhat different assumptions, and therefore the improvement over our baseline yields varies somewhat with planet mass, particularly for the lowest mass planets.

| $M/M_\oplus$ | DRM1 | DRM2 (266 days) | DRM1 / Cassan | DRM2 / Cassan (266 days) |
|---|---|---|---|---|
| 0.1 | 68 | 47 | 135 | 93 |
| 1 | 354 | 245 | 706 | 488 |
| 10 | 1554 | 1073 | 1462 | 1010 |
| 100 | 5589 | 3836 | 1040 | 714 |
| 1000 | 15467 | 10584 | 562 | 385 |

**Table 5: Comparison of estimated yields with optimized field selection and higher event rate. The first two columns show the result for the "raw" detections assuming a planet per dex², whereas the last two columns show the results using the Cassan et al. (2012) distribution function.**

## 2.3 High Latitude Infrared Survey

*Primary NIR Survey Science Objective: Produce a deep map of the sky at NIR wavelengths, enabling new and fundamental discoveries ranging from mapping the Galactic plane to probing the reionization epoch by finding bright quasars at z>10.*

In addition to dark energy and exoplanet microlensing science, WFIRST will provide a unique and powerful platform for a broad range of additional compelling infrared survey science. Some of these studies will use the data acquired for the dark energy and microlensing surveys. This archival research (AR) science is best exemplified by the optical Sloan Digital Sky Survey (SDSS) where the number of research papers now published using the archival SDSS data — originally collected to study the large-scale structure of the Universe — far exceeds the number of papers written by the original SDSS collaboration. To date, thousands of papers have been written using SDSS public data, making it one of the most successful astronomical surveys yet undertaken. WFIRST will provide a deeper infrared counterpart to SDSS (and future optical imaging surveys such as DES, HSC, and LSST), imaging thousands of square degrees to considerably greater depths, AB~26, with much higher spatial resolution. WFIRST will also obtain low-resolution infrared prism spectroscopy of every source over thousands of square degrees of extragalactic sky to AB~22.5 (1σ, per pixel, where the mean per-pixel resolving power is R~600 per 2 pixels). We describe ancillary science that will come from the High Latitude Survey (HLS) in this section, and in the following section (§2.4) we describe infrared Galactic science possible from large, dedicated low-latitude surveys with WFIRST. In what follows, we consider the HLS to consist of the wide-area, single-epoch weak lensing and galaxy redshift surveys as well as the deeper, multi-epoch supernova survey.

Besides this AR science, additional WFIRST science will come from dedicated observations, or general observer (GO) programs. We describe several potential GO programs in §2.6.

### 2.3.1 Quasars

The WFIRST HLS, designed for the cosmological measurements described in §2.5, will identify unprecedented numbers of quasars at very high redshift. Quasars are among the most luminous objects in the universe, observable to the earliest cosmic epochs. They are thought to be supermassive black holes at the centers of galaxies, converting mass into energy 20 times more efficiently than stars. They provide fundamental information on the earliest phases of structure formation in the universe and are unique probes of the intergalactic medium. The discovery of a Gunn-Peterson (1965) trough in the spectra of several quasars at redshift $z > 6$ implies that the universe completed reionization near that redshift (e.g., Fan et al. 2001), though poor sample statistics, the lack of higher redshift quasars, and the coarseness of the Gunn-Peterson test make that inference somewhat uncertain. Currently, the most distant confirmed quasar is at $z = 7.1$ (Mortlock et al. 2011), and only a dozen quasars have been confirmed at $z > 6$. Ambitious surveys will push this number to around 100 in the next few years, probably identifying one or two quasars at $z \sim 8$. WFIRST will fundamentally change the landscape of early universe investigations. Based on the Willott et al. (2010) quasar luminosity function, WFIRST will identify thousands of quasars at $z > 6$, hundreds of quasars at $z > 7$, and push out to $z \sim 10$ should quasars exist at those redshifts (see Table 6). Such discoveries will directly measure the first epoch of supermassive black hole formation in the universe, probe the earliest phases of structure formation, and provide unique probes of the intergalactic medium along our line of sight to these distant, luminous sources. The large numbers of quasars identified in the first Gyr after the Big Bang will enable clustering analyses of their spatial distribution (e.g., Coil et al. 2007, Myers et al. 2007). Optical surveys are not capable of identifying quasars above $z \sim 6.5$ since their optical light is completely suppressed by the redshifted Lyman break and Lyman-alpha forests. Though JWST will be extremely sensitive at near-infrared wavelengths, it will not survey nearly enough sky to find the rarest, most distant, luminous quasars, which have a surface density of just a few per 10,000 deg[2]. WFIRST will be the definitive probe of the first phase of supermassive black hole growth in the universe. In particular, as shown in Table 6, the depth and sensitivity of WFIRST will make it a much more capable high-redshift quasar finder than ground-based infrared programs, and it will complement Euclid by going two magnitudes deeper, over a smaller area, finding similar numbers of quasars in total but over different ranges of luminosity.

| Survey | Area (deg²) | Depth (5-sigma, AB) | z>7 QSO's | z>10 QSO's |
|---|---|---|---|---|
| UKIDSS-LAS | 4000 | Ks=20.3 | 8 | - |
| VISTA-VHS | 20,000 | H=20.6 | 40 | - |
| VISTA-VIKING | 1500 | H=21.5 | 11 | - |
| VISTA-VIDEO | 12 | H=24.0 | 1 | - |
| Euclid, wide | 15,000 | H=24.0 | 1359 | 22 |
| Euclid, deep | 40 | H=26.0 | 14 | - |
| WFIRST, HLS | 3400 | K=26.0 | 1304 | 25 |
| WFIRST, SN (wide/deep) | 6.5 / 1.8 | K = 28.1 / 29.6 | 12 | - |

Table 6: Number of high-redshift quasars predicted for various ground- and space-based near-infrared surveys, based on the quasar luminosity function of Willott et al. (2010). Note that 6.6% of the Euclid wide survey will have two or fewer dithered exposures per sky position. Such data will lack the redundancy required to robustly identify rare objects, and thus have been omitted in the above calculations.

### 2.3.2 Galaxy Clusters

Galaxy clusters have historically played a central role in cosmology, from the first indication of dark matter (Zwicky 1933) to important evidence that we do not live in an Einstein-de Sitter universe (e.g., Donahue et al. 1998). Galaxy clusters are the most massive, virialized structures in the universe, and provide some of the most compelling evidence regarding the nature and amount of dark matter. Fundamental cosmological pa-rameters combined with the initial fluctuation spectrum determine the abundance and growth of clusters of galaxies (*e.g.*, Voit 2005). The most dramatic effects are seen in the evolution of the abundance of the most massive clusters, both in an absolute sense and relative to less massive clusters. In recent years, several groups have noted that the number of the extremely massive clusters at high redshift appears to be difficult to reconcile with the Gaussian distribution of primordial fluctuations predicted by standard inflationary models with $\Lambda$CDM matter and energy content (*e.g.*, Hoyle et al. 2012, Williamson et al. 2011). The key to observational cosmology using galaxy clusters lies in making accurate measurements of both the number of clusters as a function of redshift and their total masses.

The wide-area WFIRST HLS will allow astronomers to find and characterize the dark matter and stellar contents of galaxy clusters to an unprecedented degree. Massive clusters of galaxies are rare – only a few per square degree exist at any redshift, and therefore only surveys that cover large fractions of the sky are suitable for finding the most distant and massive clusters of galaxies. The WFIRST survey will not only find high-redshift clusters but will provide several independent measurements of their masses. On the time-scale of the WFIRST surveys, the German-Russian eROSITA telescope is expected to have executed an all-sky low-energy X-ray survey, providing an additional mass measurement from the hot intracluster medium that dominates the baryonic mass of relaxed clusters. WFIRST imaging and wide-field spectroscopy will find and weigh clusters through the clusters' gravitational lensing (both weak and strong) of background galaxies, and by the stellar luminosity of their member galaxies, tracked by their red sequence galaxies (Gladders & Yee 2000), which are infrared bright. For an extragalactic survey sufficient for galaxy colors and shear maps, standard cosmology with the assumption of gaussian fluctuations would predict ~1 cluster every 5 deg$^2$ more massive than $10^{14}$ solar masses between z=1.5-2 and ~1 every 80 deg$^2$, same mass limit but z=2-2.5 (G. M. Voit, private communication, Ventimiglia, PhD thesis 2012; cf. Voit 2005). Interestingly, if the detection threshold captures clusters only a factor of two less massive, the numbers increase to 1 cluster every 0.4 deg$^2$ (z=1.5-2) and 1 every 3 deg$^2$ (z=2-2.5). Since the normalization, evolution, and shape of the cluster mass function and the cluster spatial correlation function depend independently on cosmological parameters, these data (with multiple constraints on cluster masses) will provide powerful new constraints on the growth function and the gaussianity of the initial mass fluctuations (e.g. Majumdar & Mohr 2004).

### 2.3.3 Galaxy Assembly

Over the past two decades, a basic framework of hierarchical galaxy formation has been established based on gravitational instability of baryonic structures within a dark matter dominated framework. Galaxies form at peaks of the underlying density distribution, where rapid collapse of baryonic gas results in galactic-scale star formation, further regulated by feedback processes from both star formation and nuclear black hole activity. Although this framework is able to qualitatively explain the large-scale distribution of galaxies over space and time, we still lack a basic understanding of the detailed physical processes that drive and control the assembly history of the baryonic component of galaxies and that give rise to the amazing diversity of galaxy properties. Galaxy formation and evolution appear to be a stochastic processes affected by many physical processes, particularly cosmological environment. Understanding the complexity of galaxy evolution requires large, well-defined statistical surveys that at the same time accurately measure physical properties of galaxies and their environment.

The WFIRST HLS will provide a legacy dataset to study fundamental questions of galaxy evolution. The near-infrared imaging survey will provide accurate spectral energy distribution (SED) measurements for several billion galaxies and accurate size and shape measurements for hundreds of millions of galaxies over several thousand square degrees. Specifically, though the spatial resolution will not be as exquisite as that of the Hubble Space Telescope, WFIRST will still resolve ~80% of galaxies to $H$~25 (AB). The prism survey will further measure the redshifts, SEDs, and provide coarse star formation rates for tens of millions of galaxies at $1 < z < 3.5$, corresponding to the peak era of cosmic star formation, when more than 60% of the stellar and black hole mass in galaxies was assembled. This WFIRST survey will build on the legacies of ground-based surveys such as SDSS and LSST, and extend small-area deep fields such as GOODS and COSMOS to the largest cosmological scales. Theoretical models will be tested against full, multi-dimensional distribution functions of galaxy properties, including the evolution of the galaxy luminosity and stellar mass functions, galaxy color-luminosity and size-luminosity relations, and the impact of nuclear activity and environment on galaxy evolution.

WFIRST is a uniquely powerful tool to study the assembly history of galaxies and the emergence of the Hubble sequence when combining the imaging and spectroscopic surveys (Figure 9). The WFIRST slitless prism survey will reach $H$ = 22 (AB) in continuum, corresponding to ~$L*$ galaxies at $z$ = 1.5. This survey will also detect emission lines to flux limits of 1x10$^{-16}$ erg s$^{-1}$ cm$^{-2}$ (7$\sigma$). For H$\alpha$, this corresponds to star formation rates of ~5 M(Sun) yr$^{-1}$ at the redshifts for which H$\alpha$ is shifted into the WFIRST near-infrared bandpass. The near-infrared continuum sensitivity of WFIRST is approximately two magnitudes deeper than what is currently possible with large ground-based telescopes. The rest-frame optical spectra that WFIRST will provide will not only provide redshifts, but will also spatially resolve well-calibrated diagnostics of star formation, stellar age, metallicity, stellar mass-to-light ratio and nuclear activity. Combined with high-resolution imaging, these data will track the emergence and growth of disks and bulges, constrain feedback processes that shut off star formation at later epochs, and determine the role of mergers and the origin of the morphology-density relation.

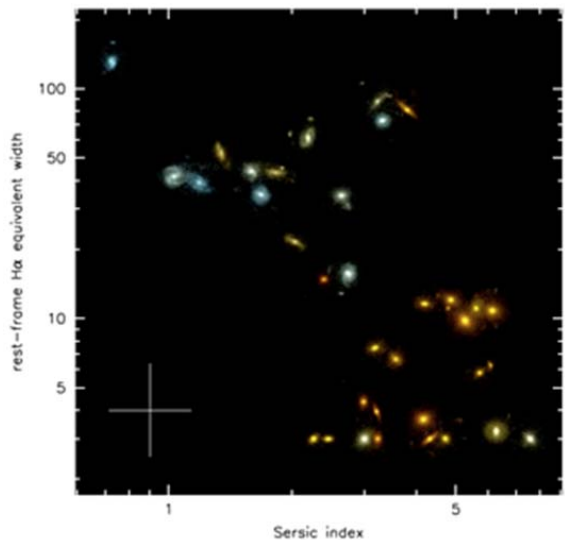

**Figure 9: Relation between star formation (quantified by H$\alpha$ equivalent width; larger equivalent width implies more star formation) and galaxy morphology (quantified by Sersic index; larger index implies earlier galaxy type) at $z$≳1, using data from the 3-D HST Treasury Program (PI. van Dokkum). 3-D HST uses HST grism spectroscopy with resolution similar to that of WFIRST to obtain deep continuum and H$\alpha$ spectroscopy of high-redshift galaxies. This combination of space-based imaging and spectroscopy reveals a striking diversity of massive galaxies at $z$≳1, ranging from large star formation spirals to compact quiescent galaxies.**

### 2.3.4 *Surveying the Stellar Halo of the Milky Way*

In addition to the extragalactic science themes, the WFIRST HLS will enable a deep-wide field map of the Milky Way's stellar halo. The observed properties of the stellar halo, such as the level of substructure, the surface brightness profile, and the chemical abundance gradient, are intimately linked to the accretion processes that built up the Milky Way.

The SDSS revealed abundant substructure in the Milky Way stellar halo through the detection of stellar over-densities in wide-field star count maps. The properties of this substructure yield robust estimates on the nature of the satellites that were disrupted to form the halo. Additionally, SDSS led to the discovery of a new class of dwarf satellites, the ultra-faint dwarf spheroidals. These galaxies are the most dark matter dominated systems known, and represent early relics of the Universe.

Deep, wide-field imaging with WFIRST will enable a new breakthrough in the characterization of the Milky Way's stellar halo. The stellar halo is old, and therefore the brightest sources in both substructure and dwarf satellites are red giants. WFIRST will be sensitive to the complete red giant branch, and also the main-sequence turnoff, in systems out to the edge of the Milky Way halo. This will lead to a new generation of high-resolution star-count maps from which the Galactic stellar density, surface brightness profile, and level of substructure will be measured. Based on the incompleteness in the SDSS due to its shallow photometric limits, WFIRST will also enable the detection of a large population of low-luminosity dwarf satellites that are thought to exist. The luminosity function of these dwarf spheroidals will provide a new test to the missing satellite problem, and follow-up kinematics will test dark matter models in systems with very high mass-to-light (M/L) ratios.

### 2.3.5 *Other Archival Research Programs using the High Latitude Survey*

Besides the few programs outlined above which will make use of the WFIRST high-latitude infrared survey, this rich data set will be of use for an immense range of additional science, far beyond the space limitations of the current report. We provide a brief list of additional programs that have been discussed by the WFIRST SDT:

- Mapping the Kuiper belt, including measuring size distributions, color distributions, and binarity;
- Studying stellar fossils, specifically cool white dwarfs, in the Galactic halo;

- Studying the properties, redshifts, SEDs and clustering of obscured AGN;
- Studies of rare, strongly lensed sources, including high-redshift quasars, double lenses, and lensed SNe;
- Finding and studying Population III pair-instability SNe in the SN survey fields;
- Infrared AGN light curves in the SN survey fields;
- Studying the shape and evolution of the faint end of the quasar luminosity function; and
- Mapping reionization with Ly$\alpha$ emitters detected in the prism survey.

## 2.4 Galactic Plane Infrared Surveys

The unprecedented wide-field near-IR sensitivity of WFIRST will also provide for a wide range of studies of our own Milky Way Galaxy. Some studies, such as finding and characterizing the lowest mass brown dwarfs, will come for free from the high-latitude cosmological surveys. Other studies, such as exoplanet searches relying on transits or astrometry (see §2.4.3), will make ancillary use of the microlensing data set. The *NWNH* report also emphasized that time should be allocated for dedicated Galactic studies; specifically, *NWNH* suggested that ~10% of the observing time be reserved for Galactic observations. In this section we outline a few potential WFIRST programs studying our own Milky Way Galaxy.

The bulk of stellar and interstellar mass of the Milky Way Galaxy lies in a thin disk of scale-heights of 300 pc and 100 pc, respectively. At the distance of Galactic center, ±300 pc subtends approximately ±2 degrees, meaning that the entire far half of the Galactic disk is packed into a very narrow strip of the sky. For this reason alone, the sensitivity and angular resolution of WFIRST will allow for major advances in ascertaining the content and structure of the Milky Way, including the detection of star forming regions, stellar clusters, red giants and even dwarf stars, both to the far side of the Galactic disk and into the densely packed Galactic center region.

With the lower mid-plane extinction in the infrared, the succession of previous generation infrared galactic plane surveys, *e.g.* COBE/DIRBE, 2MASS, MSX, Spitzer/GLIMPSE, Herschel/Hi-Gal have led to the discovery of many rare but important objects, particularly infrared dark clouds, young stellar objects, massive star outflows, Wolf-Rayet stars and Luminous Blue Variables (Churchwell & Benjamin 2009). They have also led to a continually evolving view of the overall structure of the Galaxy; a current schematic of this structure is shown in the "artist's conception" image in Figure 10. But because of the crowding in the Galactic plane, each of these surveys has hit a confusion limit. This confusion limits the distance out to which different populations can be detected, *e.g.* nearby dwarfs are lost against the glare of distant red giants, a possible inner bar to the Milky Way cannot be detected via red clump giants, and distant stellar clusters are undetectable without associated nebular emission.

### 2.4.1 *Galactic Structure*

The WFIRST Galactic plane survey will provide the most complete map yet of the structure of our Galaxy across the entire stellar disk. Previous surveys of the stellar and protostellar content of the disk of our Galaxy have been hampered by a combination of source confusion and extinction. For most lines of sight in the inner Galaxy, survey depth is limited by the effective resolution; WFIRST represents an order of magnitude improvement in survey depth compared to the best possible surveys from the ground in the near-IR. The excellent parallax measurements of Gaia will be hampered by both source confusion and more importantly, the 5x greater extinction in the longest Gaia passband compared to the longest WFIRST passband in the Galactic mid-plane. The current view of the structure of our Galaxy is a four-armed spiral with a prominent bar (Dame & Thaddeus 2008). Currently, the symmetry of our Galaxy is mostly conjectured. Little direct knowledge of the far side beyond the Galactic center exists, although the recent discovery of the Outer Scutum-Centaurus arm (Dame & Thaddeus 2011) at the distance predicted in Figure 10 provides support for the hypothesis of a grand design Milky Way. WFIRST will identify the red clump giant population, which are standard candles used to map Galactic structure, throughout the Galaxy, even through substantial extinction. This will permit the first direct measurement of the stellar distribution on the far side of the Galaxy.

Recent observations in the near- and mid-IR have added to the census of 'missing' evolved stars (*e.g.* Shara et al. 2009) in the Galaxy doubling the number of massive evolved stars (Wolf-Rayet/LBV candidates) detected (Shara et al. 2012, Mauerhan et al. 2011). WFIRST will aid in reconciling the small numbers of massive evolved stars detected with the prediction by providing a complete sample of near-IR color selected candidate WRs/LBVs through the entire Galactic disk. WFIRST will also identify young stellar clusters on the far side of the Galaxy enabling the first complete picture of star formation for the Milky Way.

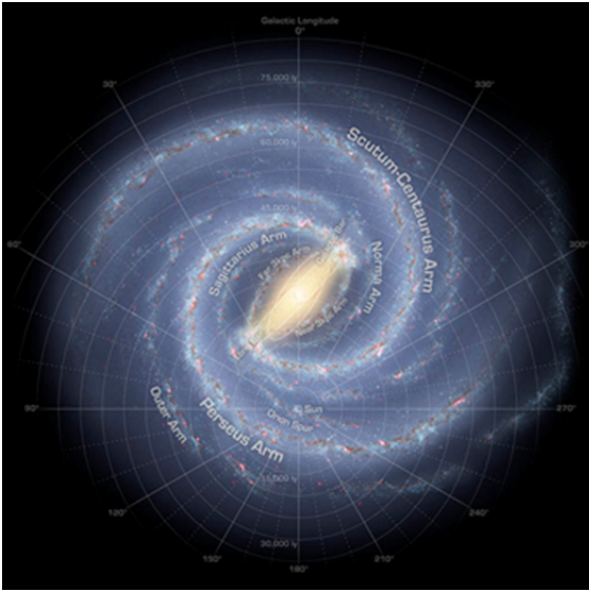

**Figure 10: Schematic of structure of Milky Way. Most structures more distant than the Galactic center have been inferred using limited data and symmetry arguments. WFIRST will provide the necessary sampling of red clump giants throughout the Galactic disk to provide the first clear picture of the structure of the entire Galaxy. Image credit Robert Hurt (IPAC).**

To fully sample the stellar and protostellar populations in the plane of the Galaxy including the flared and warped outer Galactic disk and the Galactic bulge, approximately 1600 square degrees need to be surveyed. For Galactic sources, a 3 sigma limiting sensitivity of K=24.8 (AB) is sufficient to provide a confusion limited survey of the entire Galactic plane. This can be easily accomplished with a survey depth of 5 x 60 second integrations in each band which will take approximately 4 months of time accounting for overheads due to slews, calibrations and downlinks.

### 2.4.2 Stellar and Substellar Populations in Galactic Star-Forming Regions

Studies of low mass populations (encompassing stars, brown dwarfs, and even free-floating planetary mass objects) are hindered in star forming regions due to observational and astrophysical effects. These include: the large angular sizes of the nearby molecular clouds harboring newborn and young stellar objects, the significant extinction and reddening, and the excess emission due to accretion processes and circumstellar dust. Nevertheless, it is possible to disentangle the complexities of these effects with a combination of multi-color photometry and spectroscopy. For young stellar objects, $Y$- through $J$-band is the best wavelength range for sampling the photospheric temperature, and $J$-$H$ and $J$-$K$ colors are sensitive to surface gravity. Extinction, however, is degenerate with intrinsic near-infrared colors and therefore spectroscopy is needed in order to accurately de-redden observed colors and enable comparisons to predicted colors and magnitudes (or temperatures and luminosities). WFIRST will vastly extend our knowledge of the initial mass function in young regions — where the low mass objects are brighter by many orders of magnitude than they are on the main sequence — down to and below the opacity limit for fragmentation within the molecular cloud. WFIRST will provide our deepest probes into the stellar mass function in these regions, and constrain both the threshold mass signifying star formation and also the age distribution of stars and sub-stellar objects across multiple decades in mass.

WFIRST will enable sensitive, systematic large-scale surveys of nearby star-forming regions. These areas are typically hundreds of square degrees and have been completely surveyed before only by the shallow 2MASS and WISE. UKIDSS has provided increased depth but with limited (albeit wide-area) coverage. The next steps with WFIRST will enable probes of the substellar mass function to near its bottom, over the required wide areas. JWST, by contrast, will have more depth but restricted spatial coverage. A unique capability of WFIRST will be its resolving power $R$~600 (2 pixels) prism spectroscopy, whose potential is illustrated in Figure 11.

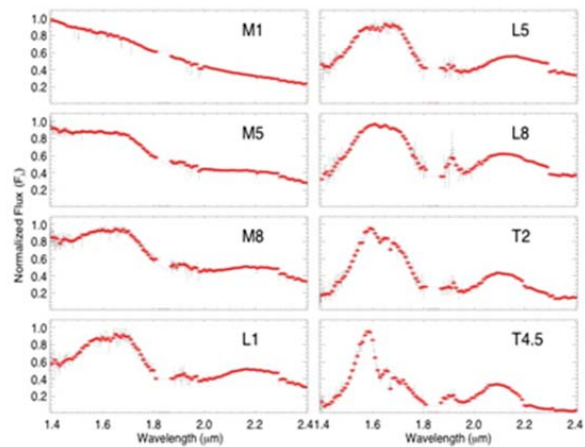

**Figure 11: Near-infrared spectra of cool stars in the wavelength range that WFIRST will access. WFIRST prism data will be very valuable for studying the coolest stars in our Galaxy.**

### 2.4.3 Other Exoplanet Science: Transits + Astrometric Searches

Besides the detection of exoplanets from microlensing surveys, WFIRST should allow additional exoplanet discoveries through both transits and astrometric searches.

Near-infrared exoplanet transits should be identifiable from the high cadence microlensing survey towards the Galactic bulge, which will entail 500 days of observations over five years. That program will obtain light curves for $\sim 3 \times 10^8$ stars with a photometric precision of $\sim 1\%$ — these numbers compare well with Kepler since WFIRST will downlink its entire field of view, while Kepler only downlinks selected stars across their field. Based on the sensitivity and observing cadence of the WFIRST microlensing survey, we predict the detection of up to 50,000 Jupiter-sized transits around main sequence stars and about 20 super-Earth transits around the brightest M-dwarfs. WFIRST is uniquely able to probe Jovian planets over all semi-major axes and the statistics on the population of hot Jupiters detected by transits relative to cold Jupiters detected with microlensing will shed light on the mechanisms responsible for planetary migration when combined with host star parameters.

While the first exoplanet discovered solely through astrometric observations remains elusive, this method of exoplanet detection has been used to confirm their existence and determine the mass of the planet. While not all orbiting planets transit their star, they do all make them wobble. Detecting this astrometric wobble simply requires a large number of data points with exquisite precision over a long temporal baseline. For instance, the 0.64 (sin $i$)$^{-1}$ M(Jupiter) planet discovered around the 5-pc distant M dwarf GJ 832 at an orbital separation of 3.4 AU and period of 9.35 years should produce a 1 milli-arcsecond amplitude astrometric wobble. Such precision is achievable with WFIRST. Specifically, assuming 0.5 arcsec resolution, the microlensing survey should detect hundreds of > 10 M(Jupiter) planets in 100 day orbits around 0.05 – 0.3 M(Sun) M-dwarfs. While the astrometric precision of a single observation is expected to be on the order of 2.5 milli-arcseconds, WFIRST will collect several thousand observations over the course of the microlensing program, providing for improved astrometric precision.

### 2.4.4 Other Galactic Archival Research Programs

In addition to the handful of Galactic programs outlined above, WFIRST will enable an immense range of additional science, beyond the space limitations of the current report. We provide a brief list of additional Galactic programs that have been discussed by the WFIRST SDT:

- Mapping the most distant star-forming regions in the Milky Way Galaxy;
- Astrometric surveys taking advantage of the wide field-of-view of WFIRST and using infrared-selected quasars as a reference frame; and
- Measuring proper motions and parallaxes for disk and bulge stars.

## 2.5 Dark Energy

*Primary Dark Energy Science Objective: Determine the expansion history of the Universe and the growth history of its largest structures in order to test possible explanations of its apparent accelerating expansion including Dark Energy and modifications to Einstein's gravity.*

The dramatic progress in astronomy of the past two decades has included several unexpected results. Nothing has been more surprising than measurements that indicate that the expansion of the universe (discovered by Edwin Hubble nearly a century ago) is accelerating. The apparent accelerating expansion could be due to: (1) a constant energy density (the "cosmological constant") that may arise from the pressure of the vacuum, (2) an evolving universal scalar field, or (3) a flawed or incomplete understanding of gravity as described by Einstein's General Theory of Relativity. Any of these possibilities has profound consequences for our basic understanding of physics and cosmology. The future of the universe will be determined largely by whatever force or property of space is causing the acceleration, making the nature of the accelerating expansion one of the most profound and pressing questions in all of science. The imperative is to distinguish between these possibilities by carrying out a careful set of measurements designed to characterize the underlying source of the so-called "dark energy" that drives the accelerated expansion of the universe. (Following common practice, we will use "dark energy" as a generic term that is intended to encompass modified gravity explanations of cosmic acceleration as well as new energy components.) Recent measurements imply that about 75% of the total mass-energy of the universe is dark energy. In other words, dark energy is most of our universe today, yet we do not know what it is. One of WFIRST's primary mission goals is to "settle fundamental questions about the nature of dark energy, the discovery of which was one of the greatest achievements of U.S. telescopes in recent years." [NWNH]

Dark energy affects the universe in two significant ways. First, the *expansion history* (or geometry) of the universe is determined by the energy density of dark energy over cosmic time. Second, the *growth of cosmic structures,* from the density perturbations we see in maps of the Cosmic Microwave Background (CMB) to the galaxies and galaxy clusters we see today, is governed by the attractive force of gravity and the repulsive effect of dark energy, which inhibits the structure growth. Within the confines of General Relativity, measurements of the expansion history and growth of structure will give consistent results. Discrepancies between these two types of measurements might indicate a breakdown of General Relativity. Only by measuring both the expansion history and the growth of structure at high precision can we distinguish among the three broad classes of explanations for the acceleration outlined above. WFIRST has been designed to measure each of these with multiple independent techniques. As described in NWNH, WFIRST "will employ three distinct techniques—measurements of weak gravitational lensing, supernova distances, and baryon acoustic oscillations—to determine the effect of dark energy on the evolution of the universe". In addition to these three methods, WFIRST adds a fourth method, redshift-space distortion (RSD), which provides an alternative measure of the growth of structure. These four methods provide complementary physical information and allow cross-checks that will be crucial to verifying the accuracy of the measurements. Recent reviews of these methods, in the context of a unified dark energy experimental program, include the reports of the Dark Energy Task Force (DETF; Albrecht et al. 2006) and JDEM Figure-of-Merit Science Working Group (FoMSWG; Albrecht et al. 2009) and the review articles of Frieman et al. (2008) and Weinberg et al. (2012); the last of these includes extensive discussion of the systematic uncertainties that affect these methods.

---

**Box 5**

"Why should WFIRST employ all three methods? Supernovae (in particular, type SNe Ia) give the best measurements of cosmic acceleration parameters at low redshift due to their greater precision per sample or per object. BAO excels over large volumes at higher redshift. Together SNe Ia and BAO provide the most precise measurements of the expansion history for $0 < z < 2$ and place significant constraints on the equation of state. Weak lensing provides a complementary measurement through the growth of structure. Comparing weak-lensing results with those from supernovae and BAO could indicate that "cosmic acceleration" is actually a manifestation of a scale-dependent failure of general relativity. Combining all three tests provides the greatest leverage on cosmic-acceleration questions. WFIRST can do all three."

From the Panel Reports--New Worlds, New Horizons in Astronomy and Astrophysics

---

Type Ia supernovae, powered by the thermonuclear explosions of white dwarf stars, all have roughly the same intrinsic explosion luminosity, providing "standard candles" that allow us to infer their distance from their apparent brightness. In more detail, there is a correlation between the peak luminosity of a Type Ia and the shape of its light curve, with fainter supernovae fading faster. Supernovae provided the first direct evidence of cosmic acceleration, a discovery that is now well buttressed both by improved supernova data and by multiple lines of independent evidence. Figure 12 presents a snapshot of the current state of supernova cosmology, based on more than 500 SN Ia from multiple surveys extending to redshifts 1.6, with results that are consistent with the predictions of a flat universe with a cosmological constant.

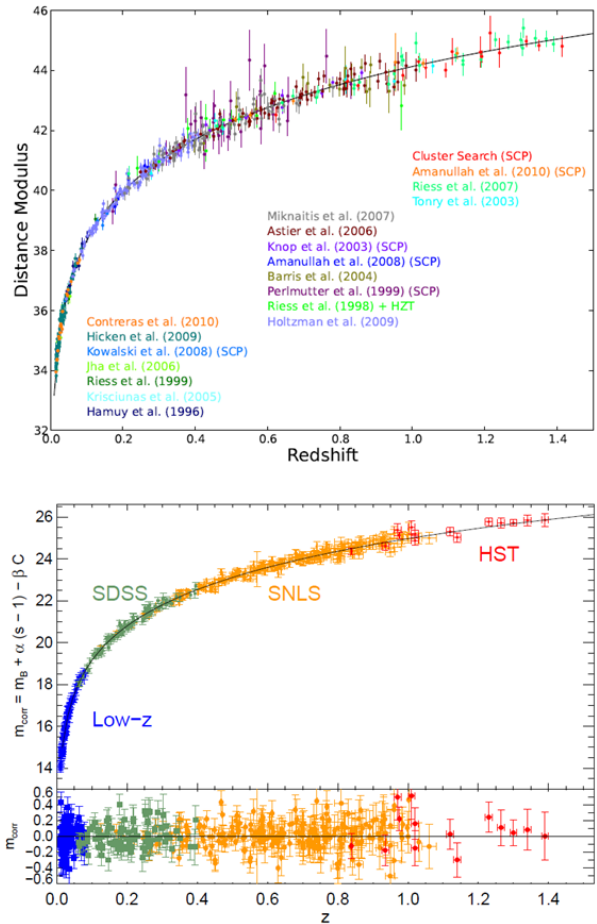

**Figure 12:** Hubble diagrams from two recent compilations of SN Ia observations, from a variety of ground-based surveys and from HST space-based observations. The upper plot is from Suzuki et al. (2012): points show binned (in Δz=0.01) measurements of the distance modulus, μ = apparent magnitude – absolute magnitude, as a function of redshift, and the curve shows the prediction for a cosmological model with a flat universe, a cosmological constant, and a matter density parameter $\Omega_m$ = 0.287. The lower panel is from Conley et al. (2011): points show the corrected apparent magnitude of individual SN, and the curve shows the prediction for a cosmological model with a flat universe, w = -0.90, and $\Omega_m$ = 0.19. Residuals from this best-fit model are shown at bottom.

The WFIRST supernova survey has two components: multi-band imaging with a 5-day cadence to discover supernovae and measure their light curves, and prism spectroscopy to confirm the Type Ia classification and measure redshifts. Our forecast for DRM1 indicates that it will discover more than 1800 spectroscopically confirmed Type Ia supernovae out to redshift z = 1.7. With the small statistical errors afforded by these large numbers, control of systematics becomes the key to the cosmological performance of the survey. Here WFIRST offers two critical advantages over present and future ground-based supernova surveys. First, the stability and sharp PSF of space-based imaging allow accurate, well calibrated photometry over the enormous dynamic range spanned by the survey. Second, WFIRST operates at near-IR wavelengths where Type Ia supernovae have more uniform peak luminosities (improving statistical performance and reducing evolutionary systematics) and where the impact of dust extinction is much smaller than at optical wavelengths. We predict an aggregate luminosity distance precision[3] of ~0.25-0.3% for the DRM1 supernova survey if we can achieve our systematic goals.

Baryon acoustic oscillations (BAO) --- sound waves that travel in the photon-baryon fluid of the early universe before recombination --- imprint a characteristic scale on the clustering of matter, which is subsequently imprinted on the clustering of galaxies and intergalactic gas. The BAO method uses this scale as a standard ruler to measure the angular diameter distance $D_A(z)$ and the Hubble parameter $H(z)$, from clustering in the transverse and line-of-sight directions, respectively. Figure 13 shows a "Hubble diagram" from recent ground-based BAO measurements at redshifts z < 0.7, with the precision of individual measurements ranging from about 2% to about 5%. While the supernova method measures relative distances, with calibration against local samples in the Hubble flow, the BAO method measures absolute distances calibrated against the pre-recombination sound horizon, which is computed using parameters that are accurately measured from the CMB. Thus, even at the same redshifts, supernova and BAO distance measurements provide complementary information about dark energy and space curvature, as well as cross-checks of systematics. The two methods are also complementary in their strength in different redshift ranges. Supernovae have unbeatable precision at low redshifts, where their systematics are also most easily controlled. The limiting precision of the clustering-based BAO measurements improves with increasing redshift because there is more comoving volume to survey at greater distances. The ability of the BAO method to directly measure $H(z)$ also improves its dark energy sensitivity at high redshift.

---

[3] We use the term "aggregate precision" to refer to the fractional error on an overall scaling factor that multiplies the value of an observable (such as luminosity distance) at all of the measured redshifts; roughly speaking, it is the error on the observable that would arise if all of the data were at the sample's median redshift.

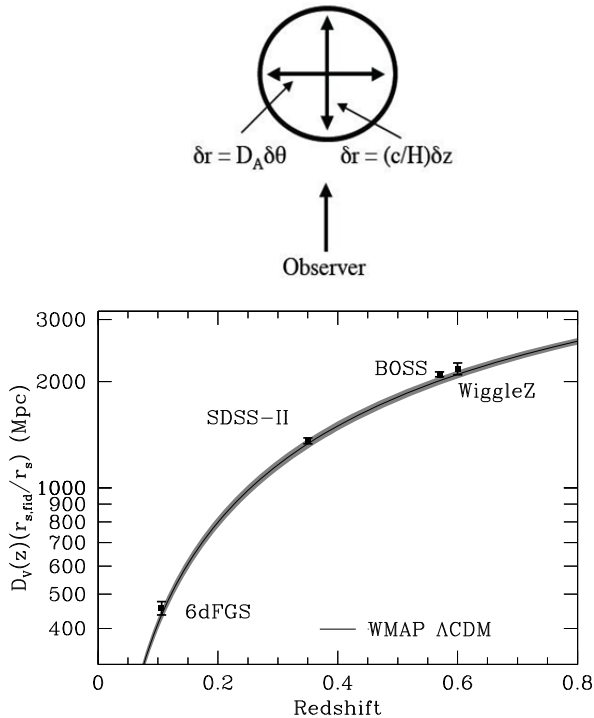

Figure 13: (Top) Illustration of the BAO method. The characteristic BAO scale, $r_s \approx 150$ Mpc, is identified in the three-dimensional clustering of galaxies in angle-redshift space. The transverse scale is used to infer the angular diameter distance $D_A(z)$ and the line-of-sight scale is used to infer the Hubble parameter $H(z)$. (Bottom) Recent BAO distance measurements, compiled by Anderson et al. (2012), compared to the distance-redshift relation predicted for a $\Lambda$CDM cosmological model with parameters constrained by WMAP CMB observations. Here the 3-d clustering is used to constrain an "effective" distance $D_V(z) \propto [D^2_A(z)/H(z)]^{1/3}$, which is scaled to the value of the sound horizon $r_{s,fid}$ computed for the fiducial WMAP parameters. Some recent analyses have also derived separate constraints on $D_A(z)$ and $H(z)$ (Chuang & Wang 2011; Xu et al. 2012).

Our current understanding of the BAO method suggests that it will be statistics-limited even for the largest foreseeable surveys; small corrections are required for the effects of non-linear gravitational evolution and non-linear galaxy bias, but these can be computed at the required level of accuracy using analytic and numerical techniques. The great challenge for BAO cosmology is to map large volumes with the sampling density required to faithfully trace structure on BAO scales. Ground-based galaxy redshift surveys should be able to do this out to redshift z=1 or slightly beyond. For the vast comoving volume available at z=1-3, the most promising approach is space-based, slitless NIR spectroscopy, measuring the redshifts of star-forming galaxies via the strong H$\alpha$ emission line. This approach is completely impractical from the ground because of the much higher NIR sky background. The WFIRST DRM1 HLS, planned for 2.5 years of observations divided nearly equally between imaging and spectroscopy, will measure the redshifts of approximately 16.5 million galaxies at redshifts $1.3 < z < 2.7$ over 3400 deg$^2$, achieving an aggregate BAO distance precision of approximately 0.5%.

When light from galaxies propagates across the Universe, its path is slightly deflected by the gravity of other galaxies, an effect called "weak lensing" (WL). Weak gravitational lensing plays a critical role in studies of cosmic acceleration because it is one of the only ways to directly measure the amplitude of clustering in the matter distribution, enabling comparisons of expansion history and structure growth that can distinguish dark energy theories from modified gravity explanations of acceleration. Weak lensing observables are also sensitive to the expansion history, with precision that can be highly competitive with supernova and BAO measurements. Weak lensing has been quantified by surveys of increasing statistical power over the last decade using both ground and space-based telescopes (see Figure 14). However, away from massive clusters the typical distortion of source galaxy shapes is only about 1%. Measuring the lensing signal with high precision, in the face of intrinsic ellipticity variations that are ~ 0.4 rms, requires enormous galaxy samples *and* exquisite control of systematic errors.

Space-based measurements offer potentially enormous advantages for weak lensing because of high angular resolution and stability of the observing platform, allowing accurate characterization of the instrumental point-spread function (PSF). The WFIRST DRM1 high-latitude imaging survey will measure shapes of approximately 480 million galaxies, with an effective source density of 40 galaxies arcmin$^{-2}$ over 3400 deg$^2$. The resulting aggregate precision in the amplitude of the WL angular power spectrum is approximately 0.3%. Crucially, WFIRST has been designed to keep systematic uncertainties in galaxy shape measurements below the level demanded by this high statistical precision, including independent shape measurements in three filters that allow cross-checks and correction for wavelength-dependent PSF issues and color gradients in the profiles of galaxy images (Voigt et al. 2011; Cypriano et al. 2010). The four-band NIR photometry from WFIRST will be combined with ground-based optical photometry to achieve robust photometric redshifts for source galaxies. Section 2.5.2 discusses WL systematics in some detail, contrasting the issues for WFIRST with those that face Euclid and LSST.

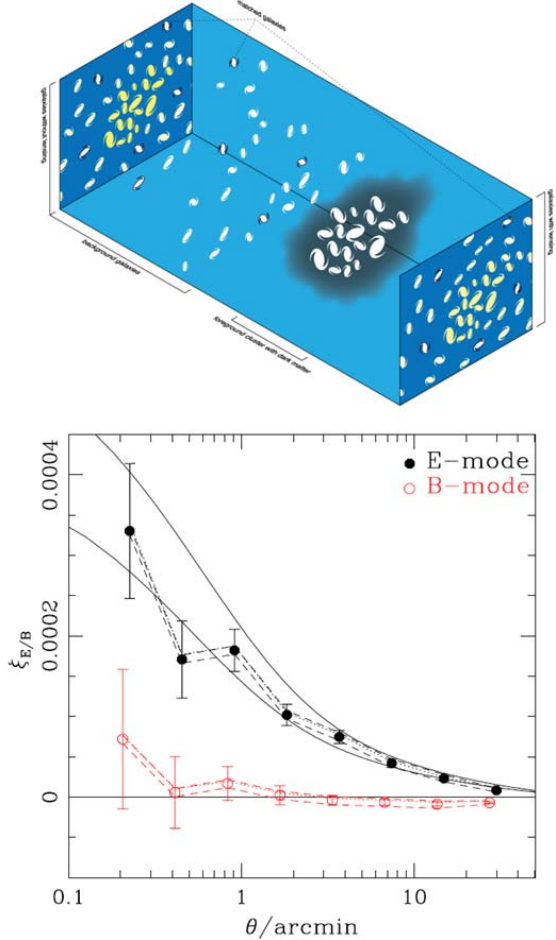

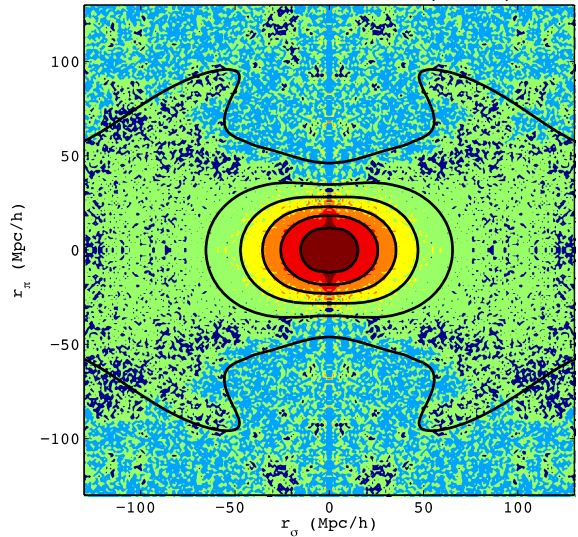

**Figure 14:** (Top) A conceptual illustration of weak lensing: a foreground mass concentration (cluster of galaxies) distorts the shapes of background galaxies, correlating their observed ellipticities. (Bottom) A measurement of the cosmic shear correlation function ξ(Θ) from the HST COSMOS survey, by Schrabback et al. (2010). Black points show the cosmological "E-mode" signal isolated from the data, and black curves show predictions from a ΛCDM cosmological model for power spectrum normalizations σ₈ = 0.7 or 0.8.

The spectroscopic galaxy redshift survey conducted for BAO also offers an alternative probe of the growth of structure. Because galaxy distances are inferred from their redshifts, which are affected by peculiar velocities as well as cosmological expansion, the line of sight becomes a preferred direction. In virialized clusters, the incoherent dispersion of galaxy velocities stretches structures along the line of sight, but on large scales the coherent inflow to overdense regions com-presses them along the line of sight (Sargent & Turner 1977; Kaiser 1987). This redshift-space distortion (RSD) can be measured statistically via the anisotropy of clustering in a galaxy redshift survey, as illustrated in Figure 15. In the linear perturbation theory regime, measurements of RSD constrain the product of the matter clustering amplitude and the clustering growth rate (specifically, the logarithmic derivative of clustering amplitude with expansion factor). An additional, but useful, complication is that the apparent anisotropy in redshift space is also affected by the adopted spacetime metric, and an underlying isotropic pattern would appear so only if one assumed a cosmological model with the correct value of the product H(z)D_A(z) (Alcock & Paczynski 1979; hereafter AP). In addition to BAO, the WFIRST redshift survey will constrain dark energy and modified gravity models via RSD growth measurements and the AP effect. Furthermore, the broad-band shape of the galaxy power spectrum provides a second "standard ruler" for geometrical measurements via the turnover scale imprinted by the transition from radiation to matter domination (TRMD).

**Figure 15:** Redshift-space distortion of galaxy clustering in the Baryon Oscillation Spectroscopic Survey (BOSS), from Reid et al. (2012). Colors show the amplitude of the measured galaxy 2-point correlation function in bins of transverse ($r_\sigma$) and line-of-sight ($r_\pi$) separation, displaying the characteristic flattening of structures along the line of sight and the elongated "fingers of God" at small $r_\sigma$. Contours show model predictions for best-fitting cosmological parameters; the unusual structure of the outermost contour reflects the impact of BAO.

Reasonable statistical forecasts suggest that RSD and AP measurements can provide tight statistical constraints on structure growth and the expansion history, substantially strengthening the results derived from BAO alone (*e.g.*, White, Song, & Percival 2009; Wang et al. 2010). The critical uncertainty in these forecasts is the level of theoretical systematics in modeling RSD and scale-dependent galaxy bias in the mildly non-linear regime. We will characterize the level of these systematics via an effective value of $k_{max}$, a maximum wavenumber to which measurements of the full 3-d galaxy power spectrum can be used to infer cosmological parameters without incurring systematic errors that degrade the statistical uncertainty of the WFIRST measurements. For our standard assumption below ($k_{max} = 0.2$ h Mpc$^{-1}$, which we consider fairly conservative), the aggregate precision of the RSD measurement in the DRM1 redshift survey is 1.4%.

### 2.5.1 Supernova Cosmology with WFIRST

The supernova technique relies on Type Ia supernovae as distance indicators to measure the expansion history of the universe and thus yield information about the acceleration history and the nature of dark energy. Type Ia's have an intrinsic luminosity spread of around 50% at blue rest-frame wavelengths, but the peak luminosity is well correlated with the light curve shape (LCS). Once this correlation is accounted for, the intrinsic dispersion in LCS-corrected peak luminosity is 10-18%, corresponding to distance errors per supernova of 5-9%. The low end of this error range is achieved by observations that measure rest-frame IR wavelengths and/or that use spectral features to define subpopulations of SNe with similar properties. Rest-frame IR observations offer two further advantages: the effects of dust extinction are drastically lower than at optical wavelengths, and the intrinsic dispersion of (uncorrected) peak luminosities is much smaller in the NIR, reducing sensitivity of the derived distance to uncertainties in the LCS measurement. Residuals from the luminosity-LCS relation are weakly correlated with the properties of the SN host galaxy (*e.g.*, Sullivan et al. 2010); this correlation is corrected for empirically in recent analyses (*e.g.*, Conley et al. 2011). Ongoing surveys of local and distant SNe have much more extensive photometric and spectroscopic coverage than the early surveys that discovered dark energy, so they are teaching us a great deal about the SN population as well as improving cosmological constraints. The design and analysis of the WFIRST SN survey will ultimately depend on lessons learned from ground-based surveys and HST/JWST observations between now and the launch of WFIRST.

The key sources of potential systematics for the WFIRST supernova survey are:

- Photometric calibration, including the effects of shifting spectral bandpasses (i.e., k-corrections), across the redshift range of the WFIRST SNe and between the WFIRST sample and local calibration samples.
- Dust extinction, including possible evolution of galactic dust properties; extinction effects must be disentangled from intrinsic correlations between SN luminosity and color.
- Possible evolution of the SN population or contamination of the Type Ia sample by other populations that are systematically less or more luminous.

WFIRST mitigates the first of these by observing from a stable, space-based platform, in three bands, with high angular resolution to separate the supernova from the host galaxy. (With the IFU spectrometer option discussed in Appendix B, k corrections would also be eliminated.) It mitigates the second by observing at NIR wavelengths and by obtaining high-cadence photometry in all three bands. It mitigates the third by obtaining spectra of all SNe used for the cosmological analysis --- employing both multicolor light curves and spectra to eliminate contamination and identify similar cohorts at different redshifts --- and by covering a wide redshift range that affords leverage for separating cosmological and evolutionary effects.

To guide the design and forecast the performance of the supernova survey, we assume that the intrinsic scatter of LCS-corrected peak absolute magnitudes is

$$\sigma_{int} = 0.10 + 0.033 \; z \; \text{mag}$$

and that the systematic error per $\Delta z = 0.1$ redshift bin is

$$\sigma_{sys} = 0.02 \; (1+z) / 1.8 \; \text{mag}$$

with no correlation of errors between redshift bins. The redshift dependence of the intrinsic scatter reflects the higher observed scatter for bluer rest-frame wavelengths. The redshift dependence of the systematic error reflects the increase of evolution systematics with redshift, the greater impact of dust at bluer rest-frame wavelengths, and the difficulty of maintaining consistent photometric calibration and k-corrections over a large redshift span. The statistical error per supernova is the quadrature sum of the intrinsic scatter and the observational error, and the statistical error for a redshift bin is reduced by $\sqrt{N_{SN}}$. The total error for a bin is the quadrature sum of the statistical and systematic errors. We also consider forecasts for an optimistic case in which the systematic errors are a factor of two smaller, *i.e.*,

$\sigma_{sys}$ = 0.01 (1+z) / 1.8 mag (optimistic).

There are two components to the WFIRST supernova survey: imaging in the three reddest bands (JHK) to discover supernovae and measure their multi-band light curves and peak magnitudes, and prism spectroscopy to confirm Type Ia identifications and measure redshifts from the 6100 Å Si absorption line. The supernova survey designed for DRM1 follows a two-tier strategy, with 18 "shallow" fields (6.48 deg$^2$) targeting SNe at z < 0.8 and 5 "deep" fields (1.80 deg$^2$) targeting supernovae to z = 1.7. Exposure times are chosen to yield S/N ≈ 15 photometric measurements at the peak of the light curve for the maximum redshift and aggregate S/N ≈ 5 for the Si feature in the co-added spectrum, at resolution R = $\lambda/\Delta\lambda$ ≈ 75.

The DRM1 supernova survey would be conducted over 1.8 years, with a five-day observing cadence (shorter by 1+z in the SN rest frame). Each set of supernova observations would occupy a contiguous 33-hour block, making the total observing duration of the supernova program (33 / (5 × 24)) × 1.8 × 12 = 6 months[4]. The breakdown of each block is:

- Low-z imaging: 300 sec exposures in 3 bands for 18 fields, 4.5 hours
- Low-z spectroscopy: 1800 sec exposures for 18 fields, 9.0 hours
- High-z imaging: 1500 sec exposures in 3 bands for 5 fields, 6.25 hours
- High-z spectroscopy: 9500 sec exposures for 5 fields, 13.2 hours.

Of course, z < 0.8 SNe will also be found in the deep fields, augmenting the low-z sample by a factor 23/18 and providing a high S/N subset that will be especially valuable for systematics tests.

Figure 16 shows a histogram of the expected number of Type Ia SNe given this survey strategy and duration. The total number is ≈ 1900, with more than 75 SNe in each $\Delta z$ = 0.1 redshift bin from z = 0.35 to z = 1.65. For z > 0.8 we only count SNe found in the deep survey, as the shallow survey will not provide adequate photometry and spectroscopy at these redshifts. Figure 17 plots the expected statistical and total errors per redshift bin, for our conservative and optimistic assumptions about systematics. The statistical errors, which include both intrinsic scatter and photometric errors, are

---

4 The plan described here is slightly inconsistent with the DRM1 operations description in §3.9, which allocates 5.4 months to SN observing, as the final scheduling iteration was done subsequent to our forecasts here.

≤ 0.02 mag in all bins except the first and last, and they are ≈ 0.01 mag in the peak sensitivity range 0.1 < z < 0.7. For our conservative systematics assumption, the systematic errors dominate the statistical errors at most redshifts, while for our optimistic assumption the systematic errors are subdominant.

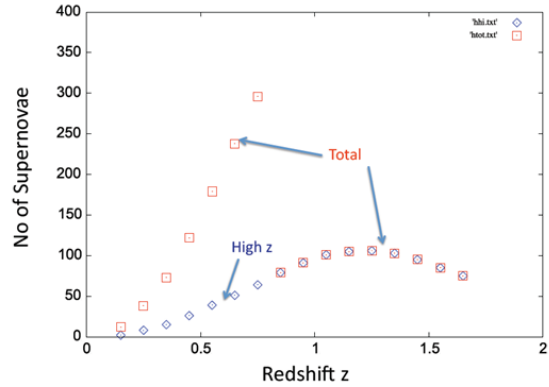

**Figure 16: Expected number of Type Ia SNe in each $\Delta z$ = 0.1 redshift bin for the DRM1 SN survey. Red squares show the total number of SNe Ia, and blue diamonds show the number from the deep fields. (At z > 0.8, all SNe are from the deep fields.)**

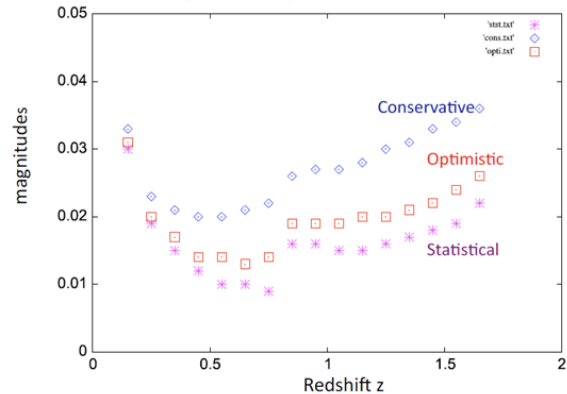

**Figure 17: Statistical errors (asterisks) and total errors in distance modulus per $\Delta z$ = 0.1 bin. The discontinuity at z = 0.8 reflects the smaller area of the deep field survey. For total errors, blue diamonds indicate results for our conservative systematic error assumption and red squares indicate results for our optimistic assumption. Fractional errors in distance are ½ the magnitude error in distance modulus (see Figure 26).**

The critical trade decision in the SN survey is the maximum redshift $z_{max}$, since for a lower $z_{max}$ one can take shorter exposures and thus cover a larger area. Figure 18 shows the result of such a trade study for the WFIRST survey, plotting the Dark Energy Task Force Figure of Merit (DETF FoM; see §2.5.4 below) for several choices of the maximum redshift of the deep field survey, with corresponding changes in the number of deep fields that keep the total observing time fixed. There are two key points to take away from this analysis. First, for either choice of systematics prior, the FoM increases with $z_{max}$, even though our forecast model includes increasing intrinsic scatter and systematic uncertainty with redshift. Second, there is a large difference in performance between our optimistic and conservative cases for systematic errors: the $z_{max}$ = 1.2 optimistic case is better than the $z_{max}$ = 1.7 conservative case.

The $z_{max}$ trend partly reflects the benefits of greater leverage from a wider redshift range, but also the fact that once the statistical errors become comparable to the systematic errors then one gains more new information by adding new redshift bins than by improving the statistics for existing bins. Our assumption that systematics are uncorrelated from bin to bin is important to the quantitative form of Figure 18; correlated systematics would reduce the FoM because separate bins would provide less independent information. If one can reduce *systematic* errors in the low-z bins by observing larger samples (enabling, for example, more finely matched cohorts or more stringent selection of SNe for the cosmological analysis), or by increasing exposure times to obtain higher S/N data, then it may be beneficial to reduce $z_{max}$. However, once the systematic error improvement is saturated, it will generally be beneficial to extend to higher redshift. This strategy question should be revisited closer to mission launch, based on what should by then be much better insight into percent-level systematics of the SN population, better information about high-redshift SN rates, and detailed simulations of WFIRST photometry and calibration. We also note that Figure 18 assumes only Planck priors in addition to WFIRST SNe and a local (z < 0.1) calibration sample of ~800 SNe. There are good arguments for optimizing each of the WFIRST probes individually, but an analysis that considers the combined performance of all probes could lead to different optimizations. For now, we take Figure 18 as justification of our choice of $z_{max}$ = 1.7 in the DRM1 supernova survey, but we specify as a requirement only the ability to conduct a survey with >100 SNe-Ia per $\Delta z$=0.1 bin to $z_{max}$ = 1.2.

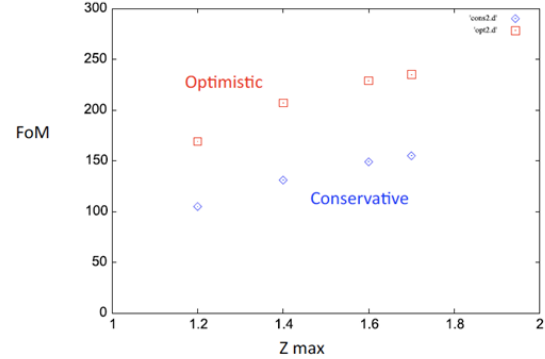

**Figure 18: DETF Figure of Merit for the WFIRST SN survey, as a function of the maximum redshift adopted for the deep field component (at fixed total observing time). Blue diamonds and red squares show forecasts for the conservative and optimistic systematics assumptions, respectively. In addition to WFIRST SNe, we incorporate Planck priors and a local calibration sample of ~800 SNe, but no other cosmological probes**.

The scientific goals of the WFIRST SN survey lead to the following requirements:

**Survey Capability Requirements**.
- >100 SNe-Ia per $\Delta z$=0.1 bin for most bins for 0.4 < z < 1.2, per dedicated 6 months
- Observational noise contribution to distance modulus error $\sigma_\mu \leq 0.02$ per $\Delta z$=0.1 bin up to z = 1.7.
- Redshift error $\sigma \leq 0.005$ per supernova
- Relative instrumental bias $\leq 0.005$ on photometric calibration across the wavelength range.

**Data Set Requirements**
- Minimum monitoring time-span for an individual field: ~2 years with a sampling cadence $\leq$ 5 days
- Cross filter color calibration $\leq 0.005$
- Three filters, approximately J, H, K
- Slitless prism spec (P130) 0.6-2 $\mu$m, $\lambda/\Delta\lambda$ ~75, 2-pixel (S/N $\geq$ 2 per pixel bin) for redshift/typing
- Photometric S/N $\geq$ 15 at lightcurve maximum in each band at each redshift
- Dither with 30 mas accuracy
- Low Galactic dust, E(B-V) $\leq 0.02$.

If something like our optimistic systematics error case can be achieved in practice, then there are significant cosmological gains from extending the survey beyond 2 years or conducting a second survey in an extended mission.

We have carried out some investigations of possible hardware options that would improve photometric calibration accuracy, based on LEDs and integrating spheres. A relative flux calibration system would provide diffuse illumination of the focal plane with precisely adjustable increments in intensity. Measuring the apparent brightness of stars as the background intensity is varied gives a direct measure of the linearity of the system response. Adjusting the introduced flux intensity from levels comparable to the sky up to that of stellar flux standards enables accurate transfer of the flux calibration for supernova photometry, measurement of the system PSF over the full dynamic range from the core to the extended tails, and other demanding applications.

Appendix B discusses a possible hardware option to the WFIRST DRMs: the addition of an integral field unit (IFU) spectrometer. This capability could be exploited by the SN survey in two rather different modes (or a mixture of the two). First is to simply replace the prism spectroscopy with IFU spectroscopy, still with the goal of obtaining moderate S/N spectra sufficient for a $5\sigma$ detection of the 6100Å Si feature. Despite the loss of multiplexing, we find that this strategy reduces the total time needed for spectroscopy because of the greater throughput of the instrument and the ability to take shorter exposures for brighter SNe (always choosing the epoch closest to maximum). Greater observing efficiency allows a larger SN survey, leading to a significantly higher FoM. The second strategy is to obtain deep, high S/N spectra that can be used (a) to create SN cohorts across redshift bins that are matched in detailed spectral properties, reducing intrinsic scatter and evolutionary systematics, and (b) to base the cosmological distance measurement on IFU spectrophotometry rather than filter photometry, enabling precisely matched rest-frame bandpasses (i.e., no k-corrections, which introduce significant systematic errors) and potentially lower calibration systematics. The reduction of intrinsic scatter alone does not justify this strategy, but reduction of systematics might. We have not included the IFU in DRM1 or DRM2 because we wanted to keep the hardware complement to the minimum required to accomplish WFIRST's primary science goals. However, an IFU would almost certainly improve the performance of the SN survey, and it would undoubtedly find uses in general observer programs.

### 2.5.2 Weak Lensing Cosmology with WFIRST

The weak lensing technique uses the shear of distant galaxies by gravitational lensing, which is determined by an integral of the tidal field along the line of sight to the source galaxy. The strength of the lensing signal as a function of the redshift of source galaxies is sensitive both to the growth of cosmic structure at intervening redshifts $0<z<z_{source}$ and to the distance-redshift relation (since the same amount of mass produces more shear on more distant sources). In addition to measurements of cosmic shear (the correlated ellipticities of source galaxies lensed by the same foreground structure), the correlation of lensing shear with foreground galaxies of known redshift $z_{fg}$ probes the relation of these galaxies to the matter field; this "galaxy-galaxy lensing" signal can be combined with measurements of galaxy-clustering to infer the amplitude of matter fluctuations. A weak lensing survey must provide wide angle sky coverage with a high density of usable sources across a wide redshift range (local to z~2), provide photometric redshifts so that the galaxies can be sliced into bins to study evolution of the signal and remove "intrinsic" (not lensing induced) shape correlations, and achieve exquisite control over coherent systematic errors in the shear signal. The latter includes both "additive" shear errors, i.e., apparent shear induced by imperfect correction for instrument effects (aberrations, anisotropic jitter, geometric distortions, etc.) that is not present in the sky, and "multiplicative" shear errors, where the calibration of the observed shear is incorrect. Some sources of error (*e.g.*, failure to recover full sampling) can lead to both additive and multiplicative errors.

The WFIRST weak lensing experiment has been designed with control of systematics foremost in mind. The thermal design and L2 orbital location of WFIRST are intended to produce a highly stable PSF. With ground-based imaging, the atmosphere is constantly changing the PSF in a way that is difficult to characterize *a priori*, while for WFIRST the PSF should change minimally with time, and its position and wavelength dependence should be described by a moderate number of parameters that characterize the optics of the telescope. With the dithering pattern planned for the HLS (§2.5.2.1) and the image combination algorithms developed by Rowe et al. (2011), WFIRST will achieve fully sampled images for shape measurements in its three reddest filters (JHK), even in the presence of defects such as cosmic ray hits. The $\geq 5$ exposures per filter are separated into two passes several months apart at different roll angles, facilitating photometric calibration embedded in the survey. WFIRST's unobstructed telescope eliminates the roll angle and SED-dependent diffraction spikes that occur in on-axis systems. The 3 filters allow internal cross-checks: the WL signal is achromatic but many systematic errors may not be, so correlating shapes measured in different bands is a powerful test. Multiple filters are also important for addressing wavelength-dependent PSF issues in galaxy shape measurements (Voigt et al. 2011; Cypriano et al. 2010); with a single shape measurement filter, the difference in PSF between galaxies and calibration stars can easily become a limiting systematic that is both difficult to detect and problematic to correct, especially in galaxies that have a radial color gradient. Robust photometric redshifts require a combination of optical and NIR photometry, to suppress confusion between the spectral breaks near 4000 Å and 1216 Å. WFIRST will measure galaxy magnitudes in Y band as well as the three shape measurement bands, which will be combined with ground-based optical photometry from LSST to obtain photo-z's for all of the WL source galaxies.

Figure 19 shows the imaging depth achieved by the WFIRST HLS, in comparison to LSST (after 10 years of operation) and Euclid. In an AB-magnitude sense, WFIRST imaging is well matched to the z-band depth of LSST. The AB magnitude limits for LSST in g, r, and i are fainter; however, a typical z>1 LSST weak lensing source galaxy has r-J or i-H color of about 1.2, so even here the WFIRST imaging depth remains fairly well matched, and of course the angular resolution is higher. The Euclid IR imaging is two magnitudes shallower than WFIRST, and it is undersampled, so it cannot be used for galaxy shape measurements. The Euclid optical imaging depth is comparable to LSST, but in a single very wide filter rather than the ~5 optical bands necessary for photometric redshift determination.

The statistical power of a WL survey depends mainly on the total number of galaxies with reliable shape measurements, whis is the product of the survey area and the effective number density of usable source galaxies $n_{eff}$ (*i.e.*, the number density downweighted by the measurement noise in the ellipticity). For the DRM1 HLS, after removing sources lost to CR hits, the total effective densities are 30, 32, and 31 galaxies arcmin$^{-2}$ in the J, H, and K bands, respectively. The total number of unique objects in all bands is 40 arcmin$^{-2}$. Galaxies generally have higher S/N in the redder bands, which leads to the increase of $n_{eff}$ between J and H, but the larger PSF size in K-band leads to a slight drop in $n_{eff}$. The area of the HLS imaging survey is 3400 deg$^2$, so with 40 galaxies arcmin$^{-2}$ DRM1 provides a total of 480 million shape measurements. The use of three shape measurement bands allows the construction of 6 auto- and cross-correlation shear power spectra, providing a great deal of redundancy. Comparison of these spectra will enable end-to-end consistency tests of the long chain of corrections that are necessary in any weak lensing analysis. Since they are based on (almost) the same set of galaxies, the differences of these spectra can diagnose even systematic errors that are well below the statistical error of the survey; the lensing signal itself should be achromatic, but many of the measurement systematics would not be.

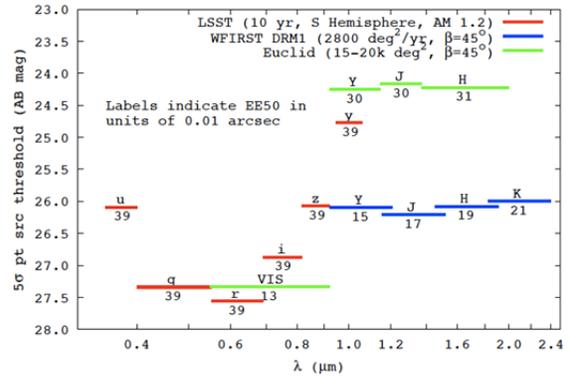

**Figure 19: Depth in AB magnitudes of the WFIRST (blue), Euclid (green), and LSST (red) imaging surveys. Labels below each bar indicate the size of the PSF (specifically, the EE50 radius) in units of 0.01 arcsec.**

Figure 20 shows the predicted cosmic shear angular power spectrum for ten tomographic bins of source galaxy photometric redshift together with the projected statistical errors for the DRM1 WL survey. Comparison to Figure 19 illustrates the dramatic advance that WFIRST would bring relative to the current state of the art, and it brings home the correspondingly dramatic improvements required in control of systematic errors. For multipole L less than a few hundred, the statistical errors are dominated by sample variance in the lensing mass distribution within the survey volume, while for high L or the extreme redshift bins the errors are dominated by shape noise, *i.e.*, the random orientations of the source galaxies that are available to measure the shear. For a given photo-z bin the statistical errors at different L are uncorrelated. Across photo-z bins, the errors at lower L are partly correlated because the same foreground structure contributes to the lensing of all galaxies at higher redshift; the errors decorrelate at high L when they become dominated by shape noise, which is independent for each set of sources. While Figure 20 displays the auto-spectra for the ten photo-z bins, there are also 10×9/2 = 45 cross-spectra that provide additional information.

The cosmological sensitivity of the WL experiment arises from the dependence of the predicted lensing power spectrum on parameters describing dark energy and structure growth. An overall increase in the amplitude of matter clustering would raise all of the model curves in Figure 20 by the same factor, while a change in the *rate* of structure growth would shift the curves relative to each other. Changes in the distance-redshift relation would also shift the curves relative to each other, as the lensing signal depends on the distance to lenses and sources. The mix of correlated and uncorrelated errors makes it difficult to read the aggregate precision of the WL measurement off a plot like Figure 20. We find that the overall measurement precision of the amplitude of the WL power spectrum (i.e., a constant multiplicative offset across all bins) is 0.29%. For an approximate scaling argument that gives a similar result, see §5.4 of Weinberg et al. (2012).

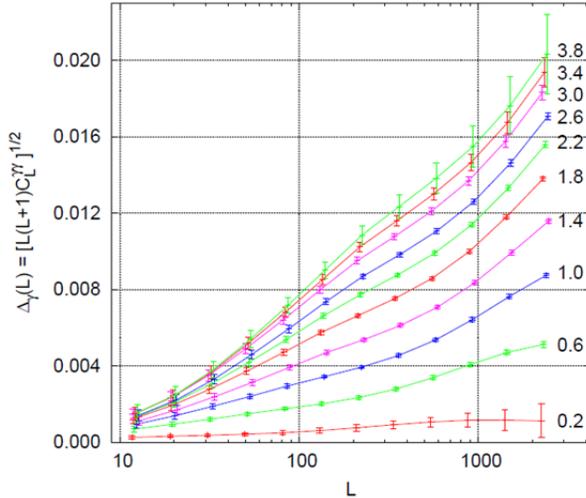

**Figure 20: The predicted cosmic shear angular power spectrum and associated statistical errors for the DRM1 WL survey, for ten tomographic bins of source photometric redshift as labeled. L is the angular multipole, and $L(L+1)C_L^{\gamma\gamma}$ is the contribution to the variance of cosmic shear per logarithmic interval $\Delta \ln L$.**

This statistical precision in turn defines the required level of control for measurement systematics. Given the possibility of an extended mission that would allow a larger area HLS, we have set our requirements for additive and multiplicative shear errors so that WFIRST shape measurement remains statistics-limited even in the event that an extended mission covers 10,000 deg². Based on the formalism of Amara & Refregier (2008), we adopt requirements of $\leq 10^{-3}$ for multiplicative bias and $\leq 3 \times 10^{-4}$ for additive bias. At a factor-of-two level, we can understand these numbers from the aggregate precision of a survey with three times the area of the DRM1 HLS, $(0.29\%)/\sqrt{3} \sim 1.7 \times 10^{-3}$. The multiplicative shear error must be smaller than this to avoid degrading the statistical errors. An additive shear error A sums in quadrature with the true shear signal S, so for a typical rms shear amplitude $S \sim 0.01$ we require $(S^2 + A^2)^{1/2}/S - 1 < 1.7 \times 10^{-3}$, implying $A < 6 \times 10^{-4}$.

The PSF ellipticity must be known well enough to ensure that we meet the additive shear error requirement, and the PSF second moment $(I_{xx} + I_{yy})$ must be known well enough to meet the multiplicative shear error requirement. The data set requirements on PSF knowledge were calculated to use 50% (in a root-sum-square sense) of the shape measurement error budget (which also includes other terms, such as residuals from the data processing algorithms and detector effects). The PSF knowledge requirements apply to all sources of PSF fluctuations in the same range of multipoles (large scales: L< $10^4$) that might be used for the WL power spectrum. For the IDRM, we carried out studies (both Fisher matrix analyses and simulation suites) to show that the number of PSF calibration stars per field is sufficient to recover the PSF with the required accuracy given the relatively small number of degrees of freedom in the optical system. We have not repeated these calculations for DRM1/DRM2, but we expect the results to be similar. This work, along with analysis of pointing stability requirements, is briefly summarized in §3.10.

### 2.5.2.1   HLS Imaging Observing Strategy

The WFIRST HLS-imaging observing strategy must support at least three modes: an imaging-only mode, used for the Y band filter; a weak lensing mode, used for the J, H, and K filters; and a spectroscopy mode (§2.5.3.1).

- All modes must support the rejection of image defects (cosmic ray hits, hot pixels, persistence artifacts) through the use of ≥3 dithered exposures (random subpixel offsets are acceptable).
- The imaging survey mode must support über-calibration – the ability to derive an overall relative calibration solution by repeated observations of objects at different times and different parts of the focal plane. Each repeat observation of an object ties together the solution at two points on the focal plane, and we require each point to be tied to any other point through a small number of such steps. The simplest implementation of this, used for the WFIRST HLS, is to have multiple passes in each filter over each part of the sky at ≥2 roll angles.

- The weak lensing survey must achieve full sampling through some combination of dithers and rolls; this can be evaluated using image reconstruction simulations (Rowe et al. 2011). Advances in image reconstruction algorithms have eliminated the JDEMΩ requirement that all of the images in a given filter have the same roll angle. However, the large-step dithers required to cover chip gaps result in random subpixel offsets due to geometrical distortions.

As a result of these considerations, for the DRM1 a tiling strategy was defined that consists of 4 exposures, each offset diagonally by slightly more than a chip gap. The sky is then tessellated with this pattern (i.e. it is repeated in both the X and Y directions spaced by the field size). This is considered to be 1 "pass" and achieves a depth of 2—4 exposures. A second pass at a different roll angle (but the same filter) achieves a depth of 4—8 exposures (90% fill at ≥5 exposures) – see Figure 21 for a graphical illustration. The J band has 5 exposures instead of 4 in one pass since we are attempting WL shape measurement in this band and it has the tightest sampling requirements.

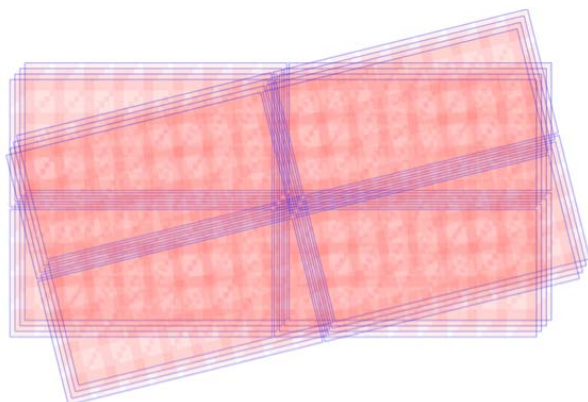

**Figure 21: Illustration of the rolled tiling strategy for DRM1 in the Y, H, and K band filters. The J-band observations have one additional exposure in one of the passes.**

For DRM2, the basic survey design is similar, except that due to the smaller chip gaps, 90% fill at ≥5 exposures (our target for the imaging and WL filters) can be achieved with one 4-exposure and one 3-exposure pass. Furthermore, due to the smaller primary mirror and consequently larger PSF, the additional exposure in J band is not required.

The exposure times for both DRMs, and the consequent survey rates, are shown in Table 7 (DRM1) and Table 8 (DRM2).

| Mode | Exposure Time | Survey Time [days per 1000 deg²] |
|---|---|---|
| Imaging – Y | 5 x 150 s | 31.3 |
| Imaging – J | 6 x 150 s | 36.8 |
| Imaging – H | 5 x 150 s | 31.3 |
| Imaging – K | 5 x 150 s | 31.3 |
| Spectroscopy | 6 x 530 s | 126.6 |
| **Total** | | 257.3 |

**Table 7: Survey Rate Parameters for DRM1. (Exposure times refer to the 10th percentile of the depth distribution, i.e. ≤10% of the sky is worse.)**

| Mode | Exposure Time | Survey Time [days per 1000 deg²] |
|---|---|---|
| Imaging – Y | 5 x 247 s | 31.5 |
| Imaging – J | 5 x 247 s | 31.5 |
| Imaging – H | 5 x 247 s | 31.5 |
| Imaging – K | 5 x 247 s | 31.5 |
| Spectroscopy | 6 x 567 s | 83.4 |
| **Total** | | 209.2 |

**Table 8: Survey Rate Parameters for DRM2. (Exposure times refer to the 10th percentile of the depth distribution, i.e. ≤10% of the sky is worse.)**

### 2.5.2.2    HLS Imaging Survey Performance

Survey depths were computed using the WFIRST ETC v10 (Hirata et al. 2012). The point source sensitivities of the surveys at 5 exposure depth are shown in Table 9. These are presented at ecliptic latitude $\beta$ = ±45° and solar elongation $\varepsilon$ = 115°, but the depth varies slowly as a function of ecliptic latitude. Typically the depth is 0.07 magnitudes shallower at $\beta$ = ±30° and 0.04 magnitudes deeper at $\beta$ = ±60°.

| Band | Depth DRM1 | Depth DRM2 |
|---|---|---|
| Y | 26.10 | 25.93 |
| J | 26.21 | 25.92 |
| H | 26.08 | 25.95 |
| K | 26.00 | 25.82 |

**Table 9: The point source sensitivities (5σ, PSF photometry) for the WFIRST HLS. Depths are in AB magnitudes and are observed-frame (not corrected for Galactic extinction). The zodiacal background was taken at ecliptic latitude ±45°.**

For both DRMs, the maximum WL shape density at fixed exposure time is achieved in the H band. The optimal shape measurement band is a balance between the galaxy spectra, which provide more photons per unit time as one goes to the red; the PSF size, which is smaller in the blue; and the zodiacal background, which is slightly blue (1.5× greater in Y band than K band, in units of photons/cm²/s/log λ). We are thus driven to carry out the WL shape measurement in the reddest band at which an acceptable number of galaxies are resolved. In order to provide color corrections and internal checks, WFIRST performs the shape measurement in both this optimal band (H) and the bands on either side (J,K).

The imaging exposure times were set to achieve $n_{eff} \approx 30$ galaxies/arcmin² for WL. For DRM1, this number of galaxies is reached internally to the H and K band filters (J band reaches 29.7 galaxies/arcmin²). For DRM2, the smaller primary mirror diameter and consequent larger PSF reduced the number of bright, resolved galaxies and forced very long integration times (*e.g.* 5× 410 s in K band). Therefore in DRM2 we required $n_{eff} \approx 30$ galaxies/arcmin² only after combining the galaxy catalogs from the 3 filters.

The WL galaxy survey yields were computed in each band according to the following criteria: (1) detection significance >18σ; (2) ellipticity measurable to $\sigma_e \leq 0.2$ (per component), where the ellipticity e is related to the axis ratio q by e=(1−q²)/(1+q²); and (3) resolved, i.e. the intrinsic effective radius of the galaxy should be >0.82× that of the PSF. These are the same criteria used for the IDRM report.

Since the number of exposures varies due to the tiling geometry, the shape densities shown are averaged over the histogram of exposure depth, with 2 exceptions. First, an exposure is considered "killed" if the galaxy is hit by a cosmic ray (CR track centerline within 3 pixels of the galaxy center). WFIRST will use the multiple read capability (114 samples in DRM1) to recover from CRs, but the recovered ramp slopes might not meet the stringent accuracy requirements for WL. A solar minimum galactic CR flux of 5 cm⁻² s⁻¹ (Barth et al. 2000) was assumed. Secondly, regions with insufficient numbers of dither positions after CR flagging (<5,4,4 in J,H,K in DRM1, and <4,4,3 in DRM2) were removed entirely.

The overall WL galaxy yield is shown in Table 10; we show both the individual shape measurements in each band (J,H,K) and the union sample. The union sample combines the shape catalogs from each band – it does *not* involve combining images from different bands to improve the S/N. We show the effective source density of galaxies $n_{eff}$, which takes into account the lower utility of a galaxy where the measurement noise is comparable to the intrinsic dispersion in galaxy ellipticities (~0.4): each galaxy receives a weight between 0 and 1 given by w = [1 + (σ_e/0.4)²]⁻¹, where σ_e is the ellipticity measurement error.

Figure 22 shows the expected redshift distributions of the source galaxy populations, for the "union" sample that has a good shape measurement in one or more of the three bands. The quartiles of the redshift distribution are 0.66/1.06/1.68 (DRM1) and 0.62/0.98/1.55 (DRM2).

| Band | WL Shape Density $n_{eff}$ [arcmin⁻²] | |
|---|---|---|
| | **DRM1** | **DRM2** |
| **J** | 29.7 | 24.1 |
| **H** | 32.4 | 26.5 |
| **K** | 30.9 | 24.9 |
| **Union** | 39.6 | 31.3 |

**Table 10: The weak lensing galaxy yields for WFIRST. The zodiacal background was taken at ecliptic latitude ±45°, and a Galactic dust column of E(B−V)=0.05 was assumed.**

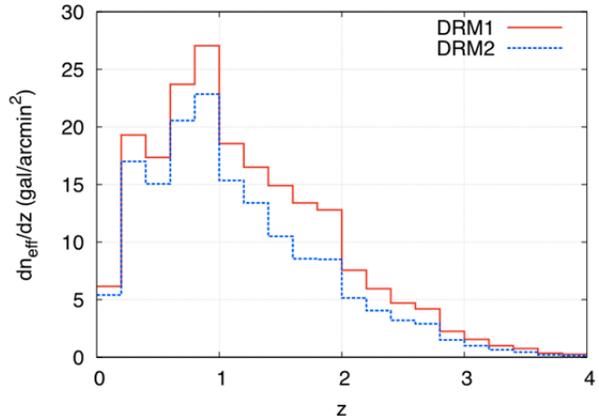

**Figure 22: Expected redshift distributions of the galaxies with shape measurements in WFIRST.**

### 2.5.2.3    Additional Systematic Errors

While photometric redshifts do not have to be precise on an individual galaxy basis, there are stringent demands on the knowledge of the *distribution* of true redshifts in each bin of photometric redshift. An error in the mean true redshift of a bin changes the predicted cosmic shear signal by a fractional amount that is similar to the fractional error in (1+z), so the requirement on knowledge of mean photo-z offsets is similar to the re-quirement on multiplicative shear calibration. There will inevitably be a fraction of photo-z outliers that populate extended tails of the true redshift distribution for a given photo-z. The existence of outliers is not by itself problematic, but their distribution must be known well in order to compute the theoretical WL power spectra for comparison to the observations. Calibration of the photo-to-z distributions requires spectroscopic redshifts for a representative sample of the WL source galaxies. The spectroscopic calibration sample should contain ≥ $10^5$ galaxies (Bernstein & Huterer 2010) spread over several fields that are observable from WFIRST, from the ground-based telescopes that provide optical photometry, and from any assets needed to complete the spectroscopic survey. Completing such a survey to the WFIRST imaging depth will be a major challenge in itself, and the optimal division between deep WFIRST spectroscopy and ground-based spectroscopy will have to be decided closer to launch. An alternative approach (Newman 2008) is to calibrate the photo-z distribution via the cross-correlation between WL source galaxies and the spatial distribution of brighter galaxies and quasars in large area spectroscopic surveys, including the WFIRST galaxy redshift survey and ground-based surveys that probe other redshift ranges, such as BOSS, BigBOSS, Subaru PFS, and HETDEX. Ultimately, WL analyses from WFIRST (and Euclid, and LSST) will likely use a combination of the direct calibration and cross-correlation approaches, but starting from a strong base of optical+NIR photometry for all source galaxies is crucial in any case.

The primary theoretical systematics for cosmic shear are accurate prediction of the non-linear matter power spectrum, including baryonic effects, and the effects of intrinsic alignments (IA) between galaxies that are physically close to one another and between galaxies and the surrounding tidal field. (These alignment signals are usually referred to as II and GI, respectively.) We anticipate that the first problem will be solved via numerical simulations on the timescale of WFIRST. The second problem is addressed by marginalizing over a parameterized description of the intrinsic alignments. Doing so without substantial loss of sensitivity requires photo-z values that are reasonably precise on a galaxy-by-galaxy basis and have moderate outlier fractions, since it is the redshift dependence of the signal that is used to separate true shear from intrinsic alignments. (In particular, intrinsic alignments of outlier galaxies cannot be removed in a model-independent way: this represents a systematic error risk if the outlier fraction is large, *e.g.* due to confusion of the Lyman-α and Balmer/4000 Å breaks as occurs without NIR photometry. The seriousness of this risk depends on the as-yet unknown amplitude of intrinsic alignments at high redshift.) If one uses galaxy-galaxy lensing and galaxy clustering in addition to cosmic shear, then there are additional theoretical systematics associated with the redshift and scale dependence of galaxy bias and the cross-correlation between the galaxy and matter fields. At present, it is not clear how much freedom must be allowed in accounting for these effects.

### 2.5.2.4 Parameter Forecasting Methodology

In our forecasts in §2.5.4 below, we will consider both conservative and optimistic assumptions about weak lensing systematics. Our conservative assumptions are as follows, separated into lists of measurement systematics and modeling systematics.

*Measurement:*

- A covariance matrix of shear calibration uncertainties $C_{ee}^{ij(obs)}$ (l)=(1+$f_i$)(1+ $f_j$) $C_{ee}^{ij}$(l) with a prior on $\sigma(f_i)$=0.001$\sqrt{N_{ph}}$ independently in each of the $N_{ph}$=10 photo-z bins. This represents the impact of a multiplicative shear bias with aggregate uncertainty over all bins of $10^{-3}$, equal to our specified technical requirement. We do not explicitly incorporate an additive shear calibration uncertainty in our forecast, but meeting the requirement that it be < 3×$10^{-4}$ would ensure that it has negligible impact.
- Photometric redshift offsets with a prior $\sigma(\Delta z_{sys})$ = 0.002(1+z) in each photo-z bin.

*Modeling:*

- We include uncertainties about galaxy bias by introducing a bias amplitude $b_g$, with $P_{gg}(k,z)^{obs}$ = $b_g(k,z)^2$ $P_{mm}(k,z)$, and a cross correlation coefficient between galaxy position and shear measurements, $r_g$, $P_{gm}(k,z)^{obs}$ = $b_g(k,z)$ $r_g(k,z)$ $P_{mm}(k,z)$. The Fisher analysis includes marginalization over an $N_{bias} \times N_{bias}$ grid of $b_g$ and $r_g$, logarithmically spaced in z and k with $N_{bias}$=5. The values at each scale and redshift come from interpolating over the grid.
- Acknowledging that our galaxy bias marginalization may not be sufficient to describe the fully nonlinear regime, we include a cutoff in multipole space for each photometric redshift bin, $l_{max}(z_i)$ = 0.132 $z_i$ h Mpc$^{-1}$ (Rassat et al. 2008, Joachimi & Bridle 2010). Galaxy position correlations with l > $l_{max}(z_i)$ are excluded from the Fisher analysis.

- Intrinsic alignment (IA) contributions to the observed shear field are modeled using a nonlinear alignment model (Hirata and Seljak 2004, Hirata et al. 2007). We marginalize over uncertainties in the amplitude of the IA auto-correlation and cross-correlation with galaxy position using an analogous approach to galaxy bias, marginalizing over 5x5 grids for two parameters $b_i$ and $r_i$ (Joachimi and Bridle 2010). As in the FoMSWG analysis (Albrecht et al. 2009) we include a prior on $\sigma(b_i r_i) = 0.003 N_{bias} \sqrt{(N_{ph}-1)}$

The first two of the modeling systematics have no effect on the cosmic shear constraints, only on the inferences from galaxy-galaxy lensing and photometric galaxy clustering; however, they add sufficient freedom to the description of galaxy bias that the galaxy-galaxy lensing measurement adds essentially no useful cosmological information. For the optimistic scenario we assume *no* shear calibration or photo-z offset uncertainties; i.e., the measurements themselves are statistics limited. We further assume that the IA signal can be removed perfectly without increasing statistical errors, and we assume that galaxy bias can be described by a single, scale-independent bias factor $b_g(z)$ at each redshift. In practice we expect some scale-dependence of bias and some departure from $r_g$=1, but this scenario represents the case in which one can model the observed galaxy clustering well enough to predict this scale dependence and cross-correlation at the required level of accuracy.

### 2.5.2.5    Survey Requirements

The scientific goals of the WFIRST WL survey lead to the following requirements. Note that some data sets required for WL are to be obtained from WFIRST, while others are possible from the ground. The data set requirements noted below apply to the space (WFIRST) data set, except for the points listed "From Ground:" Because the imaging and spectroscopic portions of the HLS are designed to cover the same area, we set requirements on the combined survey speed rather than the two survey speeds separately.

**Survey Capability Requirements**

- $\geq 1400$ deg$^2$ per dedicated observing year (combined HLS imaging and spectroscopy)
- Effective galaxy density $\geq 30$ per arcmin$^2$, shapes resolved plus photo-z's
- Additive shear error $\leq 3 \times 10^{-4}$
- Multiplicative shear error $\leq 1 \times 10^{-3}$

- Photo-z error distribution width $\leq 0.04(1+z)$, catastrophic error rate <2%
- Systematic error in photo-z offsets $\leq 0.002(1+z)$

**Data Set Requirements**

- From Space: 3 shape/color filters (J, H, and K), and one color filter (Y; only for photo-z)
- S/N $\geq 18$ (matched filter detection significance) per shape/color filter for galaxy $r_{eff}$ = 250 mas and mag AB = 23.9
- PSF second moment ($I_{xx} + I_{yy}$) known to a relative error of $\leq 9.3 \times 10^{-4}$ rms (shape/color filters only)
- PSF ellipticity ($I_{xx}-I_{yy}$, $2I_{xy}$)/($I_{xx} + I_{yy}$) known to $\leq 4.7 \times 10^{-4}$ rms (shape/color filters only)
- System PSF EE50 radius $\leq 166$ (J band), 185 (H), or 214 (K) mas
- At least 5 (H,K) or 6 (J) random dithers required for shape/color bands and 4 for Y at same dither exposure time
- From Ground: $\geq 4$ color filter bands ~0.4 $\leq \lambda \leq$ ~0.92 $\mu$m
- From Ground + Space combined: Complete an unbiased spectroscopic PZCS training data set containing $\geq 100,000$ galaxies with $\leq$ mag AB = 23.9 (in JHK bands) and covering at least 4 uncorrelated fields; redshift accuracy required is $\sigma_z < 0.01(1+z)$

### 2.5.2.6    Relation to Euclid and LSST

We conclude this section with a brief qualitative comparison between the WFIRST, Euclid, and LSST weak lensing experiments. This comparison pertains to the shape measurements – in other areas, such as photometric redshifts, only a combined program (optical + NIR) can achieve the requirements.

Where the WFIRST, Euclid and LSST surveys overlap, they will measure the shapes of many of the same galaxies, with WFIRST pushing to slightly higher median source redshifts due to making observations in the IR. WFIRST has been specifically designed to make these shape measurements with the maximum possible control and understanding of systematics. It achieves the thermal stability available from L2, eliminates atmospheric turbulence, *and* has multiple layers of redundancy built in to the instrument and survey strategy in order to achieve this exquisite systematics control.

As a ground-based project, LSST has lower angular resolution than WFIRST, and hence with similar cuts to define resolved galaxies might achieve a lower source density $n_{eff}$ than we expect from WFIRST. But due to its much larger sky coverage (the whole Southern Hemisphere), LSST will make a larger number of shape measurements. The main challenge for LSST is the time-variability of the PSF that is inevitable in ground-based observations due to atmospheric turbulence and varying mechanical and thermal conditions. This has thus far been a persistent source of systematic uncertainty in WL measurements, greatly complicating the WL analyses of data sets such as CFHTLS that are <1% the size of LSST (see, *e.g.*, Fu et al. 2008; Tereno et al. 2009; Heymans et al. 2012). The large number of visits, multiple filters, and highly redundant survey strategy of LSST offer many opportunities to mitigate these systematics, and we are confident that LSST will achieve greatly improved control of systematics relative to present-day ground-based WL surveys. The extent of this improvement is difficult to quantify given our present state of understanding of the time-variability of ground-based PSFs. WFIRST, by observing above the effects of the atmosphere and utilizing the thermal stability afforded by L2, completely bypasses these difficulties.

Experience with HST has identified the main problems in space-based lensing -- the PSF changes as the mirrors move, and the narrow field of view results in only a small number of PSF calibration stars. WFIRST solves both of these problems, by using temperature-controlled optics, avoiding thermal loads from Earth illumination and shadow, and having a wide field of view. We note that the use of an interlocking observing strategy to solve out systematic errors that correlate with position on the focal plane is a key element of both LSST and WFIRST.

Euclid's weak lensing strategy, as outlined in Laureijs et al. (2011), is to cover as wide an area as possible to as much depth as possible in order to have the minimal statistical errors on weak lensing. This has pushed Euclid to an observing strategy with only a single very wide shape measurement filter, and taking only 4 dithers at each measurement position. These are all taken consecutively, at adjacent (fraction of a chip size) positions in the focal plane, and at the same roll angle, which improves survey speed but minimizes the ability to disentangle instrument effects from true cosmic shear. Even before taking into account cosmic ray losses, this will leave Euclid with ~50% of the survey area imaged in only 3 exposures and thus marginally undersampled. WFIRST, on the other hand makes redundant shape measurements in 3 filters. The 5-6 random dithers in each shape measurement filter ensure that the image is properly sampled even when losses due to cosmic rays are taken into account. WFIRST NIR HgCdTe detectors do not suffer from a time-dependent degradation of charge transfer efficiency like HST's CCDs or those planned for Euclid. This loss of CTE is a serious weak lensing systematic (Rhodes et al. 2010) mitigated by WFIRST's choice of detectors.

The WFIRST strategy is to ensure that all steps are taken so that systematic effects in the weak lensing measurements arising from instrument and survey design are far subdominant to statistical errors in the measurement. Given the 2.5-year observing time allocated to the HLS, this necessitates a smaller survey footprint than will be available from LSST and Euclid. It is worth emphasizing that the focus on redundancy in the WFIRST approach is a deliberate choice, not a matter of hardware. In the time available to the HLS imaging survey, WFIRST *could* carry out an H-band-only survey of 14,000 deg$^2$ with 32 galaxies arcmin$^{-2}$, with similar statistics to the Euclid survey; much shorter exposures in other bands would suffice for photometric redshift information. We are pursuing the YJHK survey over a smaller area because it allows much better control and testing of systematics and because we expect that its science yield outside weak lensing is higher than that from a larger area, single-band imaging survey.

Cross-correlation of shape measurements from WFIRST, LSST, and Euclid will allow powerful tests of systematics in all three experiments. For the reasons emphasized above, we expect WFIRST to have the best internal control and tests of systematics. The cross-comparison may show that the LSST and Euclid WL measurements are in fact reliable at a level that they cannot demonstrate internally because of lower redundancy or lower angular resolution. In this case, the WFIRST comparison would allow the LSST or Euclid measurements to be used with confidence to the statistical limits afforded by their large survey areas.

### 2.5.3 *Galaxy Redshift Survey Cosmology with WFIRST*

The BAO method is, according to current knowledge of dark energy methods, "the method least affected by systematic uncertainties" (DETF; Albrecht et al. 2006), the only one that the NRC's Beyond Einstein Program Assessment Committee (BEPAC) described as "very robust." It relies on the feature imprinted on matter clustering by primordial acoustic waves at the sound horizon scale $r_s \approx 150$ Mpc (comoving). WMAP data determine this scale to 1.1%; Planck data are expected to shrink the uncertainty in $r_s$ to 0.25%. BAO appear as a localized bump in the galaxy correlation function $\xi(r)$ at $r_s$, with width of approximately 10 Mpc (FWHM, in linear theory) set by diffusion between photons and baryons before recombination. In the power spectrum (Fourier transform of the correlation function), BAO appear as a series of damped oscillations about the smooth underlying function whose broad-band shape is determined by the dynamics of cold dark matter. The robustness of the BAO method arises because the acoustic scale is large compared to the scale of non-linear gravitational evolution and because observational or astrophysical systematics generally produce smooth distortions of clustering signals, not localized features that can mimic or distort BAO.

If one estimates the power spectrum P(k) from a redshift survey of volume V that contains $N_k \approx 4\pi k^2 dk \times (V/16\pi^3)$ Fourier modes in the range k to k+dk, the fractional measurement error is

$$\sigma_P/P \approx N_k^{-1/2}\ (1+1/nP),$$

where n is the mean galaxy number density. If nP >> 1, then the measurement precision is limited by the number of modes in the survey volume, i.e., by the sample variance of the large scale structure present within the survey, and the corresponding error of the BAO distance measurement scales as $V^{-1/2}$. However, if $nP \leq 1$, then shot noise of the galaxy density field degrades the measurement of each Fourier mode, and the BAO precision is lowered correspondingly, by a factor (1+1/nP). The spectroscopic portion of the WFIRST HLS (described in detail in the following subsections) is designed to cover the same footprint as the imaging survey, which provides the unshifted positions and photometry of the galaxies whose redshifts are measured by slitless spectroscopy, and to achieve a space density of targets that yields nP close to unity at the BAO scale $k \approx 0.2$ h Mpc$^{-1}$ over most of the redshift range 1.35 < z < 2.65[5].

Non-linear evolution and non-linear galaxy bias have mild effects on BAO measurement. The most noticeable impact is to broaden the BAO bump in $\xi(r)$ as galaxies "diffuse" out of the BAO shell because of tidally induced large scale flows. This lowers the BAO measurement precision because a broader bump cannot be centroided as accurately. In Fourier space, this non-linear effect appears as a damping of oscillations at higher k, so that the matter power spectrum is approximately

$$P_m(k) = \exp(-k^2\Sigma^2_{nl})(P_{lin}(k)-P_{nw}(k)) + P_{nw}(k),$$

where $P_{nw}(k)$ is the "no wiggle" linear theory power spectrum that would arise in the absence of BAO. Here $\Sigma_{nl}$ is, approximately, the rms relative displacement of galaxy pairs separated by $r_s$. The precision of the power spectrum measurement of BAO declines as $\Sigma_{nl}$ grows because one gets less useful information from the oscillations at higher k, where the power spectrum itself is better measured.

Non-linear degradation of the BAO measurement can be substantially reduced by "reconstruction" (Eisenstein et al. 2007; Noh et al. 2009), which uses the gravity field predicted from the observed galaxy distribution to reverse non-linear displacements and restore (approximately) the linear density field. In terms of the equation above, reconstruction can reduce the effective value of $\Sigma_{nl}$ by a factor $p_{NL} \sim 2$, thus restoring information in the higher k oscillations of P(k). Reconstruction has proven highly successful in application to both observations and simulations (Padmanabhan et al. 2012 and references therein). Non-linear gravitational evolution and galaxy bias can also cause broad-band tilts of the power spectrum that lead to small displacements of the BAO peak (~0.3% at z=0). In simulation tests, reconstruction eliminates even these small shifts to within the simulation measurement precision, roughly 0.1% or better (Padmanabhan & White 2009; Seo et al. 2010; Mehta et al. 2011).

Analysis of the full galaxy power spectrum, including its anisotropy in redshift space, can provide powerful constraints beyond those from BAO alone. The most important new effect is redshift-space distortion (RSD) induced by galaxy peculiar velocities. RSD makes the redshift-space power spectrum dependent on the cosine $\mu$ of the angle between the wavevector and the line of sight:

$$P_{gal}(k,\mu) = (b+\mu^2 f)^2\ P_m(k) \times F_{nl}(k,\mu),$$

where $P_m(k)$ is the matter power spectrum, b is the large scale galaxy bias factor, and $F_{nl}(k,\mu) = 1$ in linear perturbation theory (Kaiser 1987; Hamilton 1998). Here f = dlnG/dlna is the logarithmic growth rate of matter fluctuations in linear theory; for General Relativity this is well described by $f(z) \approx [\Omega_m(z)]^\gamma$ with $\gamma = 0.55$ (Linder 2005). With a good model of the non-linear corrections $F_{nl}(k,\mu)$, the measured $\mu$-dependence of the power spectrum allows one to isolate the product of f(z) and the normalization of the matter power spectrum $P_m(k)$, typically described by the rms linear theory matter fluctuation $\sigma_{8m}(z)$ in spheres of comoving radius $8h^{-1}$ Mpc. If one assumes GR, then RSD tightens constraints on dark energy through its constraints on $\Omega_m(z)$ and $\sigma_{8m}(z)$. Additionally, RSD can test whether the rate of structure growth is consistent with GR predictions given dark energy constraints from supernovae and BAO.

---

[5] While a range of wavevectors k contribute to BAO measurements, it is standard practice to specify nP at k=0.2h/Mpc. For Hubble constant measurements H(z=2), a survey with nP(k=0.2/Mpc)=1 provides a constraint equivalent to that of a sampling variance-limited survey with 32% of the volume.

Additional information in a full $P(k)$ analysis comes from the broad-band shape of the power spectrum and the AP effect. As noted earlier, the turnover scale imprinted by the transition from radiation to matter domination provides a second standard ruler for measuring geometry, though it is more difficult to protect this smooth transition (as opposed to the sharp BAO feature) from measurement systematics such as slow calibration drifts or astrophysical systematics such as scale-dependent galaxy bias. In the absence of peculiar velocities, the AP effect would enable very tight constraints on the product $H(z)D_A(z)$ from demanding isotropy of the power spectrum (or correlation function) on scales well below the BAO scale, where the large number of Fourier modes allows much higher measurement precision. The $H(z)D_A(z)$ product is a useful dark energy diagnostic in its own right, and it allows $D_A$ information from BAO to be transferred to $H(z)$, which is more directly sensitive to the dark energy density.

The primary systematic for both RSD and the AP test is theoretical modeling uncertainty in the non-linear corrections $F_{nl}(k,\mu)$ for peculiar velocity distortions. This correction factor goes to unity at low $k$, but non-linear velocities, including the velocity dispersions of galaxy groups and clusters, can have a noticeable impact out to surprisingly large scales. Tests on N-body simulations indicate that the best analytic models for RSD are accurate to the few-percent level or better at $k \approx 0.2$ h Mpc$^{-1}$ (e.g., Tinker 2007; Reid & White 2011; Chuang & Wang 2012; Reid et al. 2012). Ultimately, of course, there is no reason that RSD modeling must be done analytically rather than by "brute force" numerics. The fundamental issue is rather how sensitive the modeling predictions are to uncertain aspects of galaxy formation physics, such as the velocity dispersions of galaxies within halos, and how effectively one can immunize RSD/AP analyses against these uncertainties by marginalizing over parameterized descriptions. As one works to a higher $k_{max}$ (or to smaller minimum scale in the correlation function), the magnitude of non-linear corrections increases, and the measurement precision from a given data set also improves, increasing the demand on the accuracy of these corrections. For our forecasts in §2.5.4 below we assume that we will be able to model the redshift-space galaxy power spectrum up to $k_{max} = 0.2$ h Mpc$^{-1}$ with uncertainties that are small compared to the ~ 1.5% aggregate statistical error. For both BAO and full $P(k)$ analysis, we assume that reconstruction or non-linear modeling can reduce non-linear degradation of the linear power information by a factor of $p_{NL} = 0.5$ in $\Sigma_{nl}$. We consider these assumptions to be conservative, as current modeling techniques are already close to achieving this level of accuracy in simulation tests. If we can in fact work to much smaller scales, then the cosmological return from the WFIRST redshift survey could be quite a bit stronger than our forecasts.

Because the galaxy redshift survey can be applied to cosmology in a multitude of ways, we will sometimes refer to these applications (BAO, RSD, AP, etc.) collectively with the acronym "GRS", parallel to our use of SN and WL for the supernova and weak lensing programs.

### 2.5.3.1 HLS Spectroscopy Observing Strategy

The spectroscopic observing strategy is similar in concept to the HLS imaging strategy as described in §2.5.2.1. The spectroscopy mode must support multiple roll angles to reduce spectral confusion and unambiguously associate an emission line with its host galaxy. Furthermore the angles must be such as to provide counter-dispersion – the roll angle should change by ~180° to suppress systematic errors in wavelength determination associated with the astrometric offset between the continuum galaxy image and the Hα emitting region. It is in principle possible to do this either by using 2 separate prisms with opposite dispersion, or by using a single prism and waiting ~6 months for the counter-dispersed image. The DRM1 design incorporates two prisms with opposite dispersion in the prism wheel, which simplifies scheduling constraints.

In both DRM1 and DRM2, the spectroscopic mode has 4 passes, each with 2 exposures offset diagonally by slightly more than a chip gap, and achieving 90% fill at ≥6 exposures. (Alternative strategies with more, shorter exposures were investigated, but they suffered a significant penalty due to read noise.)

The exposure time was set to reach a minimum detectable line flux of $10^{-16}$ erg/cm$^2$/s at the most sensitive part of the band (2.0—2.4 μm) for an exponential profile source with $r_{eff} = 0.3$ arcsec. This corresponds to ~1.2L* at z = 2.2 (Sobral et al. 2012) and reaches the density of galaxies n ≈ 2.5×10$^{-4}$ Mpc$^{-3}$ where the shot noise and galaxy clustering power are comparable on BAO scales. This resulted in 530 s exposures for DRM1 and 567 s for DRM2. The exposure times for both DRMs, and the consequent HLS rates for both imaging and spectroscopy, are shown in Table 7 (DRM1) and Table 8 (DRM2).

### 2.5.3.2    HLS Spectroscopy Survey Performance

The spectroscopy survey depths (for 6 exposures) are shown in Figure 23. Note that in DRM2 the spectroscopy bandpass is reduced to 1.7—2.4 μm. This reduces the background and allows similar extended source sensitivity with a smaller telescope, despite only 7% greater exposure time.

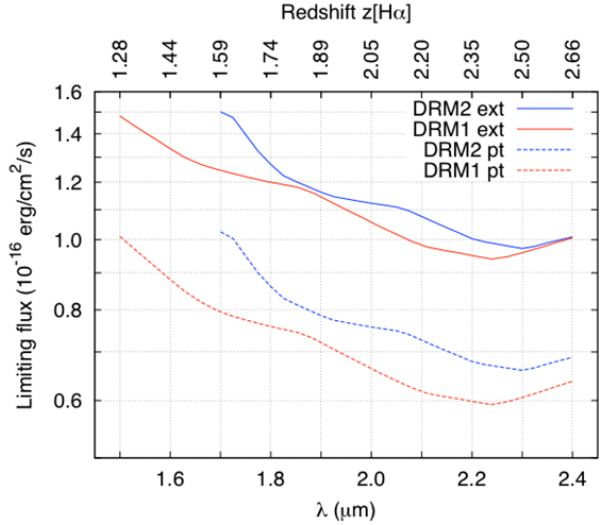

Redshift z[Hα]

**Figure 23: The emission line sensitivity for the WFIRST HLS spectroscopy survey. The dashed lines show 7σ point source sensitivities, and the solid lines show extended source ($r_{eff}$ = 0.3 arcsec, exponential profile) sensitivities. The depth is observed-frame (not corrected for Galactic extinction). The zodiacal background was taken at ecliptic latitude ±45°.**

The galaxy yields from the spectroscopy survey as a function of redshift are shown in Table 11. The total galaxy yield is 4858 deg$^{-2}$ (DRM1) or 2936 deg$^{-2}$ (DRM2). Computations assumed a threshold of 7σ (matched filter significance) for detection of the Hα emission line and 70% completeness (taking advantage of the reduced confusion from multiple roll angles and the availability of deep multicolor zero-order imaging. To be conservative we have not assumed any enhancement in S/N from the [N II] doublet, although at the WFIRST dispersion ($D_\Theta$ = 150 - 250 arcsec, where $D_\Theta = \lambda/(\delta\lambda/\delta\Theta)$ and $\Theta$ is the sky angle, the equivalent point-source resolution R=$\lambda/\delta\lambda$, is obtained by dividing by the angular size of a 2-pixel resolution element) this will always be a partial blend with Hα. The galaxy yields in the table are averaged over the depth histogram of the survey. Over 90% of the survey bounding box is observed at ≥6 exposures and hence reaches ≥80% of the mean galaxy density. The depth variations are typically on scales of order the chip gaps, which are 1.4—2.2 Mpc comoving.

| Red-shift z | DRM1 | | DRM2 | |
|---|---|---|---|---|
| | dN/dV Mpc$^{-3}$ | dN/dz/dA deg$^{-2}$ | dN/dV Mpc$^{-3}$ | dN/dz/dA deg$^{-2}$ |
| 1.35 | 3.77×10$^{-4}$ | 3990 | N/A | N/A |
| 1.45 | 3.73×10$^{-4}$ | 4123 | N/A | N/A |
| 1.55 | 3.55×10$^{-4}$ | 4075 | N/A | N/A |
| 1.65 | 3.28×10$^{-4}$ | 3874 | 2.18×10$^{-4}$ | 2578 |
| 1.75 | 3.03×10$^{-4}$ | 3673 | 2.55×10$^{-4}$ | 3087 |
| 1.85 | 2.84×10$^{-4}$ | 3502 | 2.60×10$^{-4}$ | 3207 |
| 1.95 | 2.89×10$^{-4}$ | 3629 | 2.55×10$^{-4}$ | 3194 |
| 2.05 | 2.97×10$^{-4}$ | 3774 | 2.42×10$^{-4}$ | 3075 |
| 2.15 | 3.07×10$^{-4}$ | 3930 | 2.33×10$^{-4}$ | 2990 |
| 2.25 | 3.01×10$^{-4}$ | 3881 | 2.35×10$^{-4}$ | 3036 |
| 2.35 | 2.58×10$^{-4}$ | 3346 | 2.12×10$^{-4}$ | 2742 |
| 2.45 | 2.16×10$^{-4}$ | 2806 | 1.81×10$^{-4}$ | 2356 |
| 2.55 | 1.66×10$^{-4}$ | 2156 | 1.48×10$^{-4}$ | 1922 |
| 2.65 | 1.25×10$^{-4}$ | 1631 | 1.12×10$^{-4}$ | 1457 |

**Table 11: The WFIRST redshift survey yields at β = ±45°, E(B−V)=0.05, and averaged over the depth histogram of the survey.**

A major change from the WFIRST IDRM report is the use of the Sobral et al. (2012) Hα luminosity function (based on blind narrow-band surveys), which updates the previous estimate by Geach et al. (2010). The new HαLF is unfortunately lower than the previous estimate; reasons include (1) consistent treatment of internal extinction corrections (which are applied to some HαLF results and must be undone to predict redshift survey yields); (2) improved statistics and addition of data at new redshifts; (3) redshift-averaging effects in previous determinations (this does not occur in narrow-band surveys); (4) aperture corrections; and (5) consistent treatment of cosmology corrections. While one might worry about the possibility of further downward revisions, the Sobral et al. (2012) sample contains 87 objects above the WFIRST-DRM1 detection limit at z=1.47 and 58 at z=2.23, and hence most of the relevant parameter space has been covered with small statistical uncertainties. At this point, the uncertainty in our yield forecasts may be dominated by the accumulated systematic uncertainty in the chain of corrections ([N II], aperture corrections, completeness corrections, etc.) necessary to convert narrow-band survey data to appropriate input for the WFIRST exposure time calculator.

The long spectroscopy exposures suffer a cosmic ray rate of 0.26 CRs/arcsec² (DRM1) or 0.09 CRs/arcsec² (DRM2) in each exposure. The SUTR capability will be required to flag CRs. WFIRST plans to recover from these events by fitting for the ramp slope; without such recovery there would be a penalty in S/N due to lost pixels and in additional complexity of the downstream data pipeline. The combination of 6 exposures should enable WFIRST to robustly distinguish real emission lines from any cosmic rays that were missed by the onboard flagging algorithm.

As discussed earlier, a BAO survey transitions from being dominated by galaxy shot noise to being dominated by sample variance when the product nP(k) exceeds unity at the BAO scale k ≈ 0.2 h Mpc⁻¹. Figure 24 plots nP as a function of redshift based on the galaxy yields in Table 4. Since the relevant power spectrum is that of the galaxies in the redshift survey, this plot requires a model for the clustering bias of emission line galaxies. For the lower curve we have used the same bias model adopted for the IDRM report, $b(z) = 0.9 + 0.4z$, based partly on the Sobral et al. (2010) analysis of Hα emission line galaxies at z = 0.85 and partly on the models of Orsi et al. (2010); this is also the model we use for our dark energy forecasts in §2.5.4 below. However, a recent analysis of z = 2.23 Hα emitters by the same group (Geach et al. 2012) finds substantially stronger clustering than this model predicts, consistent with $b(z) = 1.4 + 0.4z$. The upper curve shows nP values for this higher bias model, which seems more directly applicable to the WFIRST galaxy population. Even for the lower bias, we see that the DRM1 HLS achieves nP > 0.7 out to z = 2.3, though shot noise becomes more important at higher redshift. For the higher bias model, the HLS achieves nP > 1.2 out to z=2.3, and even at z = 2.65 it has nP = 0.5.

Triangles in Figure 24 show our nP vs. z calculations for the Euclid mission, where we have again adopted the Sobral et al. (2012) Hα luminosity function, and we have used the exposure times and throughput values of Laureijs et al. (2011). For the lower bias model, the Euclid measurements are dominated by shot noise at all redshifts, with nP ≤ 0.4 at z > 1. For the higher bias model, the Euclid measurements achieve nP > 0.7 out to z = 1.1 and fall to nP ≈ 0.15 − 0.3 for 1.2 < z < 2. The ground-based BigBOSS experiment

(Schlegel et al. 2011) plans to measure BAO from large galaxy redshift surveys out to z ≈ 1.6 over 14,000 deg² (and from the Lyman-α forest at z > 2). Current forecasts suggest that BigBOSS will have nP > 1 out to z ≈ 1.0 with density declining steadily towards higher redshifts.

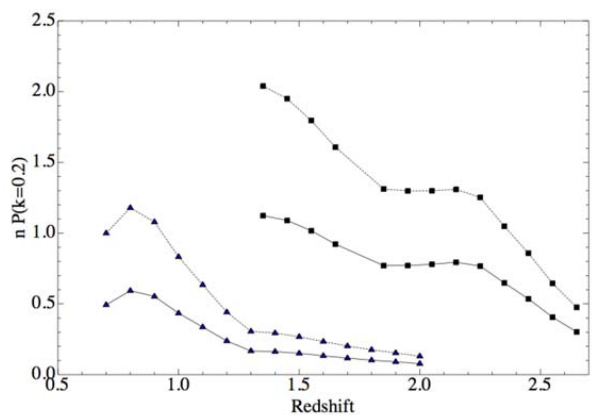

**Figure 24: Product nP of the mean galaxy space density and the amplitude of the galaxy power spectrum at the BAO scale as a function of redshift for the WFIRST HLS (black squares) and Euclid (blue triangles). Lower and upper curves show results for the lower and higher bias models described in the text; the higher bias model is supported by recent observations at z ≈ 2.2. For nP=1, sample variance and shot noise make equal contributions to the statistical error in the power spectrum, while shot noise dominates for nP << 1.**

Figure 25 compares the forecast fractional errors on the Hubble parameter H(z) from WFIRST and Euclid. This forecast uses BAO information only, not the tighter constraints that can be obtained from full P(k) modeling. We assume 50% reconstruction (i.e., $p_{NL}$=0.5). Upper and lower curves show results for the two emission line galaxy bias models described above; the higher bias leads to a significant error reduction by suppressing shot noise. At z ≈ 1.5, the WFIRST and Euclid errors are comparable, with WFIRST's somewhat lower, but they achieve these errors in different ways: WFIRST maps 3400 deg² with errors dominated by sample variance, while Euclid maps 15,000 deg² with errors dominated by shot noise. With additional observing time, WFIRST could reduce its errors by covering more area, while Euclid would likely need to observe deeper (as it is already covering the sky that is available at high Galactic and ecliptic latitudes). We cannot carry out directly comparable calculations for BigBOSS because the galaxy selection and observing modes are very different, but their publicly available forecasts (Schlegel et al. 2011, Table 2.6) indicate fractional H(z) errors comparable to the Euclid errors at z < 1.5, with errors rising rapidly at higher z. WFIRST provides significantly better constraints than either Euclid or BigBOSS over most of the redshift range that it covers.

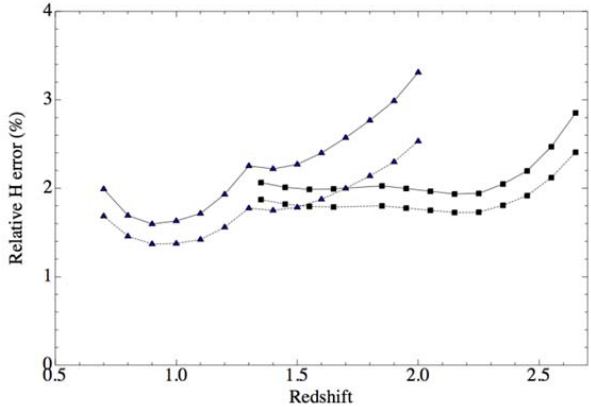

**Figure 25: Forecast errors on the Hubble parameter H(z) in bins of width Δz=0.1 for WFIRST (black squares) and Euclid (blue triangles). Upper curves show the model with weaker galaxy bias (and consequently larger shot-noise errors), while lower lines show the model with stronger galaxy bias. Forecasts in this figure use BAO information only.**

### 2.5.3.3 Galaxy Redshift Survey Requirements

The principal requirement for the BAO/RSD program is the measurement of 3D positions (RA/Dec/redshift) for a large sample of galaxies with a characterizable selection function. The 3D measured positions of galaxies need only be accurate enough not to cause significant degradation of the large-scale structure measurement at mildly nonlinear scales; for Gaussian errors the loss of S/N per mode is $\exp(-k^2\sigma^2/2)$. Thus one needs to know positions to an accuracy of better than $1/k$. The transverse requirement is trivially met, but the radial (redshift) measurement requirement is not. We have therefore set a redshift error requirement of 300 km/s rms, which corresponds to a radial error of 3.2-3.5 Mpc/h (depending on z). The degradation of S/N per mode is 20% at $k_\parallel$ = 0.2 h/Mpc.

Misidentified lines have two major effects on the power spectrum: there is a suppression of the power spectrum due to introduction of a smoothly distributed component (proportional to the contamination fraction $\varepsilon$), and an additional source of power associated with clustering of the contaminants (proportional to $\varepsilon^2$). In order to keep the degradation of the signal small we require the total misidentification fraction to be $\Sigma\varepsilon$<0.1. We have to remove the clustering of contaminants from the true signal and hence we need the total amount of such power (proportional to $\Sigma\varepsilon^2$) to be small enough that there is no significant systematic error after it is subtracted. This flows down to a requirement on the misidentification fraction per contaminant. The requirement on knowledge of the fraction of misidentified lines is driven by the RSD: the $\sim\varepsilon$ suppression of power has no effect on the BAO signal but biases the RSD measurement by a factor of 1-$\Sigma\varepsilon$. Therefore the RSD drives the requirement on $\Sigma\varepsilon$ to be known to $2\times10^{-3}$.

At a spectral dispersion $D_\Theta \approx 200$ arcsec, the redshift accuracy of 300 km/s rms corresponds to a requirement to centroid the emission line to 1.1 pixels rms (including the raw centroiding of the emission line, additional issues such as astrometric uncertainties and [NII] or continuum blending, and the mapping from line position to wavelength). This dispersion also supports the splitting of the [OIII] doublet (2900 km/s or 11 pixels), which both reduces the raw contamination fraction (by reducing the S/N for [OIII] detections) and makes [OIII] directly identifiable if the S/N is high enough to support marginal detection of the weaker doublet member. This dispersion does not split [NII] from H$\alpha$: the splitting of H$\alpha$ from the stronger [NII] feature is 900 km/s (3.2 pixels). Such splitting would not be desirable since it would reduce the S/N for the combined line detection. (Note, however, that our galaxy yields do not assume any [NII] contribution – we have treated it as margin.)

Variations in the selection function are the principal observational systematic challenge for the galaxy clustering program. For WFIRST, these include zodiacal light brightness variations, Galactic dust, photometric calibration drifts, stellar density, and variations of the PSF across the fields of view. Repeated visits to calibration fields (*e.g.* the supernova fields) will track secular changes in the photometric calibration as a function of time. The sky brightness and stellar density variations are critical for WFIRST as they vary by factors of a few over the survey area. However, these are effects that can be simulated by operations on the actual data (*e.g.* by adding sky photons or stellar traces to an image and then re-processing it); and so we expect that the associated selection function effects will be very well known. Moreover, even if one did not know the amplitudes of these effects, most of their power is in unique patterns on the sky (low-multipole modes and features associated with the tiling) that are very different from the modes used for BAO/RSD, and so remov-ing them via template projection provides a backup option.

Deeper spectroscopic measurements of a small subset of the galaxies will be required to measure contamination fractions. This requirement will be met using some combination of tile overlaps, the SN fields, and the ground assets used for WL-PZCS observations.

In summary, the requirements for the galaxy redshift survey component of the dark energy program are:

**Survey Capability Requirements**

- ≥ 1400 deg$^2$ per dedicated observing year (combined HLS imaging and spectroscopy)
- A comoving density of galaxy redshifts at z=2 of 2.9x10$^{-4}$ Mpc$^{-3}$ (see Table 11 for redshift dependence).
- Redshift range $1.3 \leq z \leq 2.7$
- Redshift errors $\sigma_z \leq 0.001(1+z)$, equivalent to 300 km/s rms
- Misidentified lines ≤ 5% per source type, ≤ 10% overall; contamination fractions known to 2×10$^{-3}$

**Data Set Requirements**

- Slitless prism, spectrometer, dispersion $D_\Theta$ = 150 - 250 arcsec
- S/N ≥ 7 for $r_{eff}$ = 300 mas for Hα emission line flux at 2.0 μm ≥ 1.1x10$^{-16}$ erg/cm$^2$/s
- Bandpass $1.5 \leq \lambda \leq 2.4$ μm
- Pixel scale ≤ 180 mas
- System PSF EE50% radius 325 mas at 2 μm
- ≥ 3 dispersion directions required, two nearly opposed
- Reach $J_{AB}$=24.0 AND ($H_{AB}$=23.5 OR $K_{AB}$=23.1) for $r_{eff}$=0.3 arcsec source at 10 sigma to achieve a zero order detection in 2 imaging filters.

### 2.5.4 *Performance Forecasts for Dark Energy*

The WFIRST supernova, weak lensing, and galaxy redshift surveys make basic cosmological measurements with complementary information content. This complementarity – in the physical quantities that are measured and in the redshift ranges that are probed with high precision – allows a combination of methods to make sharper and more informative tests of cosmic acceleration theories than any single method could do on its own. Equally important, consistency tests among methods provide cross-checks for unrecognized systematic errors, supplementing the essential systematics tests internal to each method. A major conclusion about dynamical dark energy or the breakdown of GR will be far more convincing if it can be reached by two or more largely independent lines of evidence.

Figure 26 illustrates the basic measurements from the supernova and galaxy redshift surveys. Supernovae measure the luminosity distance $D_L(z)$ in bins of redshift. Because they are calibrated to objects in the local Hubble flow, these distance measurements effectively have units of h$^{-1}$ Mpc. (More accurately, the supernova method measures distance ratios between all bins, including the local calibrators; our FoM calculations use this more accurate characterization of their information content.) The black and red error bars in Figure 26a show our forecast 1σ fractional errors on $D_L(z)$ for the conservative and optimistic assumptions about SN systematics, as discussed in §2.5.1 and shown previously in Figure 17. Over most of the redshift range, the errors per bin are ~ 1-1.5% in the conservative case and ~ 0.75-1% in the optimistic case. There are 16 bins, so one can roughly judge the aggregate precision of the data by taking the median error and dividing by $\sqrt{N_{bin}}$; for our forecasts, the aggregate precision of the luminosity distance measurement is $(\Sigma_i [\Delta \ln D_L(z_i)]^{-2})^{-1/2}$ = 0.32% for the conservative case and 0.23% for the optimistic case, with an effective redshift z ≈ 0.8. This combination assumes that the systematic errors are uncorrelated among redshift bins, and one of the key challenges for the supernova survey will be keeping correlated systematics below the level of this aggregate precision. Of course, the actual measurements over a range of redshift are more informative than a measurement of the same aggregate precision at a single redshift, as they can detect or rule out the redshift trends expected in different models.

Curves in Figure 26a show the fractional changes in the predicted luminosity distance for several models relative to a model with a flat universe, a cosmological constant, and $\Omega_m$ = 0.267. Dotted and short-dashed curves show the effect of changing the equation-of-state parameter from w = -1 to w = -0.96 or w = -1.04, while maintaining a flat universe. Long-dashed and short-dashed curves show the effect of changing the space curvature to $\Omega_k$ = ± 0.002 while maintaining w = -1. Isolated changes to w or $\Omega_k$ with all other parameters held fixed tend to produce models that are easily ruled out by CMB data. Here we have followed the strategy of Weinberg et al. (2012), where for each change to w or $\Omega_k$ we adjust the values of $\Omega_m$, $\Omega_b$, and $h$ in a way that keeps the CMB power spectrum almost perfectly fixed. (Specifically, we allow the value of h to change while fixing the values of $\Omega_m h^2$, $\Omega_b h^2$, and the angular diameter distance to the surface of last scattering to those in our fiducial model). Because we assume calibration in the local Hubble flow, the model curves converge at z=0 even though the Hubble constant changes.

The model differences illustrated by these curves provide a context for assessing the size of the projected errors, though plots like these do not capture the ability of multiple probes to break parameter degeneracies in a high-dimensional space of theories. Note that with $N_{bin}$=16, a model that skirted the top or bottom of all the error bars would be ruled out at $4\sigma$ significance.

In panel (b), black error bars show our forecast errors for BAO measurements of the angular diameter distance $D_A(z)$, nicely illustrating the complementarity of the WFIRST SN and galaxy redshift surveys. In the range 1.3 < z < 1.7 the BAO errors are similar to those for the conservative SN forecast, about 1.7% per $\Delta z$=0.1 bin. The BAO measurements continue from z=1.7 up to z=2.65, while the SN measurements continue from z=1.3 down to z=0. BAO measurements are calibrated to the sound horizon $r_s$, so they are effectively in Mpc rather than h⁻¹ Mpc. As a result, the predictions of different models diverge towards low z instead of converging as they do in panel (a), and SN and BAO distances provide distinct cosmological information even when measured with equal precision at the same redshift. At z ~ 2, $D_A(z)$ is more sensitive to a curvature change than to a (constant) w change, in contrast to $D_L(z)$ at low redshift, so the combination of the two breaks degeneracy between these parameters. For this plot we have assumed exact knowledge of $r_s$, which is a good but not perfect approximation at this level of accu-racy; our FoM forecasts below marginalize over uncertainties in $r_s$ including Planck CMB priors.

Errors on H(z) (panel c) are larger than those for $D_A(z)$ because there is one line-of-sight direction vs. two transverse directions. However, H(z) measurements are more sensitive to the dark energy equation of state because they directly probe the total energy density, which is the sum of the matter, radiation, and dark energy densities. (Recall that the critical density is $\rho_c(z)$ = $3H^2(z)/8\pi G$.) For the constant-w models shown here, the deviations in H(z) are small at these redshifts because the dark energy density is nearly constant and the matter density is higher by $(1+z)^3$. However, high redshift H(z) measurements are a powerful diagnostic of dynamical dark energy models, especially "early dark energy" models in which the dark energy density tracks the matter density during the matter dominated epoch (*e.g.*, Albrecht & Skordis 2000; Doran et al. 2001).

The BAO errors in Figure 26 were computed independently of those in Figure 25, and they are somewhat larger (~2.5% per bin in H(z) vs. ~2.0%) for two reasons. First, the computation in Figure 26 adopts a galaxy bias model from Orsi et al. (2010), which is itself slightly lower than the "lower bias" model used in Figure 25. Second, the computation here adopts $k_{max}$ = 0.2 h Mpc⁻¹ for the BAO analysis as well as the full P(k) analysis; however, because non-linearities are unlikely to produce oscillatory features in P(k), there is a substantial argument for adopting $k_{max}$ = ∞ for BAO-only analysis, as was done for Figure 25. We therefore consider the BAO-only forecasts in Figure 26 and the subsequent calculations of this section to be on the conservative side. With plausible alternative assumptions (the Geach et al. 2012 galaxy bias and $k_{max}$ = ∞), the DRM1 BAO constraints on H(z) and $D_A(z)$ could be 25-30% tighter (a factor 1.7-2 improvement in inverse variance).

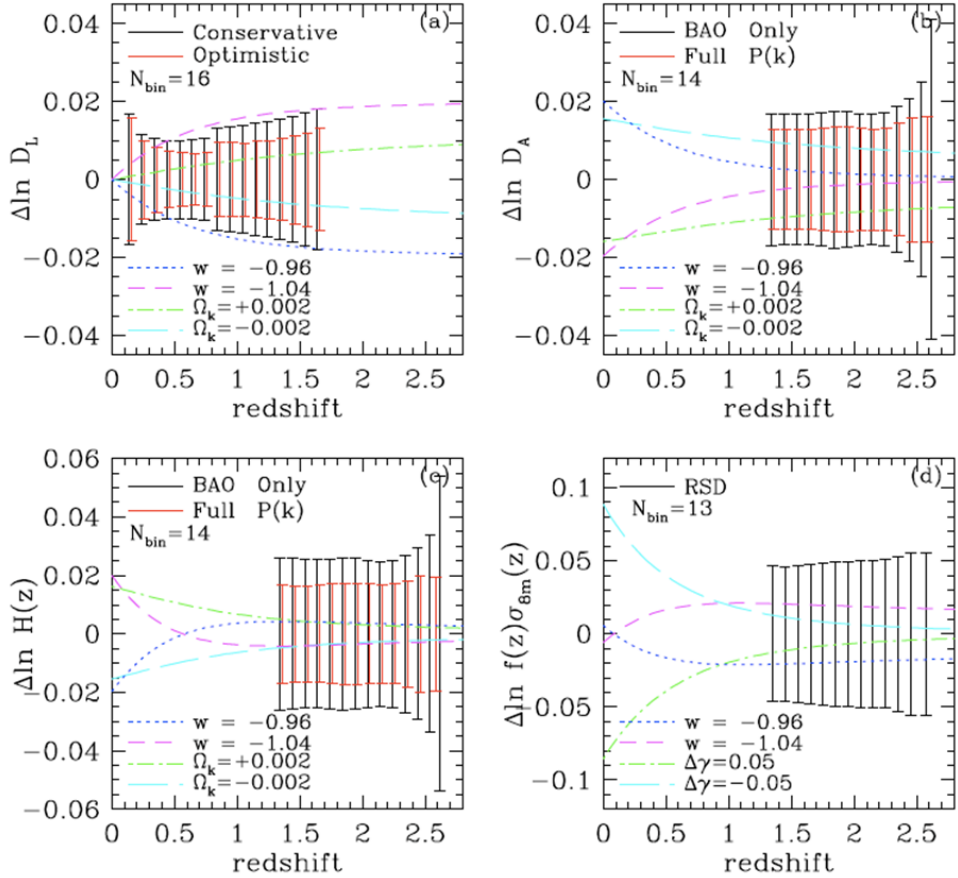

**Figure 26:** Forecast precision of measurements from the WFIRST supernova and galaxy redshift surveys. Error bars in each panel show forecast $1\sigma$ errors for the DRM1 surveys on the luminosity distance, angular diameter distance, Hubble parameter, and growth rate (a-d, respectively). Curves (computed for us by M. Mortonson) show the fractional changes relative to a $\Lambda$CDM model that arises from making the indicated change in $w$, $\Omega_k$, or $\Delta\gamma$.

Red error bars in panels (b) and (c) show forecasts that use the full redshift-space galaxy power spectrum as described in §2.5.3, assuming $k_{max} = 0.2$ h Mpc$^{-1}$ and $p_{NL} = 0.5$. Errors on $D_A(z)$ shrink by about 25%, and errors on $H(z)$ shrink by ~ 35%. The information content of a full $P(k)$ analysis is more complex than that of a BAO "standard ruler" analysis, but roughly speaking the improvement in the full $P(k)$ errors comes from a combination of the AP effect, redshift-space distortions constraining $H(z)$ through growth, and the TRMD scale providing an additional standard ruler. The aggregate $D_A$ precision from the $N_{bin} = 14$ BAO redshift bins in panel (b) is $(\Sigma_i [\Delta\ln D_A(z_i)]^{-2})^{-1/2} = 0.49\%$, while for the full $P(k)$ analysis (which has $N_{bin} = 13$) it is 0.37%, with error-weighted redshift $z \approx 1.9$. The corresponding aggregate precision for $H(z)$ is 0.72% (BAO only) and 0.48% (full $P(k)$). Combination of these measurements with the $0.25 - 0.3\%$ measurement of $D_L(z)$ at $z \approx 0.8$ provides powerful leverage on the history of dark energy and the curvature of space.

As discussed in §2.5.3, redshift-space distortions constrain the product of the fluctuation growth rate $f(z)$ and the fluctuation amplitude $\sigma_{8m}(z)$. Error bars on panel (d) show forecast errors on $\sigma_{8m}(z)f(z)$ computed by the methodology of Wang (2012), again assuming $k_{max} = 0.2$ h Mpc$^{-1}$ and $p_{NL} = 0.5$. (BAO measurements do not constrain structure growth, so there is no "BAO only" case in this panel.) Errors per bin are approximately 5%, with aggregate precision of 1.4%. Dotted and short-dashed curves show the impact of maintaining GR but changing $w$, which alters structure growth by changing the history of $H(z)$. (The convergence at low z reflects a cancellation between the effects of changing $w$ and changing $\Omega_m$ in a way that restores the CMB power spectrum.) Long-dashed and dot-dashed curves show the effect of maintaining $w = -1$ but changing the index of the growth rate $f(z) \approx [\Omega_m(z)]^\gamma$ from its GR-predicted value of $\gamma \approx 0.55$, by $\Delta\gamma = \pm 0.05$. This is a simple way of parameterizing the possible impact of modified gravity on structure growth, but we caution that the convergence of model predictions at high z, where $\Omega_m(z)$ goes to one, is to some degree an artifact of the parameterization. Other forms of GR modification could produce stronger growth deviations at high redshift, as could GR-based models with early dark energy.

We have computed similar forecasts for the Euclid galaxy redshift survey, with the depth and area assumptions described in §2.5.3 and the same assumptions about P(k) or BAO modeling used in this section. In the seven overlapping redshift bins from z = 1.3 – 1.9, the forecast errors for the two experiments are similar in both the BAO and full P(k) cases; the larger area of the Euclid survey compensates for its lower galaxy space density and consequently higher shot noise. WFIRST adds measurements in seven redshift bins from z = 2.0 to 2.65, while Euclid adds measurements in six redshift bins from z = 0.7 – 1.3. (The Euclid errors drop by ~25% below z = 1.2 because of reduced shot noise.) This comparison shows the excellent complementarity of the WFIRST and Euclid redshift surveys. In the overlapping redshift range they allow percent-level cross-checks for systematics, between surveys that adopt quite different tradeoffs between volume and space density. Because the Euclid errors are dominated by shot noise, the measurements will be largely independent even if WFIRST maps a subset of the Euclid volume, so combining the measurements will yield a factor ~2 gain in inverse variance. Outside of the overlap range, the two surveys measure distance, energy density, and growth in complementary redshift regimes, with comparable levels of statistical precision.

With current knowledge, it is hard to predict whether low redshift or high redshift data will provide the greater insight into the origin of cosmic acceleration, but the best tests will certainly come from combining precise measurements across a wide redshift range. In the range 0 < z < 0.7, the SDSS-III BOSS survey is presently mapping BAO and RSD, with a 10,000 deg[2] survey and errors limited mainly by sample variance. The range 0.7 < z < 1.3 may be covered by future ground-based surveys with facilities such as BigBOSS, DESpec, and the Subaru PFS; the Euclid measurements will be independent to the extent that they map different areas of the sky. In the range z = 2 – 4, ground-based surveys can map BAO using the Ly$\alpha$ forest (as BOSS is doing now). Very large bounding boxes can be probed this way, but observing a sufficient density of background quasars to approach the sample variance limit will be extremely challenging. Ground-based surveys of Ly$\alpha$ emitters, such as the one planned for HETDEX, can also map the z > 2 regime, though it appears infeasible to reach the volume that WFIRST will probe.

It is difficult to abstract WL constraints into a form similar to Figure 26, with errors on an "observable" as a function of redshift. First, the WL signal depends on both the amplitude of structure and the distance-redshift relation: for sources at $z_s$ and lenses at $z_l$, the quantities $\sigma_8(z_l)$, $D_A(z_l)$, and $D_A(z_s)$ all affect the shear. Second, the errors on the WL signal for two different bins of source redshift can be strongly correlated because matter in the foreground of both redshifts can lens the galaxies in both bins. For WL one obtains errors that are more nearly uncorrelated by considering different multipoles $C_L$ of the angular power spectrum. Systematic errors in shear calibration, photometric redshifts, or intrinsic alignment corrections could correlate errors across both redshift and angle.

As an approximate illustration, Figure 27 repeats the predicted curves and statistical errors on the angular power spectrum for four of the ten tomographic bins shown previously in Figure 20, along with additional curves that show the effect of systematic errors (left panel) or model parameter changes (right panel). To make the curves distinguishable on this plot, we have chosen systematic errors that exceed our requirements by more than an order of magnitude. An additive shear bias of 0.005 produces small changes at high z and high L but readily detectable changes at low z and low L, where the shear variance itself is low. A multiplicative shear bias of 0.05 or a fractional photo-z offset of 0.05 produce deviations that are comparable to or larger than the errors in each $\Delta$lnL bin for L > 100 at each redshift. Changing from a cosmological constant to w = -0.9 depresses the power spectrum by more than the single-bin error (often much more) at all L > 100, while changing the growth index by $\Delta\gamma$ = -0.10 from the GR value produces changes comparable to the single-bin errors for L > 100. Clearly either of these model variations would be detectable at very high statistical significance from the full set of measurements, though the left panel illustrates once again the tight control of systematic effects that is required to achieve the statistical error limits.

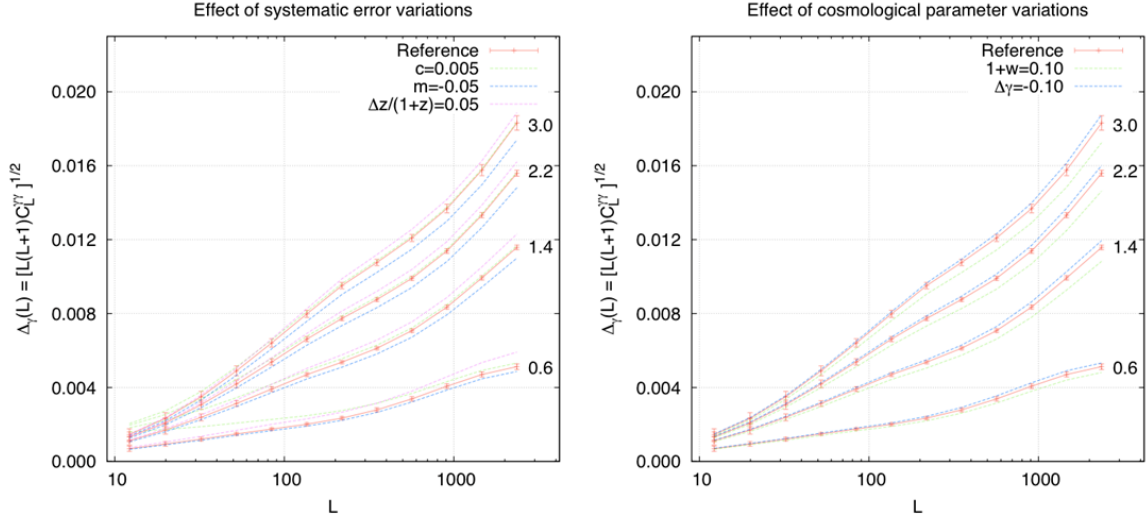

**Figure 27: Cosmic shear angular power spectrum and forecast DRM1 error bars for four of the tomographic bins shown in Figure 20, with curves showing the impact of systematic errors (left) or model parameter variations (right). In the left panel, the green curve shows the impact of an additive shear bias with magnitude c = 0.005 (more than ten times our requirement of ≤ 3 × 10⁻⁴), the blue curve shows a multiplicative bias of m = -0.05 (requirement ≤ 10⁻³), and the red curve shows a systematic photo-z offset Δz/(1+z) = 0.05 (requirement is ≤ 0.002 per bin). In the right panel, green curves show the effect of changing w to -0.9 and blue curves the effect of changing the growth index by Δγ = -0.10.**

To illustrate the combined power of the WFIRST measurements, we compute the DETF Figure of Merit (Albrecht et al. 2006), which assumes an underlying dark energy model with equation-of-state history w(a) = $w_0 + w_a(1-a)$, where a = $(1+z)^{-1}$. One can recast this equation in the equivalent form w(a) = $w_p + w_a(a-a_p)$, where $a_p = (1+z_p)^{-1}$ is a "pivot" expansion factor chosen so that the errors on $w_p = w(a_p)$ and the derivative $w_a = dw/da$ are uncorrelated. The value of $a_p$ depends on the experiments being considered, but it typically corresponds to $z_p \approx 0.5$. The DETF FoM, with the normalization conventionally used, is FoM =$[\sigma(w_p)\sigma(w_a)]^{-1}$, the inverse product of the errors on $w_p$ and $w_a$. Typically the error on $w_a$ is about ten times that on $w_p$, so the uncertainty on $w(z_p)$ is roughly $\sigma(w_p) \sim 0.3$ (FoM)$^{-1/2}$. To characterize tests of modified gravity, we compute the error $\sigma_\gamma$ on the FoMSWG parameter $\Delta\gamma$, which is the deviation from the GR-predicted value of the growth index in f(z) ≈ $[\Omega_m(z)]^\gamma$. When computing the DETF FoM we assume GR is correct. In both cases we allow non-zero space curvature, marginalizing over $\Omega_k$.

The DETF FoM scales as an inverse variance, so for a given set of experiments it increases roughly linearly with data volume in the absence of systematic uncertainties – divide all errors by √2 and the FoM doubles. It has the virtue of allowing both departure from a cosmological constant and time-variation of the equa-tion of state. However, the $w_0$-$w_a$ parameterization cannot represent scenarios in which dark energy tracks matter at early times, so the FoM tends to "underrate" the value of such high redshift measurements in testing early dark energy models. The $\Delta\gamma$ parameterization has a similar bias towards low redshift measurements, as it forces f(z) to one when $\Omega_m(z)$ goes to one. More importantly, an FoM calculation is only as good as the assumptions that go into it, especially assumptions about systematic errors and about prior information from external data. One must therefore use caution when comparing FoM forecasts computed by different groups, which may show large differences because of these assumptions about systematics or priors, even when considering similar data sets. We view the FoM and $\sigma_\gamma$ as useful tools for understanding the combined information content of the WFIRST dark energy probes, but the description in terms of basic measurements as illustrated in Figure 26 and Figure 27 is more fundamental.

In our computations, we consider both the conservative and optimistic assumptions about SN systematics and WL systematics, as described in §§2.5.1 and 2.5.2.4. For the galaxy redshift survey we consider both the BAO-only case and the full P(k) case. We actually still view the latter case as "conservative," since we expect modeling of non-linearity and galaxy bias at the one-percent level for $k_{max} = 0.2$ h Mpc⁻¹ and $p_{NL} = 0.5$ to be readily achievable on the timescale of WFIRST observations. All of our forecasts incorporate the FoM-WSG Fisher matrices that describe the priors for Planck CMB data and the results of near-term, "Stage III" experiments such as BOSS and DES.

The Venn diagram of Figure 28 shows the DETF FoM for different combinations of our conservative SN and WL and BAO-only galaxy survey scenarios. Considered individually, the SN survey gives the largest FoM contribution for these assumptions, with FoM=411 in combination with Planck and Stage III. Adding either BAO or WL pushes the FoM over 500, and the combination of all three methods gives FoM=682. Figure 29 presents the corresponding diagram for the optimistic SN and WL systematics assumptions and the full P(k) analysis of the galaxy redshift survey. The FoM for each of the three methods improves, with a dramatic change in the case of WL, where the FoM nearly triples to 581. The combined FoM for the three methods is 1370, twice that of the conservative case. We have investigated the impact of separately dropping the measurement and modeling systematics in the WL forecast and find that the modeling systematics have greater impact. This is not surprising, as we set the measurement systematics requirements for WFIRST such that they would not substantially degrade the errors (relative to cosmic shear statistical errors) of a 10,000 deg² survey. If we drop the modeling systematics but retain the measurement systematics then the WL FoM is 524 rather than 581, still dramatically improved over the FoM = 200 conservative case. Thus, the most important contribution to the WL improvement is the ability to fully exploit galaxy-galaxy lensing and photometric galaxy clustering in addition to cosmic shear.

Table 12 presents our forecasts for the combined probes systematically, showing all combinations of the conservative and optimistic SN/WL scenarios and the BAO-only and full P(k) galaxy scenarios. We have also made equivalent calculations for the Euclid experiment, with exactly the same assumptions about the WL systematics and the galaxy BAO/P(k) analysis. For the most conservative assumptions, WFIRST outperforms Euclid by more than a factor of two, FoM = 682 vs. 293, because WFIRST has a SN component while Euclid does not. Going to full P(k) analysis narrows the gap, to 774 vs. 460. Going to the optimistic WL assumptions makes a major difference because in the absence of systematics the greater area of the Euclid WL survey makes it substantially more powerful. With optimistic assumptions for all three probes, the two missions have essentially equal FoM, 1370 vs. 1376.

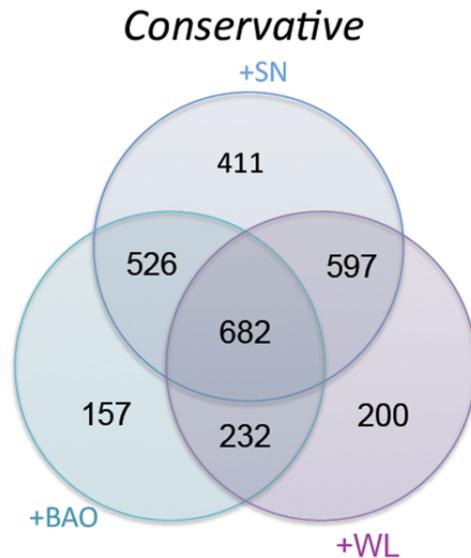

## Conservative

**Figure 28: Forecasts of the DETF FoM for different combinations of the DRM1 WFIRST probes. All forecasts incorporate priors for Planck CMB and Stage III dark energy experiments, which on their own have an FoM of 116. Outer circles show the impact of adding WFIRST SN, WL, or BAO to these individually, and overlaps show the impact of adding combinations. The FoM for all three probes combined is 682. For this figure we adopt our conservative assumptions about SN and WL systematics, and we use only BAO information from the galaxy redshift survey.**

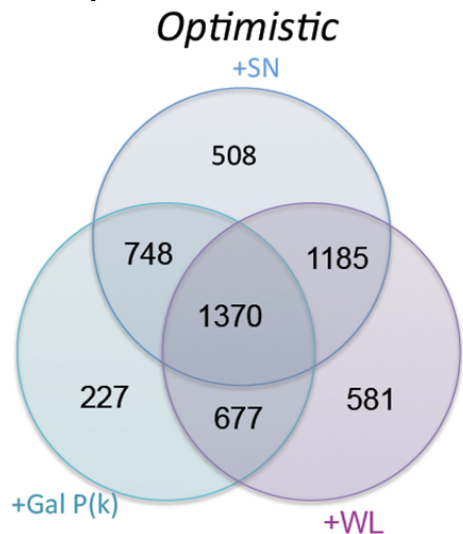

## Optimistic

**Figure 29: Same as Figure 28, but using our optimistic assumptions for SN and WL systematics and using full P(k) information from the galaxy redshift survey.**

Table 12 also lists the forecast errors on the growth index $\gamma$ for each combination of scenarios. The precision is $\approx 0.01$ for cases with conservative WL systemat-ics and ≈ 0.008 for cases with optimistic WL systematics. We marginalize over the dark energy parameters ($w_0$, $w_a$, $\Omega_m$, $\Omega_k$, etc.), which are constrained by the SN and redshift survey measurements in addition to the WL data. At z=1, the median redshift of the WL survey, $\Delta\gamma$=0.01 corresponds to a 0.3% difference in the growth rate $f(z) = [\Omega_m(z)]^{\gamma+\Delta\gamma}$, so in this sense WFIRST will achieve a sub-percent test of the growth predicted by GR + dark energy models. If we omit WL entirely but retain the full P(k) analysis of the redshift survey, then RSD yields a marginalized $\Delta\gamma$ error of $\sigma_\gamma$ = 0.032.

For our assumed $k_{max}$ = 0.2 h Mpc$^{-1}$, the RSD constraint on $\Delta\gamma$ is several times weaker than the WL constraint, but the information content of the two measurements is different in a way that this one-parameter modified gravity description does not capture. First, WL measures $\sigma_{8m}(z)$ while RSD measures $f(z)\sigma_{8m}(z)$; the direct sensitivity of RSD to the growth rate allows the combined measurements to break degeneracies in models that allow more complex $f(z)$ histories, or early dark energy, or decay of dark energy into dark matter. Second, the WL measurement is centered at z ≈ 1 and the RSD measurement at z ≈ 2, again increasing leverage on multi-parameter modified gravity models. (At z=2, the 0.032 error on $\Delta\gamma$ again corresponds to a 0.3% error on f(z).) Third and most important, GR predicts consistency between WL and RSD, but this consistency is violated in some modified gravity theories. Thus, comparison of WL and RSD growth measurements is itself a critical test of alternative gravity models.

For Euclid we forecast errors of 0.007-0.008 on $\Delta\gamma$ for our conservative WL assumptions and ≈ 0.005 for our optimistic assumptions. Because of its larger area, the Euclid WL survey will achieve stronger constraints than the WFIRST survey if it is not limited by measurement systematics. As emphasized in §2.5.2, achieving the statistical limits of a $10^4$ deg$^2$ survey imposes extremely demanding technical requirements, especially on knowledge of the PSF and its dependence on time, position, and galaxy color. We expect the high level of redundancy in the WFIRST WL experiment to play a critical role in meeting these requirements. If we omit WL, then the RSD constraints from Euclid yield $\sigma_\gamma$ = 0.027, somewhat stronger than the 0.032 from WFIRST RSD. The measurement errors on growth are similar in the two experiments, but a change in $\Delta\gamma$ has larger impact at the lower redshifts probed by Euclid. In reality, the two RSD measurements are complementary in a way that a 1-parameter model cannot capture; they provide measurements with similar precision centered at significantly different redshifts.

It is worth emphasizing that the WFIRST dark energy programs are limited largely by the observing time available in the primary mission: 2.5 years for the HLS and 0.5 years (over a two-year interval) for the supernova survey. Thus, the dark energy constraints could be significantly improved by additional observing in an extended mission. If we achieve our optimistic systematics goals for the SN survey, then its errors after DRM1 will still be dominated by statistics (see Figure 17), and continued observations will improve the $D_L(z)$ measurements. If we achieve our technical requirements for the WL survey, then it will remain statistics-limited for a survey area three times larger than that covered by DRM1. Redundancy and knowledge gained within DRM1 might demonstrate that an extended WL survey could be carried out at higher efficiency by concentrating on H-band observations. Extending the area of the spectroscopy survey will almost certainly lead to improved BAO measurements, and it will likely lead to improved RSD growth constraints and AP geometry constraints. As long as systematics remain subdominant, contributions to an inverse-variance measure like the DETF FoM should grow roughly linearly with observing time.

The motivation for an extended dark energy program will depend strongly on the state of the field near the close of DRM1. By that time, WFIRST and other experiments should have improved the precision of many cosmological measurements by an order of magnitude or more, while simultaneously extending their redshift range and strengthening tests for systematics. If results remain stubbornly consistent with GR + cosmological constant despite these improvements, then even an extended WFIRST would not increase precision enough to demonstrate a departure. If WFIRST and other experiments have shown believable hints of dynamical dark energy or deviations from GR, then it will be crucial to turn these hints into convincing evidence, one way or the other, by improving measurement precision. If there is already a clear discovery of such deviations, then there will be strong motivation to characterize them with greater precision and detail, just as particle physicists are now beginning quantitative characterization of the putative Higgs boson to understand its implications for fundamental physics. In either of the latter two cases, observations will provide guidance for the focus of an extended dark energy mission, e.g., whether to focus on high or low redshift, and on geometry or structure growth.

| SN | WL | Galaxy Redshift Survey | WFIRST FoM | $\sigma_\gamma$ | Euclid FoM | $\sigma_\gamma$ |
|---|---|---|---|---|---|---|
| Conservative | Conservative | BAO-only | 682 | 0.0101 | 293 | 0.0079 |
| Conservative | Conservative | Full P(k) | 774 | 0.0097 | 460 | 0.0072 |
| Conservative | Optimistic | BAO-only | 1119 | 0.0079 | 1238 | 0.0053 |
| Conservative | Optimistic | Full P(k) | 1195 | 0.0077 | 1376 | 0.0050 |
| Optimistic | Conservative | BAO-only | 810 | 0.0100 | 293 | 0.0079 |
| Optimistic | Conservative | Full P(k) | 919 | 0.0096 | 460 | 0.0072 |
| Optimistic | Optimistic | BAO-only | 1274 | 0.0079 | 1238 | 0.0053 |
| Optimistic | Optimistic | Full P(k) | 1370 | 0.0076 | 1376 | 0.0050 |

**Table 12: Forecasts for the combined probes investigated systematically, showing all combinations of the conservative and optimistic SN/WL scenarios and the BAO-only and full P(k) galaxy scenarios. We report the DETF FoM and the forecast 1σ error on the growth index γ.**

## 2.6 General Observer Program

Additional science will come from dedicated observations, or general observer (GO) programs making use of the unique capabilities of WFIRST. This point is repeatedly emphasized in the *NWNH* report – *e.g.*, "… the committee considers the guest investigator program to be an essential element of the mission …" (*NWNH*, pg. 207) and "As a straw-man example for the first five years ... the panel imagines ... a Galactic plane survey of one-half year, together with about one year allocated by open competition …" (EOS Panel Report, pg. 274). To give a flavor of the type of science, we now briefly discuss some of hypothetical GO programs.

### 2.6.1 *Probing the Hydrogen-Burning Limit in Globular Clusters*

Star clusters in the Milky Way have served as the primary tools to calibrate stellar evolution models and to define the color-magnitude relation of stars. These models and empirical luminosity functions are a key input to generating reliable population synthesis models, which are fundamental to many topics in astrophysics.

The characterization of star cluster properties has largely relied on establishing high-precision color-magnitude diagrams (CMDs) for these systems. To date, this work has largely focused on visible-light investigations. Yet cool, infrared stars dominate both the brightest phases of stellar evolution in old populations (red giants) and the most populated phases for a normal IMF (low mass dwarfs). Investigations of globular clusters with WFIRST will provide a new tool to establish the infrared color-magnitude relation of stars, and to sensitively map its dependency on metallicity. The resolution of WFIRST, combined with its wide field of view, is essential to this breakthrough.

Figure 30 illustrates the morphology of the infrared CMD of the globular cluster 47 Tuc, from 100 orbits at three orbit depth with the WFC3/IR camera on Hubble (Kalirai et al. 2012). The sharp "kink" on the lower main sequence is caused by collisionally induced absorption of H2. Unlike the visible CMD, the inversion of the sequence below the kink is orthogonal to the effects of distance and reddening, and therefore degeneracies in fitting fundamental properties for the population are largely lifted. The location of the kink on the CMD is also not age-sensitive, and therefore can be used to efficiently flag 0.5 M(Sun) dwarfs along any Galactic sightline with low extinction. A WFIRST two-stage survey will first establish the infrared color-magnitude relation and the dependency of the "kink" on metallicity through high-resolution, deep imaging of Galactic star clusters. Second, this relation can be applied to field studies to characterize the stellar mass function along different sightlines, the dependency of the mass function on environment, and to push to near the hydrogen burning limit in stellar populations out to tens of kpc. These low mass stars have optical-to-infrared colors of >6, and are therefore much more efficiently measured in the infrared bandpasses.

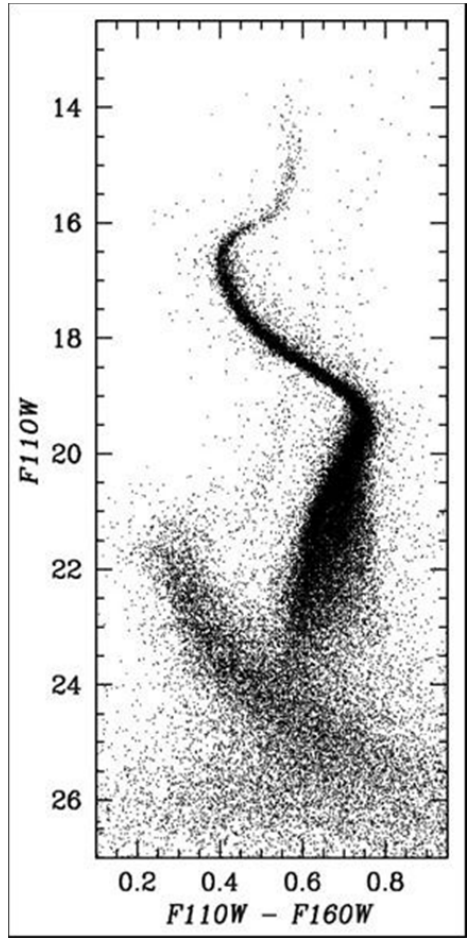

**Figure 30: Infrared color-magnitude diagram for the nearby globular cluster 47 Tuc, constructed from a three-orbit (depth) observation with Hubble (Kalirai et al. 2012). The kink in the lower main-sequence of the cluster is caused by H2 opacity. The fainter main-sequence represents stars from the background SMC galaxy.**

### 2.6.2 Probing Below the Hydrogen-Burning Limit in Open Clusters

The shape of the initial mass function (IMF) contains within it many clues to the star-formation process. One particular topic of recent interest has been the shape of the IMF at very low masses. Are brown dwarfs formed in the same way as stars — or are they generally formed in circumstellar disks or hierarchical binaries and ejected at young ages. What is the minimum mass object that can form in isolation — and what sets that mass? These questions are best answered by very deep surveys of nearby, young open clusters because dynamical evolution both preferentially ejects low mass cluster members and causes those that remain to be less well-concentrated to the cluster core. Current surveys and existing facilities cannot reach sufficient depths to answer these questions.

WFIRST will enable a uniform wide-field, deep, multi-epoch study of the nearest, young open clusters (including clusters such as Pleiades, Alpha Per, IC2391/2602). Even for the oldest of these clusters (the Pleiades, at 100 Myr), WFIRST can go deep enough to reach 5 M(Jupiter); for the youngest, WFIRST should be able to reach much lower in mass. In favorable cases, two epochs separated by 3-5 years will provide accurate enough proper motions to isolate cluster members. For other clusters, multiple epochs could be used to use variability (plus colors) to isolate cluster members. Comparison of empirical isochrones from these clusters of varying ages would then allow determination of evolutionary tracks for planetary mass objects. The spatial location of the lowest mass members relative to the stellar members would be used to constrain the physical mechanism for forming these free-floating planets.

### 2.6.3 Targeted Galaxy Cluster Programs

In addition to the galaxy cluster science that will come, essentially for free, from the high-latitude cosmological surveys (see §2.3.2), WFIRST will also enable important science from targeted follow-up of galaxy clusters. Such science could come from both deeper observations of clusters within the HLS, or from targeted follow-up of clusters identified outside that survey field.

Cluster candidates will be found through their Sunyaev-Zel'dovich (SZ) decrement on the cosmic microwave background (*e.g.*, from SPT and/or ACT) and/or from their X-ray emission (*e.g.*, found by Chandra, XMM/Newton, or eROSITA) and/or found through infrared selection (*e.g.*, found by Spitzer or WISE). WFIRST observations will efficiently confirm and characterize distant massive clusters. For example, Spitzer observations of the ~8 deg² Boötes field have identified over one hundred $z > 1$ galaxy groups-to-cluster candidates (Eisenhardt et al. 2008). Recently, that group reported one extremely massive cluster at $z = 1.75$ (Figure 32), initially identified as a candidate based on infrared imaging but confirmed with Hubble grism spectroscopy and subsequently identified as a strong X-ray source (Stanford et al. 2012), as having a robust SZ decrement (Brodwin et al. 2012), and, most impressively, strongly lensing a background galaxy (Gonzalez et al. 2012). WFIRST follow-up observations of similar rare, extreme sources will allow better characterization of their properties (*e.g.*, weak lensing masses, wider-area properties, and prism redshifts).

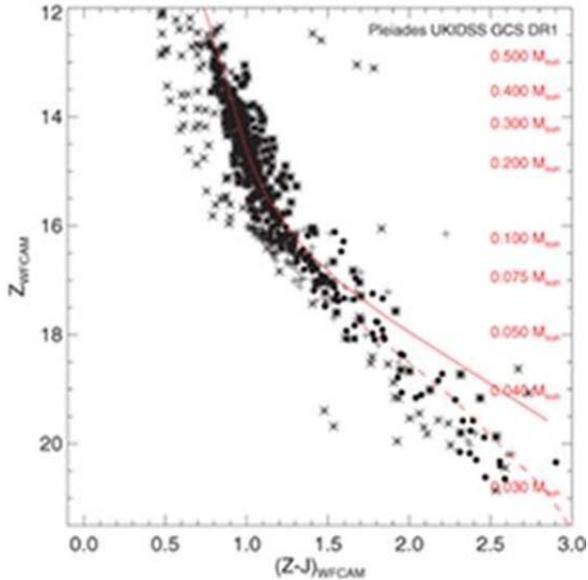

**Figure 31: Color-magnitude diagram for proper motion selected stars and brown dwarf members of the Pleiades from the UKIDSS galactic cluster survey (Lodieu et al. 2012). The solid and dashed lines are the 120 Myr Nextgen isochrone and the 120 Myr "Dusty" isochrone from the Lyon group.**

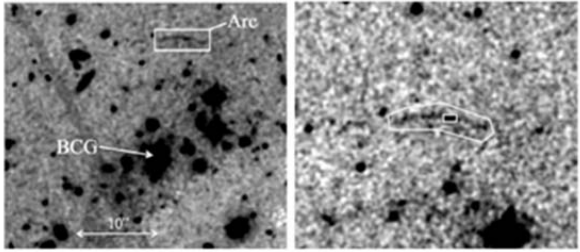

**Figure 32: Infrared images of the z=1.75 Spitzer-selected massive galaxy cluster IDCS J1426.5+3508 taken with the Hubble Space Telescope. This is the most massive cluster known at z>1.4, with intriguing cosmological implications. Images show the combined cluster center and arc (left) as well as a close up of the arc (right). Standard ΛCDM does not predict any gravitational arcs of this brightness behind z≥1.75 clusters across the full sky, let alone in the 8.5 deg2 from which this cluster was identified. [From Gonzalez et al. 2012].**

#### 2.6.4 *Nearby Galaxies*

Studies of nearby galaxies and their stellar populations have been a prime focus of previous and ongoing space missions. However, WFIRST will be the first mission that can study nearby galaxies over their full extent, at high spatial resolution, and to significant depth. This is expected to provide great advances in our understanding of stellar astrophysics, the stellar mass function, star formation and star formation histories of stellar populations, galaxy structure and substructure, galaxy interactions and accretion, galaxy formation, the interstellar medium, and globular cluster populations.

Relevant features in the CMDs of nearby stellar population include the main sequence turnoff ($M_V \sim$ 4.5), the red clump/horizontal branch ($M_V \sim$ 0.5), and the tip of the red giant branch ($M_V \sim$ -2.5). WFIRST can reach an equivalent depth of $V \sim$ 27.5 in the near-infrared in a matter of hours. Hence, these features are easily studied with WFIRST to distances of 0.4 Mpc, 2.5 Mpc, and 10 Mpc, respectively. At these distances, a 0.5 degree linear field size corresponds to 3.6 kpc, 22.5 kpc, and 86 kpc, respectively. Hence, a one-month WFIRST program can map some one hundred nearby galaxies over their full extent. For a galaxy at 10 Mpc, a single pointing will suffice to cover the luminous body of the galaxy down to the tip of the red giant branch (TRGB). For more nearby galaxies (*e.g.*, the LMC, M31, etc.), it will be necessary to mosaic a much larger region of the sky, but this can be done with shorter exposures to reach the same intrinsic depth.

WFIRST will surpass planned ground-based surveys for the study of nearby galaxies because of its superior spatial resolution (*e.g.*, compare to ~0.7 arcsec for LSST). WFIRST will resolve stars much further into the crowded inner regions of galaxies. Most background galaxies, which provide the main source of confusion in the sparse outer regions, will be spatially resolved by WFIRST. The ability to study structural and population gradients over large areas for many galaxies opens up the promise of fully constraining how the different structural components of galaxies grow and evolve over time. Observations of substructure and stellar streams in many galactic halos will tightly constrain galaxy formation models. The hierarchical galaxy assembly process is strongly stochastic, making it imperative to study a large sample of galaxies. A study of this kind will reveal what fraction of mass in present day galaxies is associated with past and current accretion events.

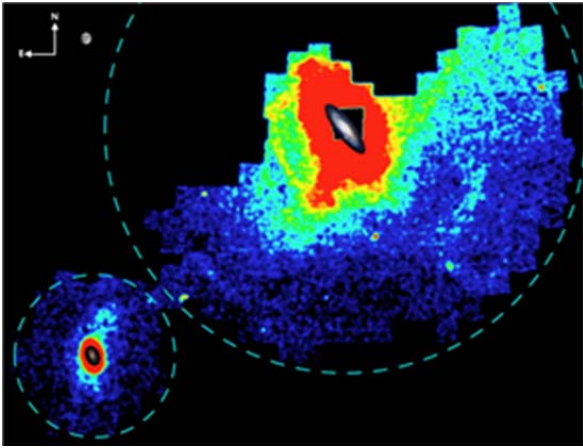

**Figure 33: Streams and tidal features in the distribution of red giant branch (RGB) stars in the halos of M31 and M33 mapped by the PAndAS collaboration (McConnachie et al. 2009). WFIRST will be able to map the halos of many nearby galaxies in similar manner, tightly constraining past accretion events and galaxy formation models. This will be only one of the many key studies that WFIRST can perform on nearby galaxies, covering an area 300 times larger.**

### 2.6.5 *WFIRST Medium-Deep Fields: Probing Early Structure Formation*

One likely WFIRST GO program with significant legacy data value would be to obtain imaging and spectroscopic observations of a series of medium deep fields, with survey areas and depths linking the wide-area, HLS to the pencil-beam, deep supernova fields. The LSST project has selected four extragalactic Deep Drilling Fields with a total area of ~40 deg$^2$. LSST will devote 10% of the total observing time to repeated imaging of these fields, reaching depths of ~27-28 AB with dense temporal sample for variability studies. A likely WFIRST GO program would be to cover the LSST deep drilling fields down to AB ~ 27 in four bands, slitless prism spectroscopy down to continuum limits of $H$ ~ 23.5 and H$\alpha$ flux limits of ~ $3\times10^{-17}$ erg s$^{-1}$ cm$^{-2}$. Such a program would require ~2 months of observing time. Figure 34 presents a recent determination of the $z \sim 8$ galaxy luminosity function based on a several HST deep fields. A WFIRST medium-deep imaging survey would have comparable depth to the HST CANDELS Legacy Project, while covering 300 times larger area. The imaging survey would be sensitive to $L*$ galaxies up to $z \sim 8$. It would also probe far down the galaxy luminosity function at lower redshifts, detecting stellar mass out to $z \sim 2$. Similarly, a medium-deep slitless prism survey would reach sensitivities comparable to the 3D-HST Treasury project with 500 times larger vol-ume, enabling redshift measurements of $L*$ galaxies at z~2.5, the peak era of cosmic star formation.

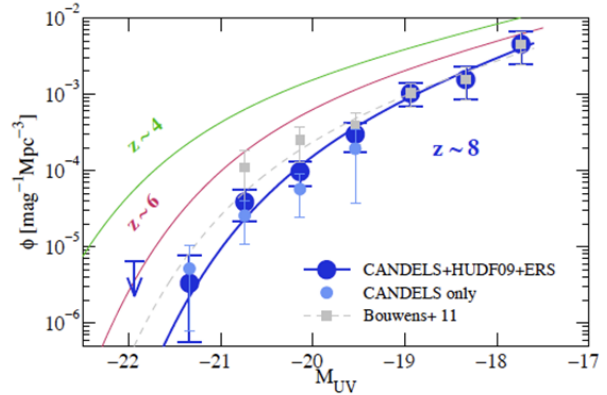

**Figure 34: Determination of galaxy UV luminosity function at z~8 based on HST surveys (Oesch et al. 2012). The WFIRST medium deep survey will reach M<-20, comparable to the CANDELS wide survey at this redshift, and cover an area 300 times larger.**

The unique combination of depth and area would enable qualitatively new extragalactic survey science. WFIRST medium-deep surveys would sample cosmic volumes comparable to the SDSS at low redshift. Such work would not only overcome cosmic variance (which has been a significant concern for current deep pencil-beam surveys), but would also allow detailed measurements of galaxy clustering from $z = 1$ to $z = 8$, mapping the emergence of large-scale cosmic structures in the Universe. WFIRST medium-deep surveys would use clustering statistics to estimate the dark-halo masses of high-redshift galaxies, and use cross-correlation to study galaxy-black hole co-evolution. The large survey volume of such an effort would reveal the rarest objects at high-redshift, including the first proto-clusters and the highest-redshift quasars. By combining with LSST time-domain observations, the WFIRST medium-deep fields could also be used to study properties of transient and variable sources, in particular AGNs. In addition, such fields would likely prove beneficial for calibrating the WFIRST dark energy HLS fields.

### 2.6.6 *Other GO Programs*

In addition to the handful of GO programs outlined above, WFIRST will enable an immense range of additional science, far beyond the space limitations of the current report (and likely beyond the current imaginations of the WFIRST SDT). We provide a brief list of additional GO programs that have been discussed by our group:

- Deep surface brightness photometry of galaxies and galaxy clusters; and
- Studying the shapes of galaxy haloes from gravitational flexion.

## 2.7    Science Requirements

The SDT and Project have developed a requirements flowdown matrix for the mission. The four top-level scientific objectives for WFIRST are:

- Complete the statistical census of planetary systems in the Galaxy, from the outer habitable zone to free floating planets, including analogs of all of the planets in our Solar System with the mass of Mars or greater.

- Determine the expansion history of the Universe and the growth history of its largest structures in order to test explanations of its apparent accelerating expansion including Dark Energy and modifications to Einstein's gravity.

- Produce a deep map of the sky at NIR wavelengths, enabling new and fundamental discoveries ranging from mapping the Galactic plane to probing the reionization epoch by finding bright quasars at z>10.

- Provide a general observer program utilizing a minimum of 10% of the mission minimum lifetime.

These objectives then drive the requirements for the observatory capabilities and design. A top-level flow-down of the WFIRST requirements is given in Figure 35. The Science Objectives above are the highest level science requirements and appear at the top of the page. The derived scientific survey capability requirements of the observatory are listed in the left-hand boxes and data set requirements in the middles boxes. The top-level Observatory design/operations parameters are listed in the right-hand boxes. The detailed discussion of the basis for the requirements is given in the preceding subsections.

**WFIRST Science Objectives:**

1) Complete the statistical census of planetary systems in the Galaxy, from the outer habitable zone to free floating planets, including analogs of all of the planets in our Solar System with the mass of Mars or greater.

2) Determine the expansion history of the Universe and the growth history of its largest structures in order to test possible explanations of its apparent accelerating expansion including Dark Energy and modifications to Einstein's gravity.

3) Produce a deep map of the sky at NIR wavelengths, enabling new and fundamental discoveries ranging from mapping the Galactic plane to probing the reionization epoch by finding bright quasars at z>10.

4) Provide a general observer program utilizing a minimum of 10% of the mission minimum lifetime

## WFIRST Survey Capability Rqts

### Exoplanet Microlensing Survey

- Planet detection capability to ~0.1 Earth mass ($M_\oplus$)
- Detects $\geq$ 1500 bound cold planets in the mass range of 0.1-10,000 Earth masses, including 150 planets with mass <3 Earth masses
- Detects $\geq$ 20 free floating Earth-mass planets

### Dark Energy Surveys
### Galaxy Redshift Survey (GRS)

- $\geq$1400 deg$^2$ per dedicated observing year (combined HLS imaging and spectroscopy)
- A comoving density of galaxy redshifts at z=2 of $2.9 \times 10^{-4}$ Mpc$^{-3}$
- Redshift range $1.3 \leq z \leq 2.7$
- Redshift errors $\sigma_z \leq 0.001(1+z)$, equivalent to 300 km/s rms
- Misidentified lines $\leq$5% per source type, $\leq$10% overall; contamination fractions known to $2 \times 10^{-3}$

### Supernova SN-Ia Survey

- >100 SNe-Ia per $\Delta z$=0.1 bin for most bins for 0.4 < z < 1.2, per dedicated 6 months
- Observational noise contribution to distance modulus error $\sigma_\mu \leq 0.02$ per $\Delta z$=0.1 bin up to z = 1.7
- Redshift error $\sigma \leq$ 0.005 per supernova
- Relative instrumental bias $\leq$0.005 on photometric calibration across the wavelength range

### WL Galaxy Shape Survey

- $\geq$ 1400 deg$^2$ per dedicated observing year (combined HLS imaging and spectroscopy)
- Effective galaxy density $\geq$30/amin$^2$, shapes resolved plus photo-z's
- Additive shear error $\leq 3 \times 10^{-4}$
- Multiplicative shear error $\leq 1 \times 10^{-3}$
- Photo-z error distribution width $\leq 0.04(1+z)$, catastrophic error rate <2%

### Near Infrared Survey

- Identify $\geq$100 quasars at redshift $z \geq 7$
- Extend studies of galaxy formation and evolution to z > 1 by making sensitive, wide-field images of the extragalactic sky at near-infrared wavelengths, thereby obtaining broad-band spectral energy distributions of $\geq 1 \times 10^9$ galaxies at z>1
- Map the structure of the Galaxy using red giant clump stars as tracers

## WFIRST Data Set Rqts

### Exoplanet Data Set Rqts

- Monitor >2 square degrees in the Galactic Bulge for at least 250 total days
- S/N $\geq$100 per exposure for a J=20.5 star
- Photometric sampling cadence of $\leq$15 minutes
- $\leq$0.4" angular resolution to resolve the brightest main sequence stars
- Monitor microlensing events continuously with a duty cycle of $\geq$80% for at least 60 days
- Sample light curves with W filter
- Monitor fields with Y filter, 1 exposure every 12 hours
- Separation of >2 years between first and last observing seasons

### Dark Energy Data Sets
### GRS Data Set Rqts

- Slitless prism, spectrometer dispersion $D_\Theta$ = 150 - 250 arcsec
- S/N $\geq$7 for $r_{eff}$ = 300 mas for H$\alpha$ emission line flux at 2.0 $\mu$m $\geq 1.1 \times 10^{-16}$ erg/cm$^2$-s
- Bandpass 1.5 $\leq z \leq$ 2.4 $\mu$m
- Pixel scale $\leq$ 180 mas
- System PSF EE50% radius 325 mas at 2 $\mu$m
- $\geq$3 dispersion directions required, two nearly opposed
- Reach $J_{AB}$=24.0 AND ($H_{AB}$=23.5 OR $K_{AB}$=23.1) for $r_{eff}$=0.3 arcsec source at 10 sigma to achieve a zero order detection in 2 filters

### Supernova Data Set Rqts

- Minimum monitoring time-span for an individual field: ~2 years with a sampling cadence $\leq$5 days
- Cross filter color calibration $\leq$0.005
- Three filters, approximately J, H, K
- Slitless prism spec (P130) 0.6-2 $\mu$m, $\lambda/\Delta\lambda$ ~75 (S/N $\geq$ 2 per pixel bin) for redshift/typing
- Photometric S/N $\geq$15 at lightcurve maximum in each band at each redshift
- Dither with 30 mas accuracy
- Low Galactic dust, E(B-V) $\leq$0.02

### WL Data Set Rqts

- From Space: 3 shape/color filter bands (J,H, and K) and 1 color filter band (Y; only for photo-z)
- S/N $\geq$18 (matched filter detection significance) per shape/color filter for galaxy $r_{eff}$ = 250 mas and mag AB = 23.9
- PSF second moment ($I_{xx} + I_{yy}$) known to a relative error of $\leq 9.3 \times 10^{-4}$ rms (shape/color filters only)
- PSF ellipticity ($I_{xx}-I_{yy}$, $2*I_{xy}$)/ ($I_{xx} + I_{yy}$) known to $\leq 4.7 \times 10^{-4}$ rms (shape/color filters only)
- System PSF EE50 radius $\leq$166 (J band), 185 (H), or 214 (K) mas
- At least 5 (H,K) or 6 (J) random dithers required for shape/color bands, and 4 for Y at same dither exposure time
- From Ground: $\geq$4 color filter bands ~0.4 $\leq \lambda \leq$ ~0.92 $\mu$m
- From Ground + Space combined: Complete an unbiased spectroscopic PZCS training data set containing $\geq$ 100,000 galaxies $\leq$ mag AB = 23.9 (in JHK bands) and covering at least 4 uncorrelated fields; redshift accuracy required is $\sigma_z < 0.01(1+z)$

### Near Infrared Data Set Rqts

- Image $\geq$ 2500 deg$^2$ of high latitude sky in three near-infrared filters to minimum depths of mag AB = 25 at S/N=5. Fields must also have deep (ground-based) optical imaging
- Image $\geq$ 1500 deg$^2$ of the Galactic plane in three near-infrared filters

**WFIRST DRM1 Design/Operations Overview**

**Key WFIRST DRM1 Observatory Design Parameters**
- Off-axis focal telescope; 1.3m diameter telescope aperture
- ≤205 K telescope optical surfaces
- Bandpass 0.6 – 2.4 μm
- Pointing jitter ≤40 mas rms/axis
- Coarse Pointing Accuracy <~3 arcsec rms/axis
- Fine (Relative/Revisit) Pointing Accuracy <~25 mas rms/axis
- ACS telemetry downlinked for pointing history reconstruction

**Imaging Mode:**
- Pupil Mask temperature: ~150 K
- Effective Area (avg over bandpass): 0.778 m²
- 6 band parfocal filter set on filter wheel, driven by microlensing, SNe, WL, plus open and blank positions
- FPA: 4x9 HgCdTe 2k x 2k SCAs, 2.5μm, ≤100K, 180 mas/pix
- FOV (active area) = ~0.375 deg²; Bandpass 0.6 – 2.4 μm
- 4 Outrigger FGS SCAs mounted to Focal Plane Assy (FPA), control Pitch/Yaw during imaging mode
- WFE is diffraction limited at 1μm
- 1 nm rms wavefront stability during the HLS imaging mode (requires 0.3 K thermal stability)

**Spectroscopy Mode:**
- SN Prism: R=75 (2-pix) parfocal, zero deviation prism
- SN Prism Effective Area (avg over bandpass): 0.750 m² (0.6 to 2.0 μm)
- 2 oppositely dispersed GRS prisms: $D_{\theta}$ = 160 - 240 arcsec, parfocal, zero deviation prism
- GRS Prism Effective Area (avg over bandpass: 0.953 m² (1.5-2.4μm)
- 6 position prism wheel provides a fail-safe open position between each of the above 3 prisms
- 2 SCAs in an independent auxiliary FGS control Pitch/Yaw during spectroscopy mode

| Name | Bandpasses (μm) |
|------|-----------------|
| Z085 | 0.732 – 0.962 |
| Y107 | 0.920 – 1.209 |
| J134 | 1.156 – 1.520 |
| H168 | 1.453 – 1.910 |
| K211 | 1.826 – 2.400 |
| W166 | 0.920 – 2.400 |
| SN Prism (P130) | 0.60 – 2.00 |
| GRS Prisms | 1.50 – 2.40 |

**Key WFIRST DRM1 Operations Concept Parameters and Constraints**
- 5-year mission life, but consumables required for 10 yrs
- Science Field of Regard (FOR): 54˚ to 126˚ pitch off the Sun, 360˚ yaw
- Roll about line of sight ±10˚, except during SNe observations where allowable roll is ±22.5˚to provide inertially fixed viewing for ~45 days near the ecliptic pole(s)
- Gimbaled antenna allows observing during downlinks
- Slew/settle times: ~16 s for dithers, ~38 s for ~0.7˚ slews

**SNe-Ia Survey (6.48 deg² monitored to z = 0.8, 1.8 deg² monitored to z = 1.7, ~6 months total)**
- A sample 2-tiered survey capability (given ~6 months dedicated time) is shown, each tier optimized for a z range
- Tier 1 (to z=0.8):18 "shallow fields" total  6.48 deg²; J134, H168, K211 (300 s @), P130 (1800 s)
- Tier 2 (z=1.7): 5 "deep" fields total 1.80 deg², J134, H168, K211 (1500 s @), P130 (9500 s)
- SNe dedicated time is distributed in a 5-day cadence over ~1.8 years to provide suitable light curve tracking and accurate host galaxy references (33 hrs of SNe field monitoring would be done every 5 days for ~1.8 years)
- SNe fields are monitored from end of one exoplanet microlensing Galactic Bulge season until the start of the 4th following exoplanet microlensing season (~1.8 yrs)
- Fields located in low dust regions ≤20˚ off an ecliptic pole (N and S fields not required)
- Sub-pixel dithers, accurate to ~25 mas, performed at each pointing

**Galaxy Shape + Galaxy Redshift Survey (~1,400 deg², 3,400 deg² total)**
- 2 imaging passes in each of JHK (shape) and Y (photo-z) filters
- Each imaging pass with the same filter is rotated from the other by ~5˚, and includes 4 exposures (150 sec each), with 5 exposures for each J band pass
- 90% of imaging field see ≥5 randomly dithered exposures (≥750 sec total) in YHK bands, ≥6 exposures (≥900 sec total) in J band
- 2 spectrometry passes in each of 2 oppositely dispersed GRS prisms
- Each spectrometry pass with the same prism is rotated from the other by ~5˚, and includes 2 exposures (530 sec each)
- 90% of spectroscopy field sees ≥6 randomly dithered exposures (≥3180 sec total)
- Zero order galaxy detection to GRS data set requirements provided in JH or JK bands by WL imaging passes

**Exoplanet Microlensing Survey (3.38 deg² monitored every 15 min, 144 days/yr, 1.2 yrs total)**
- The Galactic Bulge is observable for two 72-day seasons each year
- The short revisit cadence impacts other observing modes while exoplanet data sets are being acquired. This, in combination with the field monitoring time span required for SN and the ≥60 days required for microlensing, limits the max number of Galactic Bulge seasons useable for exoplanet  observations to six (over 5 yrs)
- In each season 9 fields are revisited on a 15 min cadence, viewing in filter W166, for light curve tracking except for one exposure every 12 hours that uses the Y107 filter for color
- Fields are revisited to an accuracy of 1 pixel rms; no precise dithers

**Galactic Plane Survey (~6 months total)**
- Uses the same survey strategy (both exposure times and tiling) as the imaging portion of the HLS
- Survey area (TBD),assumed to be the 1240 deg² region with |b|<1.72°for scheduling purposes

**Figure 35: WFIRST requirements flowdown overview**

# 3  DRM1: A 1.3 M TELESCOPE WITH A 5 YEAR MISSION

## 3.1  Overview

The WFIRST DRM1 observatory uses conventional telescope, instrument, and spacecraft architectures and hardware (see Figure 36) to meet the WFIRST science requirements. The WFIRST DRM1 payload (instrument and telescope) configuration provides the wide-field imaging and slitless spectroscopy capability required to perform the Dark Energy, Exoplanet and NIR surveys (see Figure 37 for an optical path block diagram and Figure 38 for a fields of view layout). A 1.3 m unobstructed aperture, focal telescope feeds a single instrument with a filter wheel and a prism wheel to provide an imaging mode covering 0.78 – 2.4 μm and two spectroscopy modes covering 1.5 – 2.4 μm and 0.60 – 2.0 μm. The instrument uses 2.5 μm long-wavelength cutoff HgCdTe detectors already developed for the James Webb Space Telescope and other programs. The prism wheel contains three prisms, including two identical prisms rotated 180° to disperse in opposing directions, providing redshift measurements for the galaxy redshift survey unbiased by spatial offsets between the line and continuum emission regions. The third prism provides low resolution dispersion for executing the SN program. The instrument covers the near-infrared and is optimized to provide good sensitivity down to 0.6 μm in the visible. The instrument provides a high quality point spread function, precision photometry, and stable observations for implementing the WFIRST science.

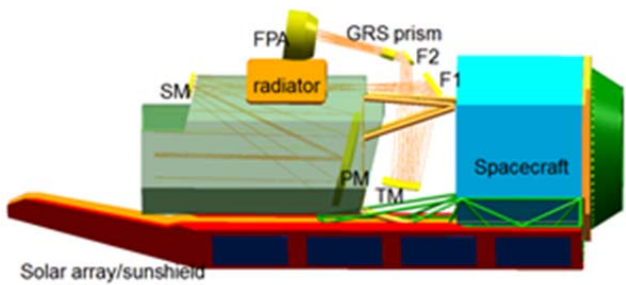

**Figure 36: DRM1 Observatory design showing the fixed solar array/sunshield with the telescope viewing normal to the top of the spacecraft. The instrument structure is not shown allowing a view of the optics and focal plane structure.**

Quality pointing accuracy, knowledge, and stability are all required to resolve galaxy shapes and precisely revisit both the Exoplanet and SN fields. Pointing to between 54° and 126° off the Sun enables the observation of Exoplanet fields for up to 72 continuous days during each of the twice yearly Galactic Bulge viewing seasons. The observatory accommodates viewing within 20° of the ecliptic poles to monitor SN fields in fixed inertial orientations for ~45 days.

The Exoplanet survey requires large light gathering power (effective area times field of view) for precise photometric observations of the Galactic Bulge to detect star + planet microlensing events. Multiple fields are observed repeatedly to monitor lightcurves of the relatively frequent stellar microlensing events and the much rarer events that involve lensing by both a star and a planet. In the latter case the planetary signal is briefly superposed on the stellar signal. Exoplanet monitoring observations are performed in a wide filter spanning 0.92 – 2.40 μm, interspersed ~twice/day with brief observations in a narrower filter for stellar type identification.

The GRS measurement requires NIR spectroscopy to centroid Hα emission lines and NIR imaging to locate the position of the galaxy image. Dispersion at R ~600 (2 pixels) enables centroiding the Hα emission lines to a precision consistent with the redshift accuracy requirement. To address completeness and confusion issues, prisms are used as the dispersing element and at least 3 roll angles, two of which are approximately opposed, are observed over ~90% of the mapped sky. The bandpass range of 1.5 – 2.4 μm provides the required redshift range for Hα emitters.

The SN measurement also requires large light gathering power to perform the visible and NIR deep imaging and spectroscopy needed to classify and determine the redshift of large numbers of Type Ia SN. Precise sampling (S/N of 15) of the light curve every five days meets the photometric accuracy requirement, and the use of three NIR bands allows measurements of SNe in the range of $0.4 < z < 1.7$, providing better systematics at low z than can be achieved by the ground and extending the measurements beyond the z ~ 0.8 ground limit.

The WL measurement requires an imaging and photometric redshift (photo-z) survey of galaxies to mag AB ~23.9. A pixel scale of 0.18 arcseconds balances the need for a large field of view with the sampling needed to resolve galaxy shapes. Observations in three NIR filters, with ≥5 random dithers in each filter, are made to perform the required shape measurements to determine the shear due to lensing (see Section 2.5.2.1), while observations in an additional NIR filter are combined with color data from the shape bands and the ground to provide the required photo-z determinations. Either the GRS or SN prisms with overlapping ground observations are used to perform the photo-z calibration survey (PZCS) needed to meet the WL redshift accuracy requirement.

The single fault tolerant spacecraft uses mature technology to support payload operations. It provides a fixed solar array/sunshield that allows operations over the full field of regard. An Earth-Sun L2 libration point orbit has been selected to provide the passive cooling, thermal stability, minimum stray light, and large field of regard needed to make these precise measurements. An Atlas V launch vehicle, with parking orbit insertion and transfer trajectory insertion control to enable eclipse-free launch opportunities virtually any day, is used to place the observatory on a direct trajectory transfer orbit to L2. The mission life is 5 years with consumables sized to allow an extension for a total of 10 years. Each of the elements of the mission is described below.

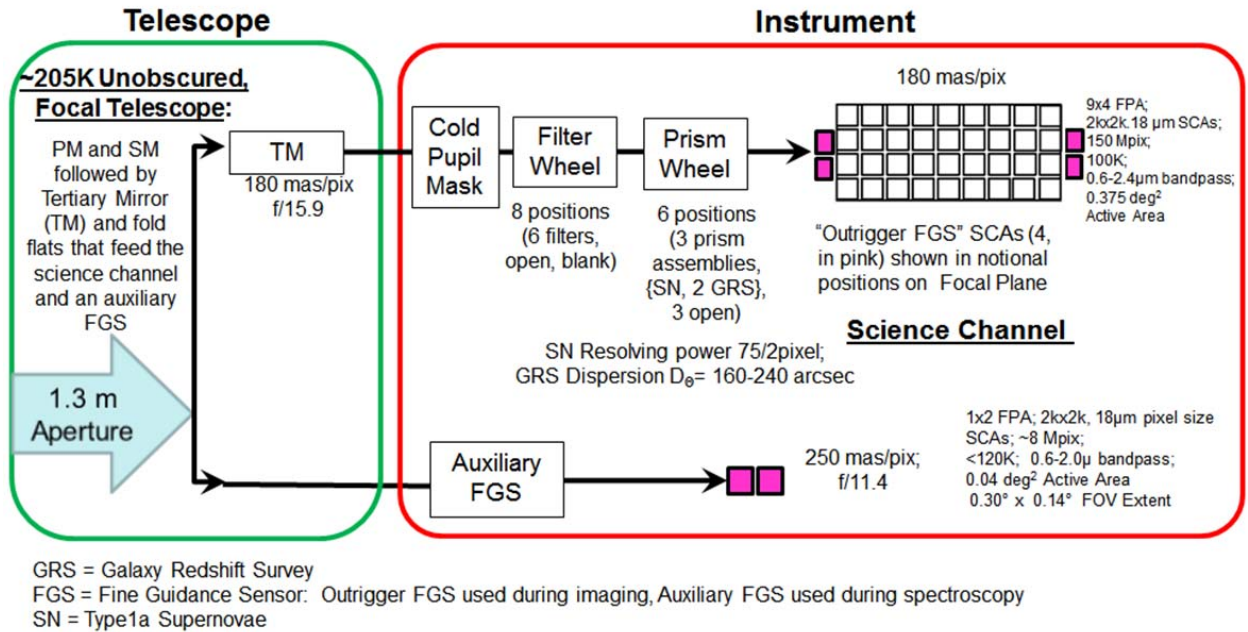

GRS = Galaxy Redshift Survey
FGS = Fine Guidance Sensor: Outrigger FGS used during imaging, Auxiliary FGS used during spectroscopy
SN = Type1a Supernova

**Figure 37: DRM1 payload optical block diagram**

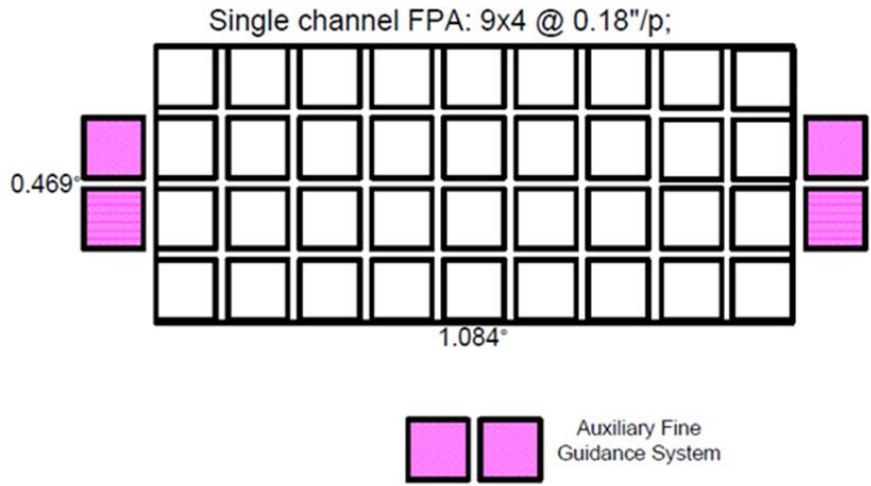

**Figure 38: Field layout as projected on the sky**

## 3.2 Telescope

The instrument is fed by a Three Mirror Anastigmat (TMA) unobstructed telescope, which offers a wide field along with a flat focal surface and good correction of low order aberrations. The design uses a focal TMA working at a pupil demagnification of 11.8 (110 mm pupil diameter with cold mask). A 1.3 meter diameter primary mirror feeds the science channel and the auxiliary guider channel. The unobstructed form was selected as the baseline for the IDRM and DRM1, a change from the JDEM-Omega baseline. Shifting the secondary mirror off-axis eliminates the diffraction pattern created by the secondary mirror and its supports. It also eliminates the need for a large secondary mirror baffle allowing the primary mirror diameter to be reduced to 1.3 m, from 1.5 m on JDEM-Omega, while still providing equivalent or better throughput. These improvements provide a lower exposure time to the same limiting flux as compared to JDEM-Omega. The field of view extent for the science channel is 0.507 deg$^2$ (0.375 deg$^2$ active area).

The Optical Telescope Assembly (OTA) reflecting surfaces are maintained below 205K to limit the in-band NIR thermal emissions to ≤ 10% of the minimum Zodiacal background. The instrument volume is maintained below 150 K to control thermal emissions. The tertiary mirror is included in the OTA, so the optical interface for the science channel is at a real pupil. The science and auxiliary FGS channels are well separated allowing access for integration of each channel in any order.

The secondary mirror has a 6 degree of freedom mechanism to adjust focus and alignment. The telescope mirrors are made from ultra low expansion fused silica (ULE) which allows a highly lightweighted, thermally stable mirror while the closed back design provides the required stiffness. The OTA structure is manufactured from low-moisture composites to minimize mass and thermal distortions while providing adequate stiffness. The telescope structure is insulated to minimize heat transfer from the solar array into the telescope and to minimize heat transfer to the instrument.

Gold mirror coatings provide good reflectance in-band and minimize thermal band emittance.

## 3.3 Instrument

The WFIRST instrument is divided into a science channel, with both imaging and spectroscopy modes, and an auxiliary fine guider channel (discussed below in 3.5). The key instrument parameters are shown in Table 13. The imaging mode is designed to a diffraction limit of 1 μm as required for WL galaxy imaging and SN S/N requirements.

The science channel consists of a cold pupil mask, filter wheel, prism wheel, and the HgCdTe focal plane array (FPA). The FPA uses 2k x 2k pixel HgCdTe detector arrays, with 18 μm pixels and a long-wavelength cutoff of 2.5 μm. The FPA is arranged in a 9x4 layout with a pixel scale of 0.18 arcseconds/pixel. An 8-position filter wheel provides 6 filters, a dark (for calibration) and a blank (open) position used during spectroscopy modes. A 6-position prism wheel is located just after the filter wheel and includes one prism for executing the SN program and two prisms for the GRS. Three blank locations in the prism wheel are provided to allow the mechanism to "fail-safe" to the open position adjacent to each prism.

The SN prism has a spectral range of 0.6-2.0 μm to cover a wide spectral and redshift range, at a nominal two pixel resolving power of 75; this varies with wavelength and rises significantly at the blue edge.

Each 4-element GRS prism consists of 3 CaF$_2$ prisms and 1 S-TIH1 prism. The GRS prisms have a spectral range of 1.5-2.4 μm and a dispersion of $D_\Theta$=160-240 arcsec. They are arranged in the prism wheel to disperse in opposing directions on the sky to reduce source confusion. The equivalent point-source resolution R=$\lambda/\delta\lambda$, is obtained by dividing by the angular size of a 2-pixel resolution element. Thus for a value of $D_\Theta$=200 arcsec, the corresponding value of R is 200/0.36=555.A description of the filter wheel and prism wheel complement is shown in Table 14.

| Mode | Wavelength Range (μm) | Sky Coverage (active area; deg²) | Pixel Scale (arcsec/pixel) | Dispersion | FPA Temperature (K) |
|---|---|---|---|---|---|
| Imaging | 0.6 – 2.4 | 0.375 | 0.18 | N/A | <100 |
| GRS Spectroscopy | 1.5 – 2.4 | 0.375 | 0.18 | 160-240 ($D_\Theta$; slitless in prism wheel, 2 copies provide counter dispersion) or R~600 (2 pixel) | <100 |
| SN Ia Spectroscopy | 0.6 – 2.0 | 0.375 | 0.18 | R=75 (2-pixel; slitless prism in prism wheel) | <100 |
| Auxiliary Guiding | 0.6 – 2.0 | 0.08 | 0.25 | N/A | <150-170 |

**Table 13: Key Instrument Parameters**

| Band | Filter Name | Min | Max | Center | Width | Dispersion |
|---|---|---|---|---|---|---|
| | | | Filter Wheel | | | |
| Z | Z085 | 0.732 | 0.962 | 0.847 | 0.230 | 3.68 |
| Y | Y107 | 0.920 | 1.209 | 1.065 | 0.289 | 3.68 |
| J | J134 | 1.156 | 1.520 | 1.338 | 0.364 | 3.68 |
| H | H168 | 1.453 | 1.910 | 1.682 | 0.457 | 3.68 |
| K | K211 | 1.826 | 2.400 | 2.113 | 0.574 | 3.68 |
| Wide | W168 | 0.920 | 2.400 | 1.660 | 1.480 | 1.12 |
| | | | Prism Wheel | | | |
| | SN | 0.60 | 2.00 | 1.50 | 1.80 | R (2 pix) ~75 |
| | GRS+ | 1.50 | 2.40 | 1.85 | 1.10 | $D_\Theta$ = ~200 |
| | GRS- | 1.50 | 2.40 | 1.85 | 1.10 | $D_\Theta$ = ~200 |

**Table 14: Filter and prism descriptions; Wavelengths are in μm. GRS prism dispersion is in units of arcsec with $D_\Theta = \lambda/(\delta\lambda/\delta\Theta)$ where $\Theta$ is sky angle.**

The instrument is kinematically mounted to and thermally isolated from the telescope structure. The instrument optics are maintained below 150 K via radiation off of the channel housings. The detectors/Sensor Chip Assemblies (SCAs) on the instrument focal plane are cooled to below 100 K via a dedicated radiator. The Sensor Cold Electronics (SCEs) that power and readout the SCAs are mounted close to, but thermally isolated from, the SCAs as they can be held at a warmer temperature. The SCEs are cooled to ~150 K via direct radiation off of their mounting plate. Panels on the side of the solar array are used as sunshields to protect the payload structure from solar illumination during the ~45 day inertially fixed SN observations.

The use of CMOS-multiplexer (readout integrated circuit) based hybrids with non-destructive readouts support noise reduction and permit electronic shutter-ing, eliminating the need for a shutter mechanism. Sample Up The Ramp (SUTR) processing is used during all observations with the raw frames that are generated every ~1.3 sec being combined during the course of the integration period to produce one output image (Offenberg et al. 2005). All detectors in the instrument are identical, simplifying the detector production and sparing.

Though not formally a part of the scientific instrumentation, there are additional imaging detectors used for fine guidance. During normal imaging operation, two of the four "outrigger" detectors located on the FPA are used for guiding (the other pair provide redundancy). When any of the disperser elements are inserted by the prism wheel, these outriggers will no longer see undispersed stellar images, so a separate auxiliary FGS with 2 SCAs provides this function at those times.

### 3.4    Calibration

The WFIRST strategy is to use ground calibration methods to the maximum extent possible, reserving on-orbit calibration to verification of the ground results and extending the calibrations where ground calibration may not be effective. To maintain the calibration requirements over the entire mission, not only are the calibrations important, but so are measurements of calibration stability. The latter will determine the need for and frequency of on-orbit calibrations. The WFIRST calibration program will place strong emphasis not only on the areas requiring calibration, but also on the verification of these calibrations, either on the ground or in orbit, using multiple techniques as cross-checks. The SN and Exoplanet fields are observed repeatedly over the lifetime of the mission, providing excellent opportunities to develop and use sky calibration standards.

All optical and detector components will be calibrated at the component, subsystem, and instrument levels. These data will be used to feed an integrated instrument calibration model that will be verified using an end-to-end payload-level thermal vacuum test. This test will involve a full-aperture (1.3 meter) diameter collimated beam that will test for optical wavefront error as well as photometry.

The exoplanet program imposes some constraints on photometric calibration, but also provides a unique opportunity to meet these calibration requirements. Proper measurements of the stellar and planetary light curves require stable relative calibration to 0.1% (relative to nearby stars in the same detector) over the course of the event. These observations use mainly a single filter. Over the course of the mission, the fields are sampled many tens of thousands of times with a random dither. Most of these observations will be of stars without microlensing events. If the star is not a variable star, then the relative calibration will be monitored during the extensive number of observations to establish stability. Slow variations across the field-of-view and over time can thus be monitored and corrected for. The absolute calibration and occasional color measurements using the bluer color filter have less stringent calibration requirements (1%) that will be met by the more stringent calibration requirements for Dark Energy that are described below.

The three Dark Energy observational methods have different calibration demands on instrument parameters and their accuracy. The SN Survey places the most stringent demands on absolute and inter-band photometric calibration. White Dwarfs and other suitable sky calibration targets will be used to calibrate the absolute flux as well as linearity of the imager over several orders of magnitude. This linearity will be tested on the ground, and verified with an on-orbit relative flux calibration system (if necessary). Observations of astronomical flux standards will be extended across the detector by means of the "self-calibration' techniques described by Fixsen, Mosely, and Arendt (2000) and employed for calibration of Spitzer/IRAC (Arendt et al. 2010). Similar techniques achieved ~1% relative calibration accuracy for the SDSS imaging data (Padmanabhan et al. 2008), despite the difficulties posed by variable atmospheric absorption. The intra-pixel response function (quantum efficiency variations within a pixel) will be fully characterized by ground testing.

For the WL Survey, the requirement for galaxy ellipticity accuracy places significant demands on both the optical and detector subsystems. The uniformity and stability of the point spread function (PSF) needs to be strictly controlled and monitored to ensure a successful mission. This drives the need to characterize the detector intra-pixel response and the inter-pixel response (capacitive cross-coupling with nearest neighbors) for both magnitude as well as spatial and temporal variations. It is likely that the combined PSF effects, including spacecraft jitter, will have some variability on time scales of a single exposure. These residual effects will be continuously monitored with the observatory attitude control system and field stars (Jurling 2012) and downlinked to provide ancillary information for the scientific data analysis pipeline.

The GRS relies primarily on the GRS prisms, which do not drive the calibration requirements for the mission. Established calibration techniques used for other space missions should be adequate to meet the relatively loose photometric and wavelength calibration requirements. Since the GRS uses the same focal plane as the more demanding imaging surveys, the small-scale flat field, nonlinearity, dark current, IPC, and intra-pixel response calibration for these surveys should be adequate for the GRS.

### 3.5    Fine Guidance Sensor

The fine guidance sensor (FGS) is used to meet the fine pointing requirements needed for the Exoplanet, WL, and SN surveys. The absolute pointing accuracy requirement of <25 milli-arcseconds rms/axis is driven by SN requirements for returning precisely to previously observed objects. WL drives the pointing stability of <40 milli-arcseconds over an exposure. The primary Outrigger FGS consists of two pair of HgCdTe detectors, a prime and redundant, located on the science channel FPA and fed through the science channel optical train, including the filter wheel. This guider is used in all observations that include imaging. An additional pair of FGS detectors, the Auxiliary FGS, is fed from a separate field at the telescope intermediate focus via an additional compact reflective relay. The Auxiliary FGS, along with the star trackers, provides pointing control for science observations when the science channel is performing spectroscopy.

### 3.6 Spacecraft

The WFIRST spacecraft has been designed to provide all the resources necessary to support the payload at L2 using mature and proven technology. The design is based on the Solar Dynamics Observatory (SDO) spacecraft, which was designed, manufactured, integrated, tested, and qualified at GSFC. The spacecraft bus design provides cross strapping and/or redundancy for a single-fault tolerant design. *Structures:* To minimize the observatory mass, the spacecraft bus is a graphite epoxy composite hexagonal structure, consisting of two modules (bus module and propulsion module) that house the spacecraft and payload electronics boxes and the propulsion tank. The spacecraft bus provides the interfaces to the payload and the launch vehicle and supports the Observatory launch mass of 2509 kg, including margin (see Table 15). It supports a multipanel fixed solar array and a sunshield to prevent the Sun from illuminating payload hardware during science observations (see Figure 36). *Attitude Control:* The spacecraft is three-axis stabilized and uses data from the payload fine guidance sensor, inertial reference unit, and star trackers to meet the coarse pointing control of 3 arcsec, and the fine relative pointing control of 25 mas pitch/yaw and 1 arcsec roll (all values RMS per axis). There are 2 sets of fine guidance sensors. The first is a pair of redundant sensors (4 total) on the science channel focal plane and the second is a pair of auxiliary guiders (2 total) for guiding during spectroscopy mode. The star trackers are used for coarsely pointing to within 3 arcsec RMS per axis of a target. After that, the FGS takes over to meet the fine pointing requirements for revisits and relative offsets. A set of 4 reaction wheels is used for slewing as well as momentum storage. The wheels are passively isolated to allow stable pointing at frequencies higher than the FGS control band. The GNC subsystem provides a safe hold capability (using coarse sun sensors), which keeps the observatory thermally-safe, power-positive and protects the instrument from direct sunlight. *Propulsion:* A hydrazine mono-prop subsystem is required for orbit insertion, orbit maintenance, and momentum dumping from the reaction wheels throughout the duration of the mission. The prop subsystem does not have any unique features for WFIRST and is well within the requirements of other recent propulsion systems that have launched. *Electrical Power:* Three fixed, body-mounted solar array panels (9.6 m²) provide the observatory power. Gallium Arsenide solar array cells operate with 28% efficiency and provide 2250 W of output power (at 0° roll and pitch) for an average orbit usage of ~1150 W. The remainder of the power subsystem is comprised of an 80 A-hr battery, sized to accommodate loads during thruster burns during cruise to L2, and power supply electronics that control the distribution of power and provide unregulated 28 Vdc power to the payload. The solar array is currently sized to provide full observatory power at EOL with 2 strings failed at the worst case observing angles. *Communications:* The communications subsystem uses S-band transponders to receive ground commands and to send real-time housekeeping telemetry to the ground via 2 omni-directional antennas as well as for ranging. A Ka-band transmitter with a gimbaled antenna will downlink stored science and housekeeping data at a rate of 150 Mbps without interrupting science operations. *Command & Data Handling:* The command and data handling subsystem includes a 1.1 Tb solid state recorder (SSR) sized to prevent data loss from a missed contact. The daily data volume is estimated at 1.3 Tb per day, assuming 2:1 lossless compression. The data will be downlinked twice daily to the NASA Deep Space Network (DSN). The C&DH/FSW provides fault management for the spacecraft health and safety as well as being able to safe the payload when necessary. *Thermal:* The spacecraft thermal design is a passive system, using surface coatings, heaters and radiators.

| | Mass (kg) |
|---|---|
| Instrument | 163.1 |
| Telescope | 649.6 |
| Spacecraft | 1005.1 |
| Observatory (dry) | 1817.8 |
| 30% contingency | 545.3 |
| Propellant | 146.0 |
| Observatory (wet) | 2509.1 |
| Atlas V lift capability | 3210.0 |
| Margin (%) | 27.9% |

Table 15: DRM1 mass breakdown

### 3.7 Ground System

The WFIRST Mission Operations Ground System is comprised of three main elements: 1) the facilities used for space/ground communications and orbit determination, 2) the Mission Operations Center (MOC) and 3) the facilities for science and instrument operations and ground data processing, archiving, and science observation planning. For each element, existing facilities and infrastructure will be leveraged to provide the maximum possible cost savings and operational efficiencies. The functions to be performed by the ground system and the associated terminology are shown in Figure 39.

The DSN is used for spacecraft tracking, commanding and data receipt. It interfaces with the MOC for all commanding and telemetry. Tracking data is sent to the GSFC Flight Dynamics Facility.

The MOC performs spacecraft, telescope and instrument health & safety monitoring, real-time and stored command load generation, spacecraft subsystem trending & analysis, spacecraft anomaly resolution, safemode recovery, level 0 data processing, and transmission of science and engineering data to the science and instrument facilities. The MOC performs Mission-level Planning and Scheduling.

The science and instrument facilities maintain instrument test beds, perform instrument & telescope calibrations, assist in the resolution of instrument anomalies, perform instrument flight software maintenance, and generate instrument command loads.

These facilities are also responsible for science planning & scheduling, supporting mission planning activities carried out by the MOC, running the Archival Research (AR) and General Observer (GO) Programs, providing Science Team, AR, and GO support, and performing EPO activities for the public and the astronomical community. Data handling involves ingesting Level 0 science and engineering data from the MOC and performing Level 1-3 data processing for the Science Teams and GO community and transmitting these calibrated data to the data archive. All data will be archived. Data search and access tools are provided to the science community that enable efficient searches and delivery of archival data that ensure interoperability with NASA data archives and the Virtual Astronomical Observatory.

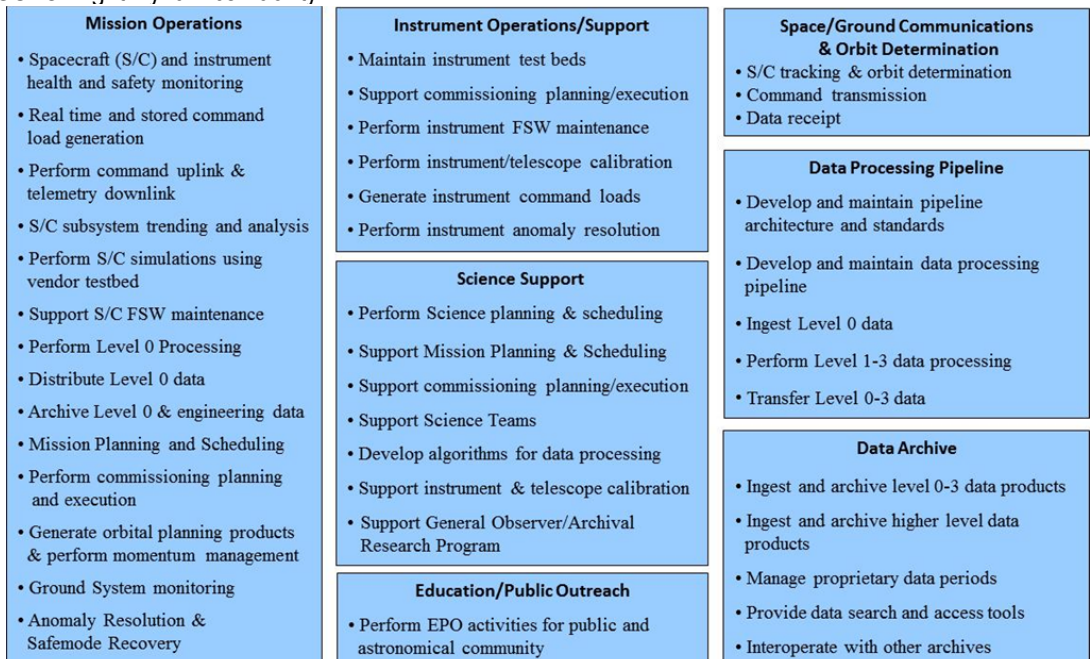

Figure 39: WFIRST Ground System functions and associated terminology

Several dedicated Science Teams will be funded over the prime phase of the mission to execute the primary dark energy, exoplanet, galactic plane and NIR survey observing programs. In this period, the GO/AR program provides funding for analysis of public data and for pointed observations selected competitively. Operations costs and grants for the GO/AR program in the primary mission are fully included in the lifecycle costs.

### 3.8   Enabling Technology

The HgCdTe near-infrared detector arrays baselined for WFIRST have extensive heritage from HST, JWST, and on-going production of commercial off-the-shelf devices for ground-based observatories. WFIRST uses a 9x4 mosaic of these detectors; heretofore the largest near-infrared focal plane built for space so far has been a 2x2 array of H2RG detector arrays. Although scaling of the focal plane from a 2x2 array to a 9x4 array is not technology development, it is considered an engineering challenge so an EDU was started on the JDEM project and work continues under WFIRST. The EDU FPA includes a 6x3 silicon carbide (SiC) mosaic plate incorporating HgCdTe detectors, the Sensor Cold Electronics (SCE), including a repackaged version of the SIDECAR ASIC, and associated mounting hardware (see Figure 40). The full mosaic plate will be populated with flight-like detectors and qualification testing of the assembly will be completed during Phase A, mitigating the technical risk of scaling of a large NIR mosaic.

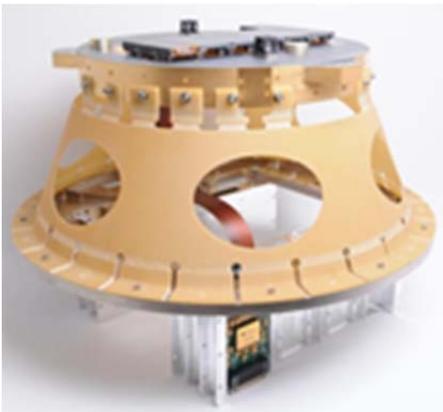

**Figure 40: The WFIRST EDU focal plane assembly**

Anticipating the need to demonstrate acceptable detector performance (in order to meet the scientific requirements) and yield (at an acceptable cost), the JDEM/WFIRST project initiated a detector demonstration program over 4 years ago. To date, two lots have been built, a more experimental one under the Detector Technology Advancement Program (DTAP), and a more WFIRST-specific lot under the FPA EDU detector build. These lots have provided test data showing improved detector performance over previous designs and these results are being used to guide mission design and optimization.

### 3.9   Concept of Operations

WFIRST is a tightly constrained mission: each of the WFIRST science objectives has unique constraints involving the field of regard, cadence, and S/C roll angles. We have therefore constructed an existence proof of a possible DRM1 observing schedule. **This is only an existence proof: the actual observing schedule will be determined at a time closer to launch.**

For the "existence proof" exercise, the following constraints were assumed:

- The total available observing time is 60 months.
- The supernova program is carried out over a period of 24 months with 22.5% duty cycle (total 5.4 months dedicated to supernovae), and the SN field is located in a continuously viewable region near an ecliptic pole (notionally $b = -37°$, $l = 270°$).
- The exoplanet microlensing program is carried out during ~ 2 month "seasons" when the Galactic bulge is in the field of regard. In order to measure the stellar microlensing curve, we require observations over the full season, except that brief interruptions for interleaved SN observations are acceptable. (These are not preferred because planetary events that occur during the SN interruptions are lost, but we accept this loss when there is no alternative.)
- The HLS program requires imaging in multiple filters as well as spectroscopy. The visits to each field are organized into two passes, during which observations are taken in each filter and each of the two GRS prism dispersion directions. Other drivers on the HLS program are to: (i) reduce zodiacal background; (ii) maximize overlap with the LSST footprint; and (iii) make as much of the footprint as possible visible from telescopes in both hemispheres. Since the zodiacal background is highest in the Ecliptic, these drivers compete. For this exercise, we have prioritized (i) and (ii), and done the best possible on (iii). WL observations are not allowed during the first 1 month of the survey. In order to enable repeat observations of each field to track drifts in calibration, the two passes on each field are separated by >1 month. Observations of regions with dust column exceeding E(B–V)=0.065 (Schlegel et al. 1998) are not allowed.

- The Galactic plane survey uses the same survey strategy (both exposure times and tiling) as the imaging portion of the HLS. The footprint is not yet defined; for scheduling purposes we have assumed for DRM1 the 1240 deg² region with $|b|$<1.72°. These need not be the final regions, but they are designed to ensure that all Galactic longitudes are accessed during the scheduling. This takes 5.3 months.

- The GO program uses time not allocated to other programs, and has a requirement of ≥10% of the total observing time. All celestial objects must be accessible at some point during the GO period.

- An observing strategy must be possible for survey start at any time during the year in order to avoid launch window constraints. (This is evaluated by considering 24 survey start times spaced by 0.5 months each.)

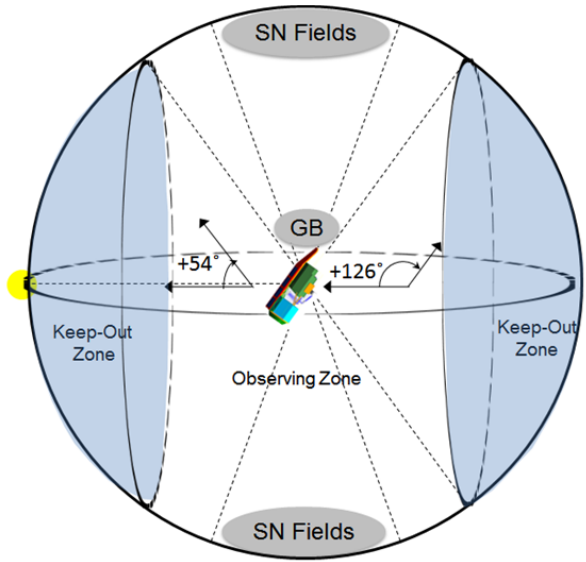

**Figure 41: WFIRST central line of sight field of regard. The white area in the center indicates the viewable sky within the allowable solar angle viewing constraints while the blue areas are outside of the allowable viewing constraints.**

For scheduling purposes, we assumed a 28.8 month HLS (3404 deg²) and 6 microlensing seasons.

Due to the large area for DRM1, the HLS footprint could not be contained within the LSST footprint while meeting other criteria and enabling other observing programs. Therefore a portion of the area (524—551 deg²) was moved to regions north of the Ecliptic but with Dec < 30°. Such regions are formally visible from Southern hemisphere sites (including LSST) but not part of the planned LSST footprint. They would also be visible with the Subaru/HSC imager, expected to come online within the next year. The portion of the area visible at elevation ≥30° from Hawaii (latitude 20° N) varies from 1585—1618 deg², depending on the time of survey start.

The available GO time is 6.1—6.7 months; the variation reflects the amount of microlensing time that collides with the supernova program, which depends on the phasing of survey start relative to the microlensing seasons (sometimes one season is interrupted and for survey start on February 21, a second season is partially interrupted).

The observing sequence for the case of science observations starting on May 21 is shown in Figure 42. Note that in this case the exoplanet microlensing seasons are not interrupted. The survey footprint, with 541 deg² north of the Ecliptic, is shown in Figure 43.

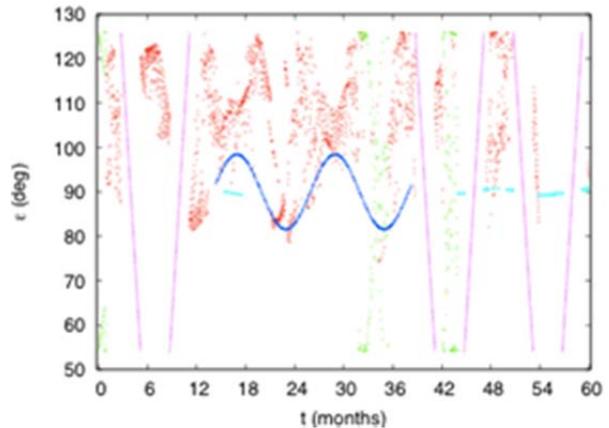

**Figure 42: The example WFIRST-DRM1 survey. The horizontal axis shows time $t$ from the start of observations, and the vertical axis shows the angle between the line of sight and the Sun (ε). The survey programs are color-coded: red for the HLS; green for the Galactic Plane survey; dark blue for the supernova survey; magenta for the exoplanet microlensing survey; and light blue for the GO program.**

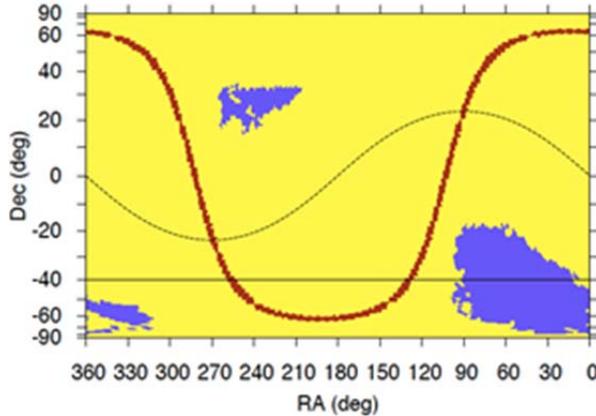

**Figure 43: The example DRM1 survey footprint for the case of observations beginning on May 21, in equal-area projection. The HLS region is shown in purple, and the Galactic Plane in red. The dashed line shows the Ecliptic, and the solid line shows the southern limit of accessibility at ≥30° elevation from Hawaii. The portion of the HLS north of the Ecliptic is not part of the planned LSST footprint.**

The distribution of the sky brightnesses for the HLS is shown in Figure 44. Despite the constraints on avoiding the Galactic plane and the dedication of large contiguous seasons to the exoplanet microlensing program, it was possible regardless of start date to keep the worst-case sky brightness to <1.07 times the reference value used for S/N calculations, and the median to <0.93 times the reference value.

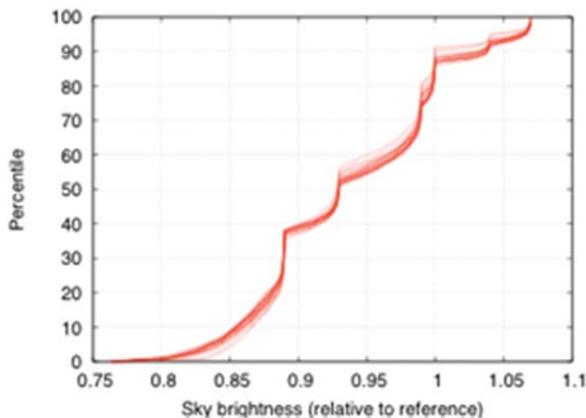

**Figure 44: The sky brightness distribution for the example WFIRST-DRM1 HLS. Each curve shows a different survey start date, and the vertical axis shows the cumulative probability of sky brightness lower than the specified value. The sky brightness is measured relative to the reference value used for S/N calculations ($\beta$=45° ecliptic latitude, $\varepsilon$=115° Sun angle, 1 AU at the mean of seasonal variations).**

### 3.10 Integrated Modeling

The project has conducted an early integrated modeling activity, using the IDRM as of the June 2011 interim report as the point design. The intent was to try to quantify what additional hardware may be required to meet the requirements for imaging point spread function stability and ellipticity. For DRM1, the main change from IDRM is in how the galaxy redshift spectroscopy is implemented; instead of fixed, opposed dispersion spectrometers, DRM1 uses a prism wheel, inserting opposed dispersers sequentially. The imaging mode is very similar in IDRM and DRM1. The optical train is a folded three mirror anastigmat with a cold pupil interface between the telescope and instrument, and a filter wheel just after the pupil. Therefore the imaging mode stability analysis conducted for IDRM and described below is very applicable to DRM1.

The modeling process uses three basic inputs, a detailed finite element mechanical model, a thermal math model, and a sensitivity matrix derived from the optical design, which includes both the effects of rigid body and surface deformations to model the net effect on both the wavefront error and the optical line of sight.

The line of sight stability is studied using a model of reaction wheel noise, which at Sun-Earth L2 is the dominant mechanical perturbation. As described in §3.6, the spacecraft includes passive vibration isolators on each reaction wheel. For each wheel, the line of sight stability is compared to the 25 milli-arcsec rms single axis requirement as a function of wheel speed. The initial results before optimization did not meet requirements, but the unstable structure was identified and with modest (~10kg) additional structural mass (behind the primary mirror and on the spacecraft panels to which the reaction wheels are mounted), and the instability was reduced, such that the requirement is met until very high reaction wheel speeds (>3000 rpm) are assumed. Figure 45 shows the worst reaction wheel results, with the rms line of sight error in milli-arcsec on the vertical axis and wheel speed (rev/sec) on the horizontal axis.

The wavefront stability is studied both for jitter (reaction wheel noise) and thermoelastic effects. We assumed a fast slew from one corner of the field of regard to the opposite corner, a motion of > 50 degrees, resulting in a somewhat different solar illumination on the solar array-sunshield. Including the model of thermal control heaters, the transient response was studied for several 150 sec science integration periods and compared to the requirement of 1.5 nm rms (TBR) wavefront stability during the HLS which would include weak lensing science images. Analysis of this transient case is still in progress but the results are expected to be < 1 nm for the very first 150 sec period, showing margin against the requirement.

We have also undertaken a wavefront sensing study to understand what opportunities the hardware allows for self-calibration (Jurling 2012). The linear optical model captures the field dependence of wavefront aberrations in a nonlinear optimization-based phase retrieval algorithm for image-based wavefront sensing. This approach allows joint optimization over images taken at different field points and does not require separate convergence of phase retrieval at individual field points. Because the algorithm exploits field diversity, multiple defocused images per field point are not required for robustness. Furthermore, because it is possible to simultaneously fit images of many stars over the field, it is not necessary to use a fixed defocus to achieve adequate signal-to-noise ratio despite having images with high dynamic range. This allows high performance wavefront sensing using in-focus science data. This approach demonstrated sub-thousandth-wave wavefront sensing accuracy in the presence of noise and moderate undersampling for both monochromatic and polychromatic images using 25 high-SNR target stars. Using these high-quality wavefront sensing results, we are able to generate upsampled point-spread functions (PSFs) and use them to determine PSF ellipticity to high accuracy in order to reduce the systematic impact of aberrations on the accuracy of galactic ellipticity determination for weak-lensing science.

Overall no exceptional or additional hardware has been found necessary to meet the required stability levels. Thermoelastic analysis and optical modeling showed that standard passive multilayer insulation and software controlled heater zones are adequate.

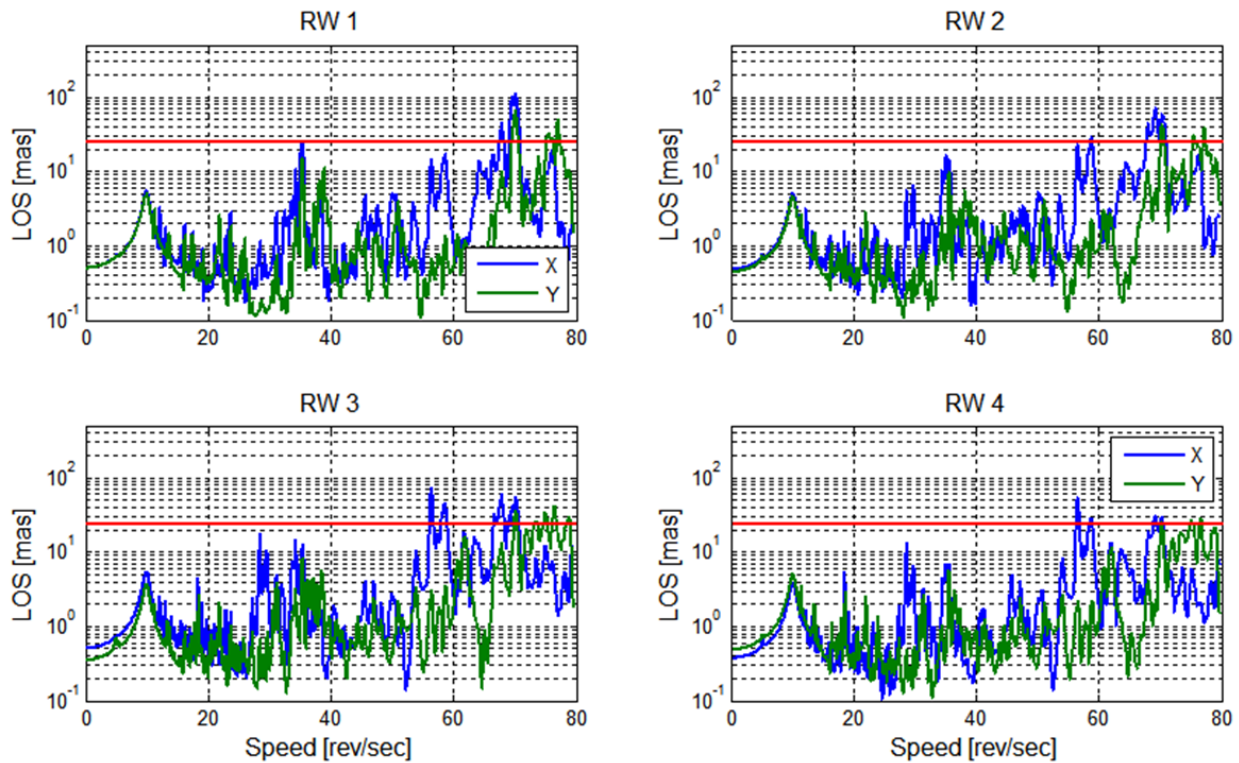

**Figure 45: Line of sight integrated modeling result, after optimization. Line of sight stability requirements (25 milliarcsec rms, red line) are met up to ~70 rev/sec reaction wheel speeds.**

## 3.11   Cost & Schedule

The WFIRST DRM1 configuration differs in two important respects from the JDEM-Omega reference mission configuration submitted to Astro2010. First, it incorporates an off-axis telescope instead of an on-axis telescope (as did the Interim Design Reference Mission). Second, it reduces the number of instrument channels from 3 to 1, eliminating the two spectrometer channels and performing spectroscopy with a moveable disperser. The WFIRST mission was estimated at $1.61B by the NWNH independent cost team. A Lifecycle Cost Estimate (LCCE) for the DRM1 mission is in process and is being developed using the same techniques (grassroots, parametric, and analogy) applied for the IDRM. Project Management, Systems Engineering, Mission Assurance, Integration & Test, and Public Outreach will be estimated using a grassroots approach and validated against analogous missions. Pre- and Post-launch Science will be projected based on the expected size of the WFIRST Science Announcement of Opportunity (AO). This estimate will include support for the AO-selected science teams in the mission development phase as well as the 5 years of operations. Additional funds for the General Observer and Archival Research Program will be included in the Science estimate. The payload and spacecraft will be estimated primarily using parametric estimates. These estimates will be constructed using master equipment lists (MELs) and historical cost databases and are adjusted for mass and complexity factors.

The development phase of WFIRST DRM1 spans 79 months, from preliminary design through launch (phase B/C/D). This development phase is preceded by 12 months to fully develop the baseline mission concept (Phase A) and several years of concept studies (Pre-Phase A). The observing phase (Phase E) of the mission is planned for five years. The development schedule is shown in Figure 46. The estimate is at a 70% Joint Confidence Level (JCL). The 70% JCL schedule allocates seven months of funded reserve to the WFIRST development.

The build-up and integration philosophy of the WFIRST observatory is based on the well-established practice of building small assemblies of hardware, thoroughly testing them under appropriate flight environments, and then moving on to the next higher level of integration with those assemblies. The WFIRST telescope and instrument will be developed and individually qualified to meet mission environments. The critical path of the mission is through the development of the instrument. The instrument is qualified prior to integration with the telescope. Following ambient check-out, the entire payload is tested at temperature and vacuum at GSFC to verify the end-to-end optical performance. The spacecraft is then integrated to the payload, and checked at ambient. Following a successful ambient checkout, a complete observatory environmental test phase is performed, including a repeat of the optical test with the fully integrated observatory. Upon successful completion of the observatory environmental test program, the observatory is readied for shipment to the KSC, where the launch campaign is conducted.

The parallel build-up of all of the mission elements, allows substantial integration activity to occur simultaneously, increasing the likelihood of schedule success. Because all of the major elements of the observatory (telescope, instrument, and spacecraft) are located at GSFC approximately two years before the planned launch, there is considerable flexibility in optimizing the schedule to compensate for variation in flight element delivery dates. Over the two year observatory I&T period, the Project will have flexibility to reorder the I&T work flow to take advantage of earlier deliveries or to accommodate later ones. Should instrument or telescope schedule challenges arise there are options to mitigate the schedule impacts by reallocating the payload level environmental test period. Should instrument or spacecraft challenges arise, there are options to modify the workflow and pull other tasks forward to minimize risk and maintain schedule. Additionally, the payload design includes access to the instrument volume even when attached to the spacecraft, allowing late access to the instrument during observatory I&T. The WFIRST observatory I&T flow is very achievable, given the planned schedule reserve and the opportunities available for workaround.

Fifty-four months are allocated to complete the WFIRST instrument, from the start of preliminary design, through the delivery of the instrument, not including funded schedule reserve. Additionally, the schedule includes seven months of funded reserve, further increasing the likelihood of executing the plan. An engineering development unit FPA is under development, with thermal vacuum testing planned for later this year. Thus even before Phase A commences, a large mosaic NIR focal plane will have been demonstrated, mitigating the risk of the element that this element would drive the mission critical path.

Early interface testing between the Observatory and ground system is performed to verify performance and mitigate risks to schedule success. Prior to payload integration, interface testing between the spacecraft and the ground system is performed. Immediately following payload integration to the spacecraft, end-to-end tests are performed, including the payload elements. These tests are performed numerous times prior to launch to ensure compatibility of all interfaces and readiness of the complete WFIRST mission team.

The DRM1 configuration of WFIRST requires no new technologies, has an implementation strategy that is conservative, proven and amenable to workarounds, and has a schedule based on continuously retiring risk at the earliest possible opportunity. This WFIRST DRM1 mission is executable within the cost and schedule constraints identified in this report, and is consistent with the New Worlds, New Horizons finding that WFIRST "…presents relatively low technical and cost risk making its completion feasible within the decade…".

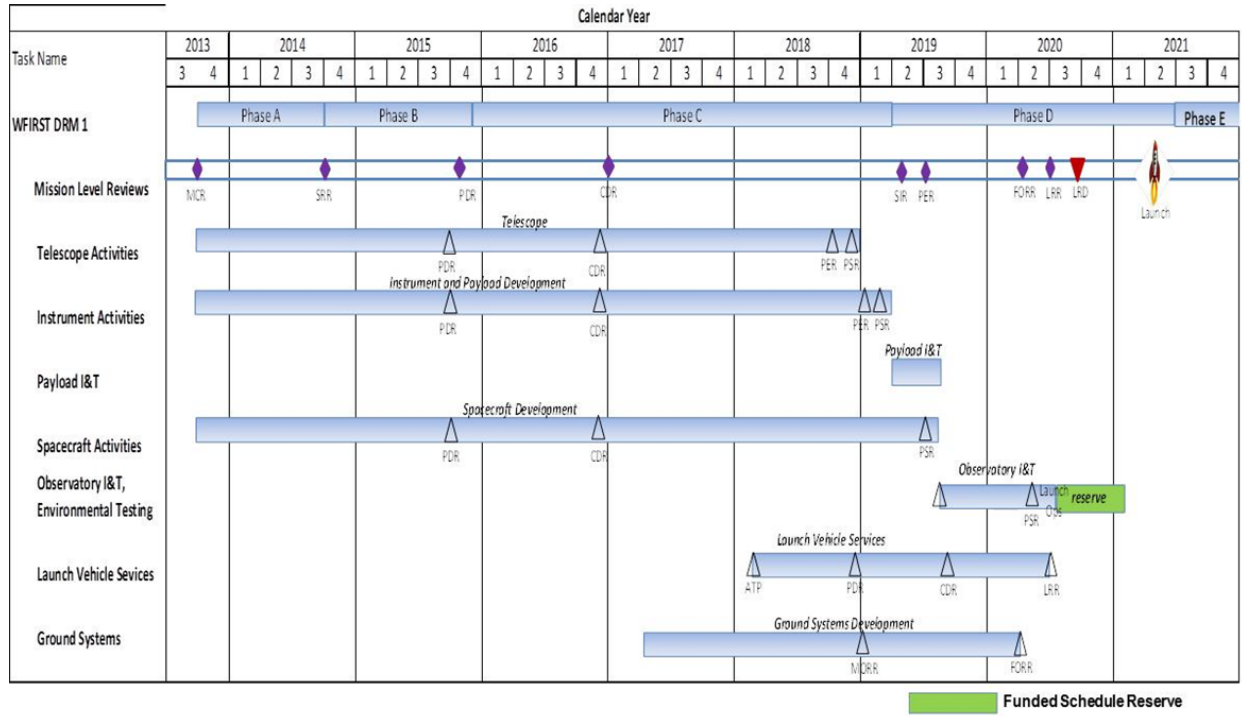

**Figure 46: DRM1 mission schedule**

# 4 DRM2: A 1.1 M TELESCOPE WITH A 3 YEAR MISSION

## 4.1 Introduction

In December 2011 the charge to the Science Definition Team was augmented, calling for two distinct design reference missions. The first was to continue along the path outlined in the interim design reference mission. The second DRM was to take into account the capabilities of ESA's Euclid mission, NASA's JWST mission, and the NSF/DOE LSST project, eliminating duplication, with an eye to reducing the cost of the mission.

In January 2012 the report of the NRC's *Committee to Consider a Plan for US Participation in Euclid* recommended a modest level of US participation in the Euclid project. In that report, the committee explicitly reported that WFIRST was not duplicative. In February and March of 2012 the SDT examined the question of duplication of capability at great length. The conclusions of those discussions are reported here.

WFIRST complements JWST in the same way that a wide-angle lens complements a telephoto lens. WFIRST has five times the number of imaging infrared detectors and roughly one eighteenth the collecting area. It covers a solid angle 100 times larger than that covered by JWST.

WFIRST and LSST are likewise complementary. LSST is an optical telescope, and is limited by atmospheric image blurring. WFIRST is an infrared telescope and is limited by diffraction. It goes to roughly the same depth as LSST (but with greater angular resolution) despite a very much smaller collecting area. And by virtue of working at wavelengths a factor of two longer than LSST, it goes to higher redshift.

The baseline program for Euclid (Laureijs et al. 2011) involves 6.25 years of imaging at optical wavelengths and infrared spectroscopy. It does not include an exoplanet microlensing program and such supernova discovery and photometry as it carries out will be incidental to its weak gravitational lensing program. The progressive degradation of charge transfer efficiency in Euclid's CCDs drives it to finish its weak lensing work before beginning other projects. Euclid will not have a general observer program.

Euclid's infrared imaging is subordinated to its optical imaging. Its sixteen infrared detectors spend less than a quarter of their time doing direct imaging, and then with 0.30" pixels, passing up the opportunity to do diffraction limited imaging.

There *was* substantial overlap between the Euclid infrared spectroscopic program and that of the WFIRST IDRM. This has been eliminated in both of the DRMs described here by shifting WFIRST's infrared spectroscopic program further to the red and carrying it out with longer integrations over a smaller field. The WFIRST GRS program is distinguished from that of Euclid in measuring the history of cosmic acceleration at earlier times than Euclid can reach and emphasizing full sampling of structure within the survey region over sparse sampling of a large volume.

The goals of the WFIRST and Euclid weak gravitational lensing programs are similar, but the approaches are quite different. Cosmological weak lensing measurements are likely to be limited by systematic errors rather than by numbers of galaxies observed. The systematic errors in the WFIRST approach are likely to be smaller than those of Euclid.

By virtue of carrying out weak lensing measurements at 3 different wavelengths with multiple passes at different roll angles in each filter, WFIRST permits extensive cross calibration that is not possible with Euclid's single filter. WFIRST's detectors do not suffer from charge transfer inefficiency. The pointing requirements for WFIRST will be less stringent than for Euclid. And because galaxies are less irregular in the infrared than in the optical, shape measurements will be more accurate. Finally, since WFIRST and Euclid are likely to be affected by systematic errors that are different in origin, they can be used to cross calibrate each other insofar as they observe the same parts of the sky.

The SDT was pressed to eliminate various observing programs from the WFIRST mission. As might be expected, each program was vociferously and persuasively defended by the individuals most closely associated with it. But in every case there was much broader support within the SDT. In the end the SDT decided not to eliminate any of the constituent programs.

The SDT was more successful in reducing the cost of the WFIRST mission. Eliminating specific programs produces little cost saving, since the attendant changes in hardware would be minimal. By contrast, substantial savings can be achieved by reducing the size of the spacecraft and the duration of the mission. The SDT's response to the call for a second DRM is a mission with a 1.1 meter diameter primary mirror and a baseline mission lifetime of 3 years.

This version of WFIRST (DRM2) cannot carry out all of the goals set forth in NWNH in its 3 year baseline mission. But it can carry out a substantial fraction of these in 3 years and all of them in an extended mission if it is outfitted with H4RG-10 infrared detector arrays rather than H2RG-18 arrays. Per unit time it would be almost as effective as DRM1, which continues to be based on H2RG-18 detectors. *But DRM2 can carry out a substantial fraction of the science of NWNH only with H4RG-10 16 Mpixel arrays.*

Since *not all* of the goals of NWNH can be carried out with a 3 year mission, it would be left to a time allocation committee to decide whether all of the programs should be cut be uniformly or if one or another were more compelling in light of intervening developments.

The decrease in the size of DRM2 permits an alternative configuration for the solar panels and sunshade which in turn allows for additional cost reduction.

We first describe the hardware differences between DRM2 and DRM1. Since the observing programs are very nearly the same we restrict ourselves to describing ways in which they must be modified to accommodate the smaller size and shorter duration of DRM2. The most serious of these are for the exoplanet microlensing for which the ability to turn relative masses into absolute masses depends strongly on the time span over which observations can be obtained.

## 4.2 Overview

In developing the DRM2 concept, the SDT and the Project Office began with the DRM1 concept and began looking for ways to both reduce the overall lifecycle cost and eliminate overlap between WFIRST and Euclid per the direction for the DRM2. There are many aspects of the DRM2 design and implementation which are unchanged from DRM1; these include:

- telescope design form (folded unobstructed TMA)
- single science channel
- focal prism wheel and filter wheel (including prisms and filter set (Table 14)
- auxiliary guider
- pixel scale (0.18"/pixel)
- gold mirror coatings
- field of regard (Figure 41)

In our effort to reduce the overall lifecycle cost, mass reduction was a key component due to rising launch vehicle costs for NASA payloads. Recent (vintage 2012) additions to the catalog of vehicles include the Falcon 9 series and Antares. The Falcon 9 has been deemed a technically feasible and lower cost alternative to the Atlas V for the launch of the WFIRST spacecraft.

There is little cost savings obtained by eliminating payload hardware as the hardware is shared across the science techniques. Therefore, we looked at ways to reduce the cost of the existing hardware including reducing the size of the telescope and using the higher pixel count H4RG detectors, allowing fewer detectors, a smaller focal plane, and a shorter optical focal length. Phase E was shortened to 3 years, eliminating 2 years of mission operations cost. This also allowed us to use a selective redundancy approach on the spacecraft, instead of the fully redundant DRM1 design, reducing the number of spacecraft boxes and reducing the overall mass. We also reevaluated the payload/spacecraft interface and determined that the smaller telescope allowed a side viewing configuration. This more compact design is structurally more efficient and also reduces the offset between center of pressure and center of gravity, allowing fewer momentum dumps throughout the mission. The side viewing telescope design is enabled by the use of a deployed solar array/sunshield similar to that of WMAP (see Figure 47).

Table 16 provides a comparison of hardware aspects that are changed for DRM2. Each of the observatory hardware elements is described below in more detail below.

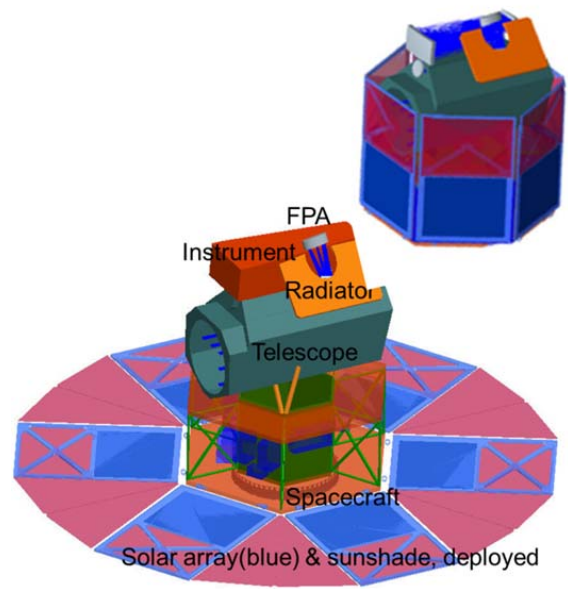

**Figure 47: DRM2 Observatory design showing both the stowed and deployed solar array/sunshield and the side viewing telescope. The instrument focal plane assembly can just be seen above the orange instrument housing.**

| | DRM1 | DRM2 | Rationale |
|---|---|---|---|
| **FPA type** | H2RG | H4RG | Additional pixels in H4RG allow fewer detectors |
| **FPA layout** | 9x4 H2RG | 7x2 H4RG | Maximize field of view |
| **Telescope aperture (m)** | 1.3 | 1.1 | Reduce telescope cost and size |
| **Active field of view size ( deg²)** | 0.375 | 0.585 | Maximize field of view with shorter mission |
| **GRS bandpass (μm)** | 1.5 - 2.4 | 1.7 - 2.4 | Does not duplicate Euclid |
| **Focal length (m)** | 20.63 | 11.46 | Scales with pixel size |
| **Optical axis relative to fairing axis** | parallel | perpendicular | More compact, fewer momentum dumps |
| **S/C redundancy** | full | selected | Reduces cost |
| **Sunshield type** | fixed | deployable | Shields "side-viewing" payload |
| **LV** | Atlas V | Falcon 9 v1.0 | Lower cost |

**Table 16: Differences in hardware configuration between DRM1 and DRM2**

### 4.3 Telescope

The smaller aperture DRM2 telescope is oriented sideways in the fairing; this lowers the center of mass and, in combination with the deployable solar array, allows significant mass savings while meeting launch stiffness requirements. The more compact arrangement requires additional fold mirrors before and after the cold stop at the exit pupil. Figure 48 shows a ray trace of the DRM2 payload.

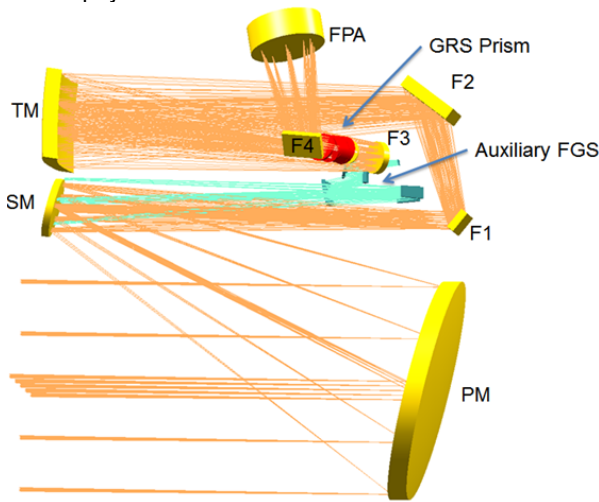

**Figure 48: DRM2 payload optical ray trace**

### 4.4 Instrument

The instrument is very similar to DRM1, with the principal difference being the switch to 4096 x 4096 pixel H4RG (10 μm pixel size) sensor chip arrays (SCAs). The bandpass of the galaxy redshift survey is shifted to 1.7-2.4 μm to probe a higher redshift portion of the expansion history than does Euclid. The optical block diagram (Figure 49) and channel layout (Figure 50) are similar to DRM1 (Figure 37 and Figure 38) but the larger number of pixels in the H4RG SCAs allow fewer detectors while providing a larger active area.

### 4.5 Calibration

There are no significant differences in the calibration approach for DRM2.

### 4.6 Fine Guidance Sensor

DRM2 eliminates the outrigger chips which are dedicated to fine guidance in imaging mode in DRM1 and instead uses the guiding window function implemented on the HnRG SCAs. Pointing data (centroid history of guide stars) will have to be supplied by the instrument electronics to the attitude control system. When spectroscopy data is taken (*i.e.* when prisms are inserted using the prism wheel), the auxiliary guider is used, just as in DRM1.

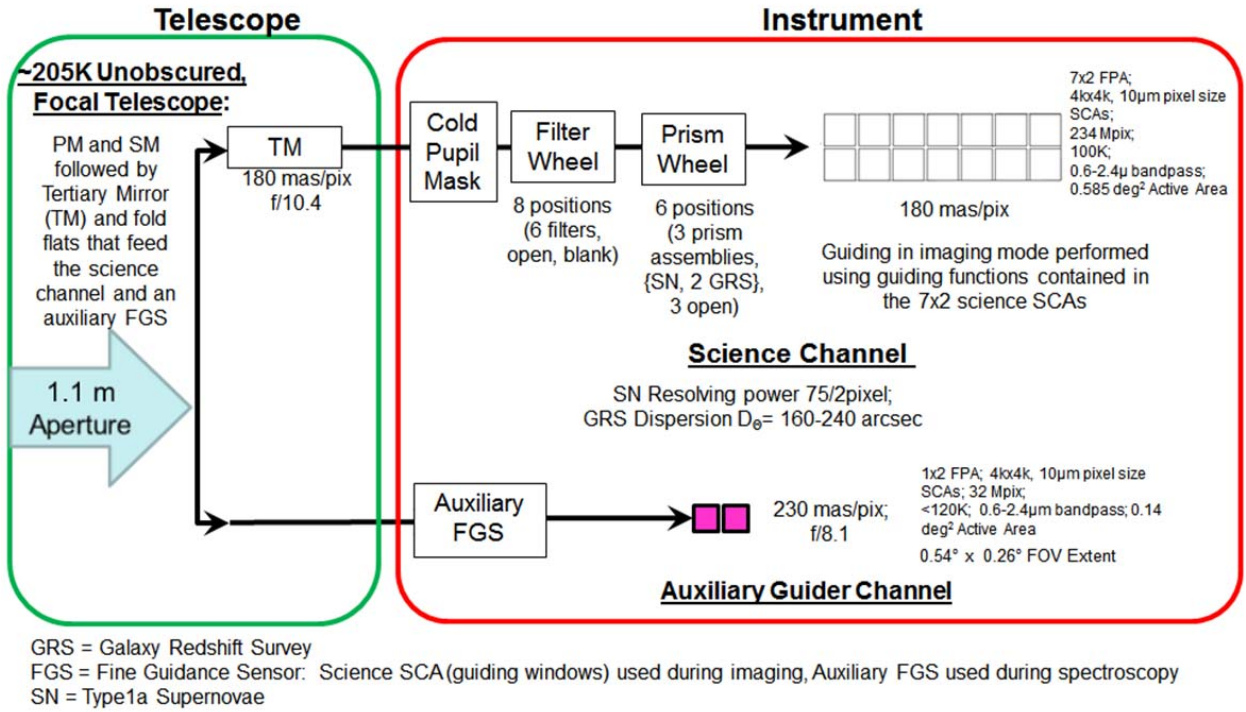

**Figure 49: DRM2 payload optical block diagram. Compare to Figure 37 for DRM1.**

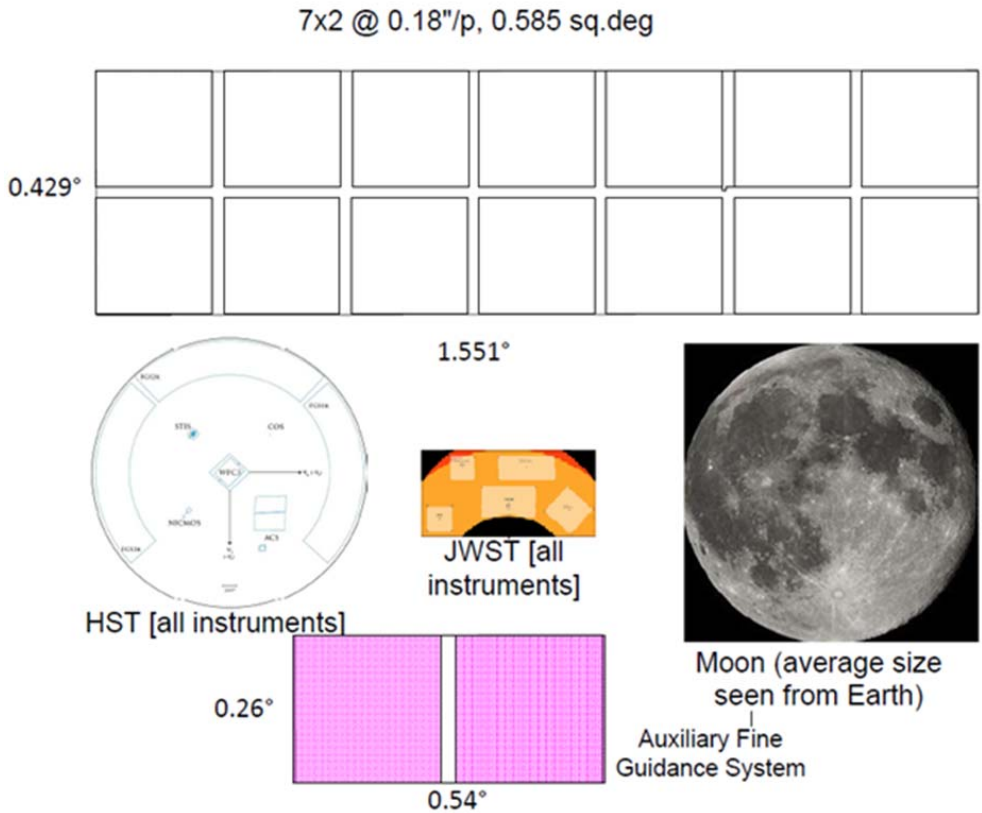

**Figure 50: Field layout as projected on the sky with field of view comparisons with the Moon, JWST and HST.**

This approach has several advantages and some potential drawbacks. In practice, we would perform a trade study to validate this approach and, if selected, perform risk reduction activities to retire the potential risks described below during preliminary project phases. Prior testing has shown no degradation to noise or other science imaging performance when this functionality is used.

Advantages:

- No outrigger devices are required; this reduces the number of working devices required and the size of optical elements due to reduced overall field of view.
- Much larger field of view available for guiding (factor of 14/2); this means either more stars are available at the same sampling interval, or one could increase the sampling interval using brighter stars.

Disadvantages:

- Instrument S/W must split the guiding data window outputs from the science data stream; with outrigger FGS devices this is not necessary.

### 4.7 Spacecraft

The spacecraft design for DRM2 follows the same design philosophy and heritage of the DRM1 spacecraft. The key differences between the two spacecraft are the deployable solar array/sunshield and the selected redundancy used as the DRM2 design life is only 3 years vs. 5 years for DRM1. *Structures:* The spacecraft structure is wider than the DRM1 spacecraft to accommodate the side viewing DRM2 payload. This configuration provides a lower center of gravity than the DRM1 spacecraft and provides a lower CP-CG offset that will reduce the required number of momentum dumps. The spacecraft supports a WMAP-like deployed solar array/sunshield that prevents the Sun from illuminating payload hardware during science observations (see Figure 51). This design has the solar array stowed against the spacecraft for launch and then deploys the array after separation from the launch vehicle. The smaller payload and more compact spacecraft offers significant mass savings (see Table 17), enabling the DRM2 observatory to launch on a Falcon 9 vehicle. *Attitude Control:* The attitude control system of the spacecraft is unchanged from DRM1. The redundancy is maintained in the reaction wheels, inertial reference unit and the star trackers. *Propulsion:* The hydrazine mono-prop subsystem is retained for DRM2, but the fully redundant system on DRM1 is reduced to a single string system. *Electrical Power:* The six deployable solar array panels (9 m$^2$) replace the fixed, body mounted array on the DRM1 spacecraft and provide a maximum of 2250 watts (at 0° roll and pitch) of output for a current best estimate average orbit usage of ~1100 W. Multi-layer insulation panels span between the solar array panels acting as the sun shield for the payload. The solar array is currently sized to provide full observatory power at EOL with 2 strings failed at the worst case observing angles. The power distribution unit design is the same as the DRM1 box but only a single unit is flown. *Communications:* The DRM2 communications subsystem is identical to the DRM1 design. *Command & Data Handling:* The command and data handling electronics retains the same design as the DRM1 subsystem but only a single C&DH unit is flown. *Thermal:* The spacecraft thermal design is a passive system, using surface coatings, heaters and radiators, consistent with the DRM1 design.

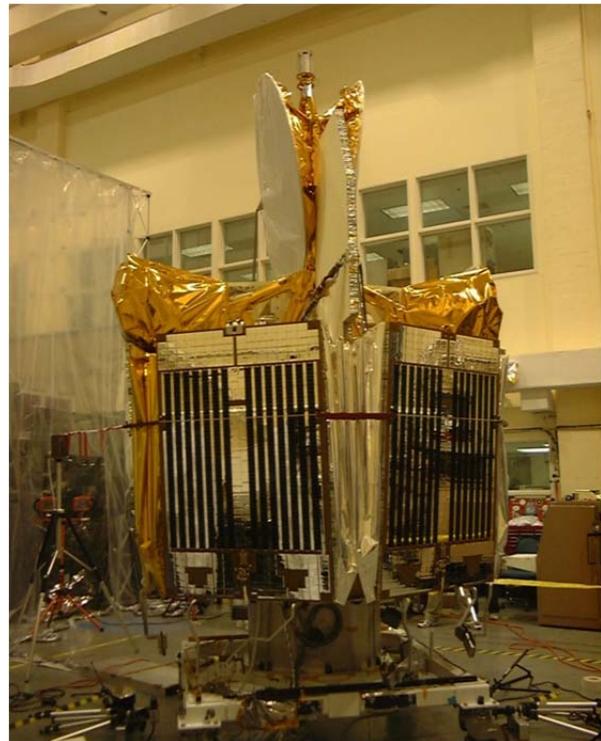

**Figure 51: Picture of the WMAP spacecraft with its deployable solar array/sunshield stowed. This design is the heritage for the DRM2 solar array/sunshield concept.**

| | Mass (kg) |
|---|---|
| Instrument | 169.8 |
| Telescope | 479.9 |
| Spacecraft | 787.1 |
| Observatory (dry) | 1436.8 |
| 30% contingency | 431.0 |
| Propellant | 112.0 |
| Observatory (wet) | 1979.8 |
| Falcon 9 v1.0 lift capability | 2530 |
| Margin (%) | 27.8% |

**Table 17: DRM2 mass breakdown**

### 4.8 Ground System

The Ground System design for DRM2 is identical to the DRM1 design.

### 4.9 Enabling Technology

For DRM2, the mission configuration makes it desirable to incorporate the next generation of higher pixel density near-IR detectors. The WFIRST Project has examined the possibility of using a 4K x 4K pixel, 10 µm pixel pitch detector arrays instead of the baseline H2RGs (2K x 2K pixel, 18 µm pitch). This improvement could potentially lower the cost-per-pixel by a factor of ~3 because the size of the die is similar but the pixel count is increased by a factor of 4. The smaller resulting focal plane assembly (for the same number of pixels) would also result in simplifications to the mechanical and thermal subsystems as well as simplifying integration and testing.

The WFIRST Project has completed a small test lot of these devices to examine the feasibility of this approch (see Figure 52). These devices use the H4RG 10 µm pixel pitch (H4RG-10) ROICs and mechanical mounting that have been developed for the USNO/NRL JMAPS Project. The mount and ROIC have been taken to TRL6 for JMAPS. The part that is new is putting an HgCdTe near-IR detector layer on the hybrid device instead of the Si detector layer used for JMAPS. Out of a total of 6 devices constructed, 5 are functional, and 4 perform well enough to merit detailed characterization. Initial testing on these devices is very promising, with the exception of a higher readout noise than desired. The cause of this is understood and mitigations are being studied, with the goal of incorporating some minor design changes into the next test lot. After the test lot demonstrates a successful "recipe", the next step is to construct a yield demonstration lot with a single recipe that validates the yield models (and therefore the flight development cost models). After this yield demonstration lot, the flight lot production can proceed. The H4RG-10 near-IR detectors are currently at TRL4-5, and the Project has developed a plan to achieve TRL6 before the PDR.

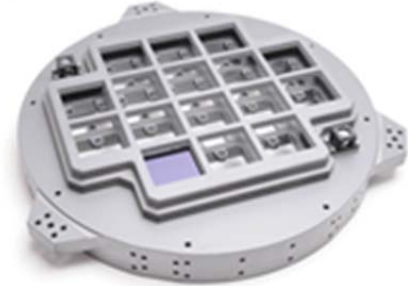

**Figure 52: A WFIRST developed EDU mosaic plate for an H4RG focal plane. One of the WFIRST developed H4RG detectors is shown in the lower left corner.**

### 4.10 Concept of Operations

We present here an "existence proof" of an observing strategy for DRM2. As for DRM1, this is only an example – the actual time allocation will be determined closer to launch. The "existence proof" strategy presented here tries to do as much of each of the WFIRST observing programs as will fit in the 36-month mission.

Relative to the DRM1 case, the following changes are made to the survey constraints for DRM2:

- The total available time is reduced to 36 months.
- The supernova program duty cycle is reduced to 15% (3.6 months spread over a total of 24 months).
- The HLS program was required to fit entirely within the LSST footprint, to avoid the need for obtaining additional (non-LSST) optical imaging coverage and the complexity in data processing due to a heterogeneous data set.
- The Galactic plane survey region is reduced to the 1000 deg$^2$ with $|b|$<1.39°, and the exposure times are set to 5×200 s per filter (instead of 5×247 s for the HLS), for a total of 3.4 months.

For scheduling purposes, we assumed a 16.4 month HLS (2390 deg$^2$) and 4 microlensing seasons. This was selected as an example of a particularly constrained program.

A program was constructed beginning at any time of the year. In every case, given the smaller HLS on DRM2, the HLS footprint was successfully contained within the LSST footprint. The portion of the area visible at elevation ≥30° from Hawaii (latitude 20° N) varies from 997—1021 deg², depending on the time of survey start. The available GO time is 3.6—4.0 months; the variation reflects the amount of microlensing time that collides with the supernova program, which depends on the phasing of survey start relative to the microlensing seasons (1 season is interrupted, and sometimes a second season is partially interrupted).

An example of such a strategy for the case of observations beginning on May 21 is shown in Figure 53. The HLS and Galactic plane footprints are shown in Figure 54.

Note in Figure 53 the interrupts of the microlensing program during the $t$ = 2.7—5.2 and 8.7—11.2 months seasons due to overlapping supernova observations.

The distribution of the sky brightnesses for the HLS is shown in Figure 55. Despite the constraints on avoiding the Galactic plane and the dedication of large contiguous seasons to the microlensing program, it was possible regardless of start date to keep the worst-case sky brightness to <1.06 times the reference value used for S/N calculations, and the median to <0.94 times the reference value.

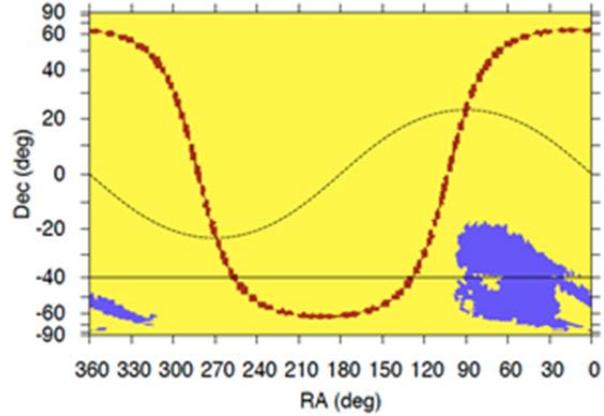

Figure 54: The example DRM2 survey footprint for the case of observations beginning on May 21, in equal-area projection. The HLS region is shown in purple, and the Galactic Plane in red. The dashed line shows the Ecliptic, and the solid line shows the southern limit of accessibility at ≥30° elevation from Hawaii.

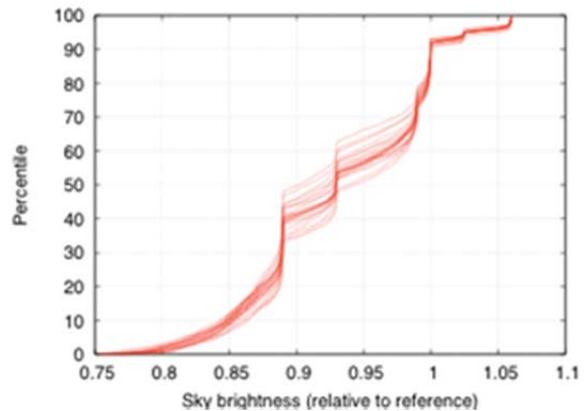

Figure 55: The sky brightness distribution for the example WFIRST-DRM2 HLS. Each curve shows a different survey start date, and the vertical axis shows the cumulative probability of sky brightness lower than the specified value. The sky brightness is measured relative to the reference value used for S/N calculations (β=45° ecliptic latitude, ε=115° Sun angle, 1 AU at the mean of seasonal variations).

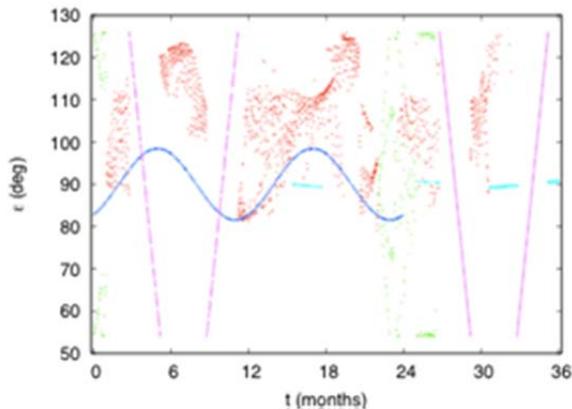

Figure 53: The example WFIRST-DRM2 survey. The horizontal axis shows time $t$ from the start of observations, and the vertical axis shows the angle between the line of sight and the Sun (ε). The survey programs are color-coded: red for the HLS; green for the Galactic Plane survey; dark blue for the supernova survey; magenta for the microlensing survey; and light blue for the GO program.

## 4.11 Cost & Schedule

The WFIRST DRM 2 concept is an innovative design that draws upon the mechanical configuration heritage from NASA Goddard Space Flight Center's Wilkinson Microwave Anisotropy Probe (WMAP) and builds on the JDEM Probe design developed in 2009. The three-channel WFIRST mission was estimated at $1.61B by the NWNH independent cost team. In 2009, NASA studied a reduced capability near-infrared survey

mission, the Probe, and based on an independent cost assessment in 2010, the Probe cost was approximately 40% lower than the configuration evaluated by NWNH. The optimizations that were applied to the Probe have been incorporated into DRM2 and these changes represent a significant cost savings compared to the original JDEM Omega or WFIRST baselines. DRM2 has gone a step further than the Probe in simplification, utilizing a single optical channel and single instrument/focal plane in the DRM2 payload. This should result in further cost reduction. With the initial concept for DRM2 now established, the Project will develop an LCCE for the DRM2 mission. Project Management, Systems Engineering, Mission Assurance, Integration & Test, and Public Outreach will be estimated using the same techniques as DRM1. Additionally, NASA HQ has requested that an Independent Cost and Schedule Estimate (ICE/ISE) be performed by the Aerospace Corporation. The ICE/ISE is expected to be completed in September 2012.

The development phase of the DRM2 WFIRST mission spans 72 months, from preliminary design through launch (phase B/C/D). This development phase is preceded by 12 months to fully develop the baseline mission concept (Phase A) and several years of concept studies (Pre-Phase A). The observing phase (Phase E) of the mission is planned for three years. The development schedule is shown in Figure 56. The estimate is at a 70% Joint Confidence Level (JCL). The 70% JCL schedule allocates 6 months of funded reserve to the WFIRST development.

The build-up and integration philosophy of the WFIRST DRM2 observatory follows the same practices as the DRM1 design. Fifty months are allocated to complete the WFIRST instrument, from the start of preliminary design, through the delivery of the instrument. Additionally, the schedule includes 3 months of funded reserve, further increasing the likelihood of executing the plan. With about a year and half between the delivery of the WFIRST instrument and launch, and given the additional flexibility that is inherent in the I&T flow, the WFIRST observatory I&T program has a high probability of executing within budget and schedule. Overall, the schedule developed for WFIRST DRM2 is very robust and achievable.

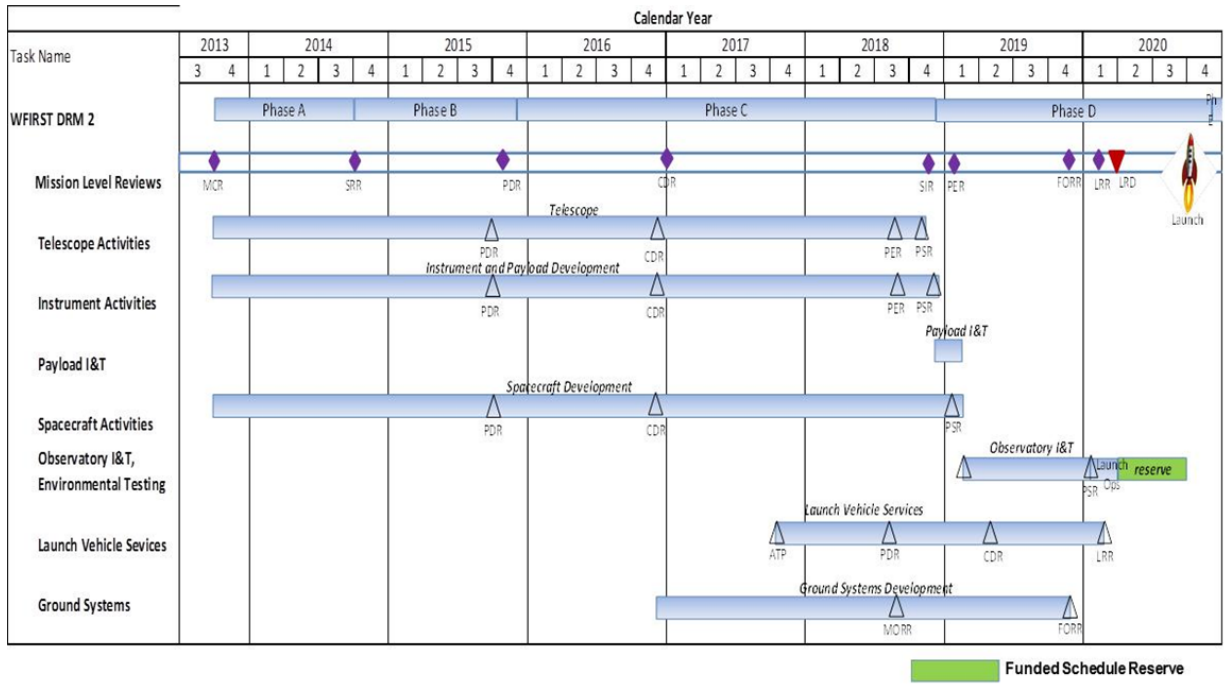

**Figure 56: DRM2 mission schedule**

# 5   CONCLUSION

This report describes the work of the WFIRST Science Definition Team and Project Office defining the WFIRST mission. The SDT has found that the mission objectives given in the NWNH Decadal Survey report are achievable. The top-level science requirements have been refined and lower level requirements derived.

Significant effort by the Project Office and the SDT has gone into refining the JDEM-Omega design concept that was specified as a template for the mission in NWNH. Three changes in what is now called DRM1 are particularly noteworthy. WFIRST now has a 1.3 m unobstructed off-axis telescope design. The imaging performance is superior to the 1.5 m obstructed design and there is adequate flight heritage to baseline such an approach. WFIRST now carries out imaging and spectroscopy separately with one instrument rather than in parallel with three. This both simplifies the design and permits greater program flexibility. Finally, the red limit of WFIRST's sensitivity has now been pushed redward to 2.4 microns from 2.0 microns. This has a positive net effect on **all** of the science goals for WFIRST, but in particular it extends the range of redshifts over which WFIRST measures the history of cosmic acceleration and permits the discovery of quasars to z=10.

The Project Office and the SDT have also produced a second design reference mission, DRM2, that goes a long way toward satisfying the science goals of NWNH at lower cost. DRM2 differs in three important respects from DRM1. It has a 1.1 m primary mirror rather than 1.3 m, it has a three year mission lifetime rather than five years, and it incorporates Hawaii 4RG-10 detectors rather than Hawaii 2RG-18 detectors. The first two of these reduce the cost of the mission, in part by allowing the use of a less expensive launch vehicle and by reducing the required level of hardware redundancy. The last of these recovers much of what is lost in going to a smaller primary and a reduced mission lifetime. But the Hawaii 4RG-10 detectors are not yet at an appropriate technical readiness level and timely development is needed to mature this technology for the DRM2 concept to be viable. Furthermore, a three year mission would not be long enough for DRM2 to achieve all of the NWNH goals.

The augmentation of the charge to the SDT midway through its report was not accompanied by an augmentation in the resources available to it. Most of its effort was therefore focused on developing DRM2. But at least some of the innovations adopted for DRM2 might be incorporated into DRM1, with savings in cost and improved performance. The most effective and efficient mission might lie somewhere in between. For either DRM, the scope of essentially all of the planned science investigations is limited by the available observing time. The mission designs have no expendables that would limit the lifetime to less than 10 years, and either form of WFIRST would be able to achieve substantially more in an extended mission.

Midway through the work of the SDT, Euclid was selected as ESA's M2 mission, now scheduled for launch in 2020. Where the Euclid mission is narrowly focused on measuring weak gravitational lensing and baryon acoustic oscillations, WFIRST has a very much broader program, with these elements absorbing less than half of its nominal mission. Moreover the approach to these two program elements is fundamentally different. Euclid maximizes sky coverage at the expense of redundancy and control of systematic errors. WFIRST emphasizes redundancy to control systematics. Weak lensing observations are carried out in three filters, not just one, and in multiple visits, permitting cross calibration. The measurement is made in the near IR instead of the visible, with more photons, and no effects from time-dependent detector charge transfer efficiency. The WFIRST galaxy redshift survey goes substantially deeper than Euclid's, measuring each Fourier mode near the BAO scale with S/N ~ 1, and it makes its measurements of BAO and structure growth at earlier times in cosmic history.

If anything the case for WFIRST is yet more compelling than it was when the decadal survey was released 2 years ago. It will measure the acceleration history of the universe, find the most distant quasars, map the Milky Way to the galactic center and beyond, complete the demographic census of exoplanet systems, and give astronomers from across the discipline the opportunity to carry out wide field infrared observations that have heretofore been impossible. It will enable extraordinary scientific advances with a mission that is technologically straightforward and moderate in cost.

The members of the Science Definition Team thank NASA for the opportunity to participate in the development of this mission. We hope that the present report conveys our excitement about and enthusiasm for the WFIRST mission.

**Appendix A    Microlensing Simulations: Ingredients, Methodologies, and Uncertainties**

Two sets of simulations were used in this report to estimate the yields of the WFIRST designs. The first uses the Manchester-Besancon microLensing Simulator (MaBμLS) described in detail in Penny et al. (2012). MaBμLS uses the Besancon galactic model (Robin et al. 2003) to predict the population of source and lens stars, and thus the microlensing event rates. We scale the resultant source star densities to match Red Clump Giant number counts from OGLE observations (Nataf et al. 2012), and the event rates to match the event rates measured from Red Clump Giants by the EROS, MACHO, and OGLE collaborations (Popowski et al. 2005, Sumi et al. 2006, Hamadache et al. 2006). The second simulation uses the methodology of Bennett & Rhie (2002), updated to apply to an IR survey (Bennett et al. 2010b). This method adopts the Holtzmann et al. (1998) luminosity function for the sources, scaled by the Red Clump Giant number counts from OGLE observations (Nataf et al. 2012), and uses preliminary 'all star' event rates from MOA-II observations (Sumi et al., in preparation).

The basic ingredients that are required for estimating the yield of a microlensing exoplanet survey are (1) a model for the population of source stars, (2) a model for the event rates and event parameter distributions for the microlensing events caused by the foreground host star lenses, (3) a model for the photometric precision for a given source star given its magnitude and including all relevant noise sources, (4) a model for the probability distribution of planets as a function of mass and semi-major axis, (5) simulations of planetary microlensing events and a criteria for the detection of the planetary signals.

We can then combine the microlensing event rate and source star density to a given magnitude with microlensing event simulations, to model the events that would be seen in a given survey, estimate their sensitivity to a planetary companion of a given mass and semi-major axis, and thus determine the rate at which planets would be detected across the inner Galactic bulge.

Unfortunately, for the regimes of interest for WFIRST, nearly all of these input model assumptions are relatively poorly constrained by empirical data, leading directly to relatively large uncertainties in the final yields. We will discuss the most important of these here, highlighting where the two simulation methodologies used to estimate yields for this report differ.

*Source star surface densities*

The magnitude distribution of source stars in the fields of interest has not been measured to the faint magnitudes that will be probed by WFIRST. Source star densities are therefore scaled by the observed number density of red clump giant stars seen in the field by the OGLE-III survey (Nataf et al. 2012). At low $|b|$, the extinction is too high to allow red clump giant star counts in the OGLE-III data, so we linearly extrapolate the counts from higher latitude fields. The two different simulations use somewhat different assumptions for the extinction and source star luminosity function, but we have adopted normalizations such that these differences are minimized.

*Microlensing Event Rates*

The microlensing event rates used here are estimated using measurements of the microlensing optical depth. There are two sets of microlensing optical depth measurements available in the literature. The first set, which we will call "Red Clump Giant" or RCG optical depths, are based on estimates that considered only events that appeared to have red clump star sources (Popowski et al. 2005; Sumi et al. 2006; Hamadache et al. 2006). A somewhat similar method is to calculate the optical depth and event rate with all events in which the light curve fit can determine the source star brightness. This has been done by the MACHO (Alcock et al. 2000) and MOA-I surveys (Sumi et al. 2003) prior to the current MOA-II analysis. We call these the "all-star" optical depths. The all-star optical depth analyses have yielded microlensing optical depths that are slightly larger than the microlensing optical depths from the red clump giant analyses. The reason for this is not completely understood, but has been reproduced by some Galactic models (Kerins, Robin & Marshall 2009).

The preliminary analysis of the 2006-2007 MOA-II data by Sumi et al. 2011 determined the optical depth and event rate toward each of the 22 MOA fields. The MOA analysis included 474 total events, and 90 of these appear to have red clump giant source stars. A preliminary analysis of microlensing optical depth and event rates has been carried out with both the full sample and the red clump giant sample. The new MOA analysis finds this same pattern as previously, with the red clump giant optical depth about 21% smaller than the optical depth measured in the MOA-II all-star sample for the 4 central MOA-II fields. Again, the reason for this difference is poorly understood, but this difference is consistent with previous measurements.

The primary advantage of the RCG optical depths is that the target stars are all resolved, and RCG are nearly standard candles and thus direct tracers of the Bulge density distribution. Originally, it was also thought that the events observed at the location of red clump giants would be unblended. However, this has turned out not to be the case. Both the MACHO (Popowski et al. 2005) and EROS groups (Hamadache et al. 2006) have found significant contamination due to lensing of main sequence stars that are blended with the red clump giants, but they argued that the effects of this blending nearly cancel. The OGLE group had somewhat higher quality light curves and was able to distinguish events with red clump giant sources from the blended imitators by means of light curve fits (Sumi et al. 2006), although there are some ambiguities. The other disadvantages of the RCG optical depths are that they have large uncertainties due to the small number of RCG events, and are determined in fields that are further away from the fields of interest for WFIRST.

The advantage of the all-star optical depths is that they are directly probing the event rates for the fainter stars that are more relevant to the WFIRST surveys. However, the fact that the higher optical depths are not well understood is a cause for concern. One thing that might contribute to this difference is the assumption that the Baade's Window luminosity function (Holtzman et al. 1998) applies to these inner fields at latitudes of $|b| \le 2$. If there are more main sequence stars per red clump stars at low latitudes than in Baade's Window, this would yield a higher apparent all-star optical depth, because the Holtzman luminosity function has been used to estimate the number of source stars. While such an error would artificially increase the all-star optical depth, it won't affect the predicted WFIRST event rate, because the same Holtzman luminosity function is also used to estimate the number of source stars, so these errors approximately cancel.

Ultimately, we have not used the all-star optical depths for our baseline estimates because this analysis is still preliminary. However, based on their experience with event rate and optical depth calculations for MACHO (Alcock et al. 2000; Popowski et al. 2005), OGLE (Sumi et al. 2006), and MOA-I (Sumi et al. 2003), the MOA group feels that the all-star analysis is likely the most reliable.

*Microlensing Event Simulations*

Both sets of simulations use the same requirement for the detection of the planetary signal, namely an improvement of $\Delta\chi^2 > 160$ in the fit to an event with the planetary perturbation relative to a single lens fit without the perturbation. This value is relatively conservative. For the free-floating planet yields, we use a higher threshold of $\Delta\chi^2 > 300$, and in addition we require at least 6 points that deviate by more than 3 sigma above baseline, in order to ensure that the event can be well characterized.

Even at fixed detection threshold and source magnitude, the detection efficiencies of the two simulations differ significantly. Although we have not identified the exact cause of these discrepancies, we believe that these are due to differences in the input assumptions used to simulate the ensemble of microlensing events. In particular, the distribution of lens properties, including the location of the lens stars relative to the source stars, the mass function of the lenses, and the detailed kinematics of the lenses, differs significantly between the simulations. Also, the detailed accounting of the noise sources also differ significantly between the two sets of simulations.

For planets near the peak of the detection sensitivity, the resulting difference in the raw yields is a factor of ~2 after normalizing to the same assumptions for the source number density and event rate, and thus we have scaled our final numbers by a factor of 1.47 in order to bring the two simulations into better agreement. For planets near the edge of the survey sensitivity, in particular planets with very small or very large separations, and very low mass planets, the discrepancies are substantially larger. This is at least partially due to the relatively strong dependence of number of detections with $\Delta\chi^2$ in these regimes. Thus, small differences in the assumptions and approximations needed to make these predictions result in large changes in the estimates of the number of detected planets. Predictions for the yield of habitable planets also suffer from these uncertainties, but are also sensitive to additional assumptions, such as the mass-bolometric luminosity relationship for stars in the bulge and disk, the age and metallicity of the stars in the bulge and disk, and the precise definitions for the mass and semi-major axes boundaries of the habitable zone. Therefore, the yields of habitable planets are even more uncertain, and we are unable to provide robust estimates for them at this time.

## Appendix B    Integral Field Unit Option

The knowledge that can be extracted from any survey is limited by the uncertainties in the data. These uncertainties can be statistical in nature, in which case additional data points (*e.g.* more SNe) will reduce the final error, or through systematic errors, which, by definition, cannot be reduced through the accumulation of additional, similar data. WFIRST is a powerful survey telescope, and will collect such large and deep surveys that we expect the final data products to be systematics limited. The only way to increase the science return in such a case is to reduce the systematic errors.

Systematic errors can be created by the instrument, and these are addressed by design and calibration. They can also be astrophysical in nature. For example, it is known that not all Type Ia SN are identical, and that sub-classes exist with different light curves. Ignoring this subtlety will eventually limit the strength of the scientific conclusions that can be drawn from a large data set that cannot remove this systematic uncertainty.

To address this issue, the SDT considered the potential inclusion of an Integral Field Unit Spectrograph (hereafter, IFU) to more accurately classify the observed SNe and reduce the systematic errors in the analysis.

The baseline supernova program uses a prism in a prism wheel to obtain slitless spectroscopy of each supernova. An IFU uses a compact splayed arrangement of mirrors to slice a small image (including, *e.g.*, a supernova, its host galaxy, and some background galaxies) into separate elements that each get dispersed. The resulting data cube of flux at each position and wavelength has many times higher signal-to-noise than a slitless spectrum with the same exposure time – or, equivalently, significantly reduced exposure times for the same signal-to-noise. In contrast, a slitless spectrum includes contributions from the full sky in each spectral bin, making it more difficult to isolate the faint SNe signal.

From a science perspective, the signal-to-noise gain with an IFU is quite important. The supernova program that can be accomplished with the lower signal-to-noise slitless spectroscopy is only sufficient to recognize a supernova as "Type Ia" and provide its redshift. With the dramatically higher signal-to-noise of an IFU the spectral features of the supernova can be used to:

1. Distinguish intrinsic color variations from the effects of dust (Chotard et al. 2011). Currently, these two sources of reddening and dimming are not distinguished at high redshift, so the mix of these two effects is assumed to stay constant over the redshifts studied. For the precision measurements of $w(z)$ it would be important to separate them, since both are important corrections in the distance modulus calculation. This systematics control remains important even in the redder observer wavelengths (out to 2.4 microns) that WFIRST can reach.

2. Improve the "standard candle" calibration of the Type Ia supernova. Bailey et al. (2009) showed that with spectral feature ratios of sufficient signal-to-noise the magnitude dispersion of SN Ia distance modulus in the rest-frame optical can be reduced from 0.16 mag to 0.12 mag dispersion. This is as good as the improvement in dispersion using restframe H band observations. With IFU spectroscopy, however, this distance modulus improvement can be obtained over a large redshift range (beyond $z = 1.7$), while the restframe H band photometry is only available to redshift $z = 0.25$ (and J band only to $z = 0.6$) even with an instrument observing out to 2.4 microns.

3. Compare the detailed composition and physical state of high-redshift supernovae to that of low-redshift supernovae. Type Ia supernovae are not all identical, but we can find spectroscopic matches of subsets of SNe Ia. If surprising cosmologies are inferred from supernova distance measurements over a range of redshifts it will be important to show that the effect is not simply an artifact due to a population of SNe Ia that is demographically drifting from one distribution of these spectroscopic subsets to another. With IFU spectroscopy it is possible to obtain the signal-to-noise sufficient to distinguish the different matching spectroscopic subsets.

4. Remove the K-correction systematics from the measurement. The slitless prism-based program uses only three filters over the whole redshift range studied, introducing the need for K corrections. The low signal-to-noise slitless prism spectra must then be combined to statistically remove any systematic biases in these K corrections (although this approach has not yet been studied to see what systematics will remain). With IFU spectrophotometry providing the lightcurves there would be no K corrections at all.

While the control of systematic uncertainties is the primary motivation for considering an IFU, its shorter exposure times can also be used to improve the depth of the survey and therefore the Figure of Merit. An example six-month-observing-time program has been developed (similar to the supernova program studied in detail by the JDEM Interim Science Working Group, see jdem.lbl.gov/docs/JDEM_ISWG_Report.pdf) that yields significantly improved FoM over the slitless spectrograph six-month-observing-time program described as the baseline program above. The systematic uncertainties discussed in the FoM section of this document may be significantly reduced with the IFU spectroscopy. One study has found that the FoM for the supernova measurement combined with WL and BAO will increase by 20% (see Figure 57).

*SDT Conclusion*

While the inclusion of an additional channel clearly involves some increases in complexity, there are potential practical advantages to using an IFU instead of a slitless prism approach. The IFU is expected to be less demanding in its pointing and stability requirements compared to the baseline slitless approach, and it is expected to eliminate the need for certain calibration instruments. However, we have not included the IFU in DRM1 or DRM2 because we wanted to keep the hardware complement to the minimum required to accomplish WFIRST's primary science goals.

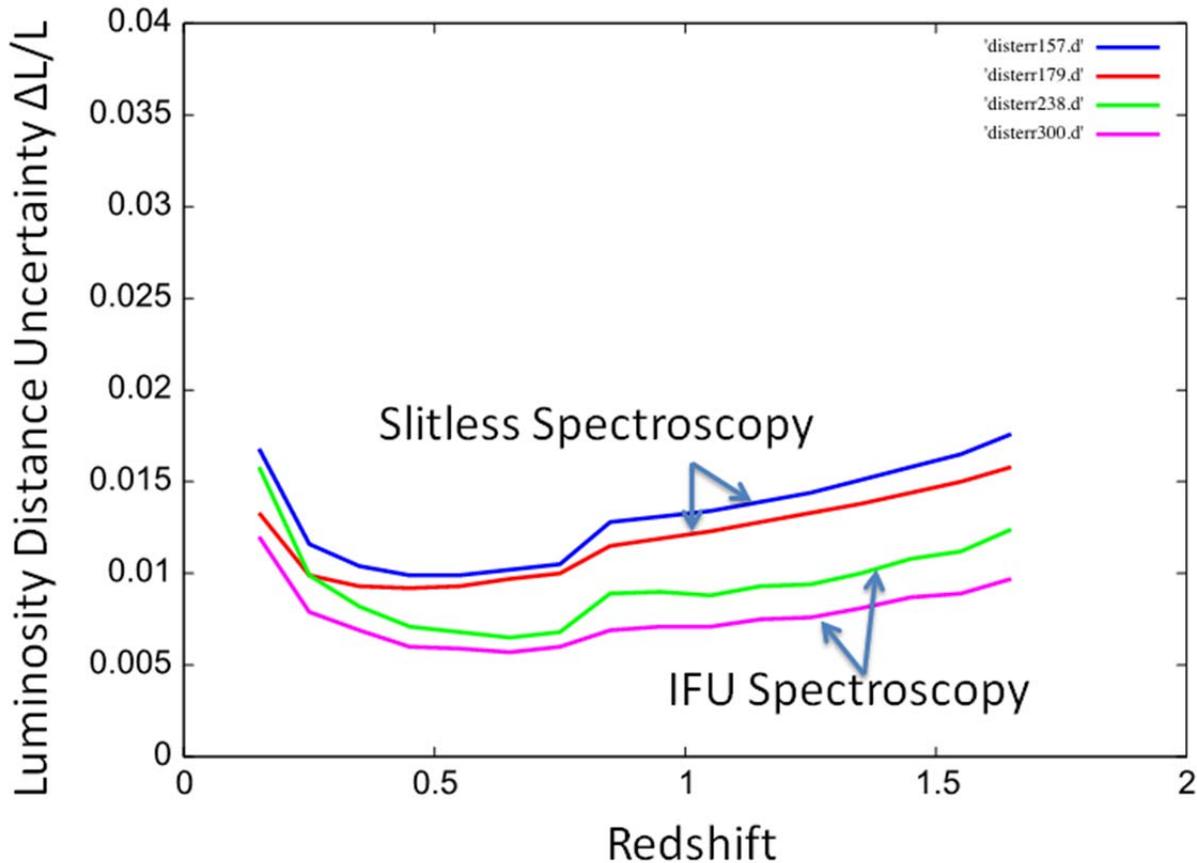

**Figure 57: Projected fractional errors on the luminosity distances in the WFIRST measurements of the expansion history of the universe using Type Ia supernovae as calibrated candles. The upper two curves are for the program described in Section 2.5.1 using slitless spectroscopy (DRM1 and DRM2 are essentially the same) and the lower curves are for spectroscopy with the IFU spectrometer as discussed in this Appendix. In each pair of curves, the upper curve corresponds to the conservative systematic errors and the lower curve to the optimistic systematic errors in Section 2.5.1.**

## Appendix C    References

**Appendix D    Acronym List**

| | |
|---|---|
| 2MASS | 2 Micron All Sky Survey |
| A&A | Astronomy and Astrophysics |
| ACS | Advanced Camera for Surveys |
| ACS | Attitude Control System |
| AJ | Astronomical Journal |
| ANGST | ACS Nearby Galaxies Survey Treasury |
| AO | Announcement of Opportunity |
| ApJ | Astrophysical Journal |
| ApJL | Astrophysical Journal Letters |
| ApJS | Astrophysical Journal Supplement Series |
| ASIC | Application Specific Integrated Circuit |
| AU | Astronomical Unit |
| BAO | Baryon Acoustic Oscillations |
| BEPAC | Beyond Einstein Program Assessment Committee |
| BigBOSS | Big Baryon Oscillation Spectroscopic Survey |
| C&DH | Command and Data Handling |
| $CaF_2$ | Calcium Fluoride |
| CCD | Charged Coupled Device |
| CDR | Critical Design Review |
| CFHTLS | Canada – France – Hawaii Telescope Legacy Survey |
| CMB | Cosmic Microwave Background |
| CMOS | Complementary Metal Oxide Simi-conductor |
| COBE | Cosmic Background Explorer |
| DE | Dark Energy |
| DES | Dark Energy Survey |
| DETF | Dark Energy Task Force |
| DOF | Degree of Freedom |
| DRM | Design Reference Mission |
| DRM1 | Design Reference Mission #1 |
| DRM2 | Design Reference Mission #2 |
| DSN | Deep Space Network |
| DTAP | Detector Technology Advancement Program |
| E(B-V) | Extinction (B-V) |
| EDU | Engineering Development Unit |
| EE | Encircled Energy |
| EELV | Evolved Expendable Launch Vehicle |
| ELG | Emission Line Galaxy |
| EOL | End of Life |
| EOS | Electromagnetic Observations from Space |
| EPO | Education and Public Outreach |
| ESA | European Space Agency |
| FGS | Fine Guidance Sensor |
| FoM | Figure of Merit |
| FoMSWG | Figure of Merit Science Working Group |
| FOR | Field-of-Regard |
| FOV | Field-of-View |
| FPA | Focal Plane Array |
| FSW | Flight Software |
| FY | Fiscal Year |
| GHz | Gigahertz |
| GLIMPSE | Galactic Legacy Infrared Mid-Plane Survey Extraordinaire |
| GNC | Guidance Navigation and Control |
| GO | General Observer |
| GP | Galactic Plane |
| GRS | Galaxy Redshift Survey (includes BAO & RSD) |
| GSFC | Goddard Space Flight Center |
| HgCdTe | Mercury Cadmium Telluride |
| HLS | High Latitude Survey |
| HST | Hubble Space Telescope |
| I&T | Integration and Test |
| IA | Intrinsic Alignment |
| ICE | Independent Cost Estimate |
| IDRM | Interim Design Reference Mission |
| IFU | Integral Field Unit |
| IMF | Initial Mass Function |
| IR | Infrared |
| IRAS | Infrared Astronomical Satellite |
| JCL | Joint Confidence Level |
| JDEM | Joint Dark Energy Mission |
| JWST | James Webb Space Telescope |
| KDP | Key Decision Point |
| KSC | Kennedy Space Center |
| L2 | Sun-Earth $2^{nd}$ Lagrangian Point |
| LCCE | Lifecycle Cost Estimate |
| LRD | Launch Readiness Date |
| LRG | Luminous Red Galaxy |
| LSST | Large Synoptic Survey Telescope |
| mas | Milli-Arc-Seconds |
| Mbps | Megabits per Second |
| MCR | Mission Concept Review |
| MEL | Master Equipment List |
| MNRAS | Monthly Notices of the Royal Astronomical Society |
| Mpc | Megaparsec |
| MOC | Mission Operations Center |
| MPF | Microlensing Planet Finder |
| NASA | National Aeronautics and Space Administration |
| NIR | Near Infrared |
| NIRSS | Near-Infrared Sky Surveyor |
| NRC | National Research Council |
| NWNH | New Worlds, New Horizons in Astronomy and Astrophysics |
| OTA | Optical Telescope Assembly |
| PASP | Publication of the Astronomical Society of the Pacific |
| PDR | Preliminary Design Review |
| Photo-z | Photometric Redshift |
| PSF | Point Spread Function |
| PSR | Pre-Ship Review |
| PZCS | Photo-Z Calibration Survey |
| QE | Quantum Efficiency |
| QLF | Quasar Luminosity Function |
| QSO | Quasi-Stellar Object (Quasar) |
| RCG | Red Clump Giant |
| RFI | Request for Information |
| RMS | Root Mean Square |
| RSD | Redshift Space Distortion |
| S/C | Spacecraft |
| S/N | Signal/Noise |
| SCA | Sensor Chip Assembly |
| SCE | Sensor Cold Electronics |
| SCG | Science Coordination Group |
| SDO | Solar Dynamics Observatory |
| SDT | Science Definition Team |
| SDSS | Sloan Digital Sky Survey |
| SED | Spectral Energy Distribution |
| SiC | Silicon Carbide |
| SIM | Space Interferometry Mission |
| SIR | Systems Integration Review |
| SINGS | Spitzer Infrared Nearby Galaxies Survey |
| SN | Supernova |
| SNe | Supernovae |
| SNR | Signal to Noise Ratio |
| SRR | System Requirements Review |
| SSR | Solid State Recorder |
| SUTR | Sample Up The Ramp |
| TBD | To Be Determined |
| Tbits | Terabits |
| TBR | To Be Resolved |
| TMA | Three Mirror Anastigmat |
| TRMD | Transition from Radiation to Matter Domination |
| UKIDSS | UKIRT Infrared Deep Sky Survey |
| ULE | Ultra Low Expansion Fused Silica |
| US | United States |
| vDC | Volts Direct Current |
| VISTA | Visible and Infrared Survey Telescope for Astronomy |
| WFIRST | Wide-Field Infrared Survey Telescope |
| WISE | Wide-field Infrared Survey Explorer |
| WL | Weak Lensing |
| WMAP | Wilkinson Microwave Anisotropy Probe |